# COMPUTATIONAL DESIGN OF SURFACES, NANOSTRUCTURES AND OPTOELECTRONIC MATERIALS

By

KAMAL CHOUDHARY

A DISSERTATION PRESENTED TO THE GRADUATE SCHOOL
OF THE UNIVERSITY OF FLORIDA IN PARTIAL FULFILLMENT
OF THE REQUIREMENTS FOR THE DEGREE OF
DOCTOR OF PHILOSOPHY

To my parents and teachers

## ACKNOWLEDGMENTS

First and foremost, I would like to convey my sincere gratitude to my advisor Prof. Susan B. Sinnott for making me a part of her research group and introducing me to various fascinating research areas of computational material science. Her continuous kind and patient support, suggestions and helpful discussions led me to overcome various obstacles in meeting my research goals. I have been also fortunate to get expert guidance from Drs. Simon R. Phillpot and Richard Hennig as secondary advisors. I thank other members of my supervisory committee: Drs. Scott Perry, Michele Manuel, Jennifer Andrew, and Youping Chen for offering their time and support. My special thanks to Drs. Asis Banerjee, David Tanner, Anthony Brennon and Paul Holloway for helping me relate the theoretical work with the experiments.

I also thank each member of the Florida Laboratory for Materials Engineering Simulation (FLAMES) group, both past and present, for their continuance support and guidance. In particular, I am thankful to Drs. Tao Liang, Aleksandr Chernatynskiy, Eric. Bucholz, Travis Kemper, Kiran Mathew and Anuj Goyal for the numerous scientific discussions, which were vital in molding my research.

TABLE OF CONTENTS

LIST OF TABLES

LIST OF FIGURES

Abstract of Dissertation Presented to the Graduate School
of the University of Florida in Partial Fulfillment of the
Requirements for the Degree of Doctor of Philosophy

COMPUTATIONAL DESIGN OF SURFACES, NANOSTRUCTURES AND OPTOELECTRONIC MATERIALS

By

Kamal Choudhary

August 2015

Chair: Susan B. Sinnott
Major: Materials Science and Engineering

Properties of engineering materials are generally influenced by defects such as point defects (vacancies, interstitials, substitutional defects), line defects (dislocations), planar defects (grain boundaries, free surfaces/nanostructures, interfaces, stacking faults) and volume defects (voids). Classical physics based molecular dynamics and quantum physics based density functional theory can be useful in designing materials with controlled defect properties. In this thesis, empirical potential based molecular dynamics was used to study the surface modification of polymers due to energetic polyatomic ion, thermodynamics and mechanics of metal-ceramic interfaces and nanostructures, while density functional theory was used to screen substituents in optoelectronic materials.

Firstly, polyatomic ion-beams were deposited on polymer surfaces and the resulting chemical modifications of the surface were examined. In particular, S, SC and SH were deposited on amorphous polystyrene (PS), and $C_2H$, $CH_3$, and $C_3H_5$ were deposited on amorphous poly (methyl methacrylate) (PMMA) using molecular dynamics simulations with classical reactive empirical many-body (REBO) potentials. The objective of this work was to elucidate the mechanisms by which the polymer surface modification took place. The results of the work could be used in tailoring the incident energy and/or constituents of ion beam for obtaining a particular chemistry inside the polymer surface.

Secondly, a new Al-O-N empirical potential was developed within the charge optimized many body (COMB) formalism. This potential was then used to examine the thermodynamic stability of interfaces and mechanical properties of nanostructures composed of aluminum, its oxide and its nitride. The potentials were tested for these materials based on surface energies, defect energies, bulk phase stability, the mechanical properties of the most stable bulk phase, its phonon properties as well as with a genetic algorithm based evolution theory of the materials to ensure that no spurious phases had a lower cohesive energy.

Thirdly, lanthanide doped and co-doped $Y_3Al_5O_{12}$ were examined using density functional theory (DFT) with semi-local and local functional. Theoretical results were compared and validated with experimental data and new co-doped materials with high efficiency were predicted. Finally, Transition element doped $CH_3NH_3PbI_3$ were studied with DFT for validation of the model with experimental data and replacement materials for toxic Pb were predicted.

CHAPTER 1
INTRODUCTION

## 1.1 General Introduction

The history of human evolution is generally classified based on advancements in material science, as in the Stone Age, Bronze Age, and Iron Age. One of the main challenges in advanced material design is that materials behave differently at microscopic length scales compared to macroscopic scale preventing a generalized and unified understanding of materials. While much information has been obtained regarding the macroscopic behavior of materials [1], information on the atomistic behavior of different materials is yet to be as well understood. Nonetheless, at present, new materials are being fabricated with atomic level engineering. Engineering materials generally contain imperfections that may occur unintentionally during processing or intentionally to make them more suitable for particular applications. This is because material imperfections or defects may have drastic effect on the mechanical, optoelectronic, thermal and magnetic properties of materials. A fundamental understanding of these imperfections can play vital role in optimizing the performance of the materials. These imperfections can be classified as:

**Point defects.** are created when an atom is missing or irregularly placed in the lattice structure. Point defects include lattice vacancies, self-interstitial atoms, doping/substitution impurity atoms, and interstitial impurity atoms.

**Linear defects.** are created when groups of atoms lie in irregular lattice positions and are commonly called dislocations.

**Planar defects.** are created when plane of atoms have interruption in their stacking sequence. Planar defects include grain boundaries, stacking faults and external surfaces. Nanostructured materials exploit the advantage of surface/volume ratio of planar materials.

**Volume defects.** are created due to non-periodicity of three dimensional portion in materials for instance-precipitates, dispersants, inclusions and voids. Volume defects can play important role in mechanical and chemical activity of materials.

Atomic scale modeling can provide key insights into processes related to the fabrication of thin films [2], electronic devices [3,4] and nanostructured materials [5,6]. Fortunately, with the help of classical and quantum mechanical methods such as empirical potential based molecular dynamics simulations and density functional theory calculations, respectively [7], it is now becoming possible to obtain a thorough understanding of the way in which defects, dopants, and microstructure on the atomic scale influence the properties and behavior of materials.

Thin film materials [2] are generally fabricated by depositing material-layers ranging from nanometers to micrometers in thickness. They are used as thermal, optical, or chemical coatings as well for making integrated electronic devices. Thin film applications such as ‘touchscreen’ electronic devices have benefitted from interfacial design that has optimized their mechanical, thermodynamic and opto-electronic properties. This design has included input from atomic-scale material modeling.

The concept of nanotechnology [5,6] was seeded by renowned physicists- Richard Feynman in his talk “There's Plenty of Room at the Bottom” [8] in which he described material design by manipulation of individual atoms. Some of the common forms of nanostructured materials include nanoparticles, nanowires and nanotubes and all are characterized by their unique surface to volume ratios.

Optoelectronic materials [9,10] are generally used to source, detect and control light. For designing new optoelectronic material elementary particle information (such as electrons, neutrons) coupled with atomic information can be great importance. Lasers, LEDs, and solar cells are some of the practical optoelectronic materials that are of great commercial importance. Subsitutional defects in optoelectronic materials and semiconductors play viral role in their performance.

### 1.2 Objectives and Directions

Atomistic calculations and simulations [7,11,12] are useful tools for material design as well as to explain fundamental physical phenomena, especially when combined with experimental measurements [13]. The primary focus of this dissertation is to elucidate the applicability of atomistic calculations and simulations, including density functional theory and classical molecular dynamics, to surfaces, interfaces, nanostructures, and optoelectronic materials. Computational predictions are validated where this is possible.

The remainder of thesis is organized as follows: a brief overview of the methodology and technical details are given in Chapter 2. Surface modification of amorphous polymers polystyrene (PS) and poly-methylmethacrylate (PMMA) are given in Chapter 3. Development of Al, $Al_2O_3$, AlN classical empirical potential under charge optimized many body potential and its application to metal ceramic interfaces are given in Chapter 4. The same potential is used to elucidate mechanical behavior of Al/$Al_2O_3$/AlN nanostructures in Chapter 5. Density functional theory based selection of dopants in $Y_3Al_5O_{12}$ and $CH_3NH_3PbI_3$ for optoelectronic applications are given in Chapter 6. Conclusions are given in Chapter 7.

# CHAPTER 2
# METHODOLOGY AND TECHNICAL DETAILS

## 2.1 General Introduction

Broadly, computational physics/material science [7] can be viewed as solving ordinary and/or partial differential equations governing the physics of a particular material phenomenon. Once the governing equations are established, various theories could be applied to the structure and imposed condition on the material based on classical physics, electrodynamics, fluid mechanics, quantum mechanics, quantum field theory and statistical physics. As the analytical solutions for complex differential equations may not be available, numerical methods are generally employed to solve the system of equations. In this section, a basic overview of the tools and methodologies used in the thesis are given, the details of which may be found from the references.

The governing equation of a material science problem can be formulated by defining the Hamiltonian of the system:

$$H = \frac{p^2}{2m} + V \tag{2-1}$$

which relates the contributions from the kinetic energy, defined in terms of momentum p and mass m, and potential energy term (*V*). A significant challenge in material science is determining the exact form for the potential energy term, whose exact form is unknown/ intractable for complex materials. For real materials, the *V* term can be theoretically determined based on the consideration that materials are composed of atoms that interact with one another via sub-atomic particles, the electrons. The atomistic description is generally determined with classical physics, and the electronic description with quantum physics.

In classical physics, system behavior is described by position $(r)$, velocity $(\dot{r})$ and acceleration $(\ddot{r})$ using Newton's equation of motion:

$$F(r_i) = -\frac{\partial V}{\partial r_i} = m\vec{a} = m\frac{d^2 r_i}{dt^2} \qquad (2\text{-}2)$$

where, *F* is force, *V* is the potential energy, *m* is mass, *a* is acceleration , $r_i$ is Cartesian/fractional coordinates array and *t* is time. In quantum mechanics, position, velocity and acceleration cannot be determined with certainty (Heisenberg's uncertainty principle). Rather the system behavior is described with wavefunction and energies (energy levels) associated with it. The main governing equation for quantum mechanics is Schrodinger equation (time-independent):

$$H\Psi(r) = E\Psi(r) \qquad (2\text{-}3)$$

where, $H \equiv -\frac{\hbar^2}{2m}\nabla^2 + V(r)$, *H* is the Hamiltonian, $\Psi(r)$ is the wavefunction which depends on the coordinates *r*, *E* is the energy associated with the wavefunction. The probability of finding a particle at position *r* is given by $|\Psi(r)|^2$. As mentioned above, these methodologies lead to partial differential materials which can be solved analytically (for very simple systems) and numerically (using high-performance computing facilities) for complex materials. In this thesis, numerical method based computational solutions with highly optimized software was used.

The atomistic modeling procedure can be summarized as: 1) build structure to mimic real experiments; 2) calculate energies and forces (classical physics) or wavefunction and energy levels (quantum mechanics) for a static/dynamic system of materials and 3) analyze the results to predict materials behavior.

## 2.2 Building Structures of Materials

Computationally materials' structure information is generally built with the help of the atomic coordinates and system size. Few common techniques that are relevant to this theses for building materials are discussed below:

### 2.2.1 Crystalline Materials.

An ideal crystal structure (for metals, ceramics, polymers or their combination) is defined as the repetition of identical structural unit in 3-dimenstional (D) space. The periodicity is defined as the lattice vectors and the identical structural units are termed basis sets. Crystal structures can be classified from their symmetry, named as space groups. For instance, is a face-centered cubic $\left(Fm\bar{3}m\right)$ aluminum metal structure can be defined with its lattice vectors of 4.05 Å in 3 dimensions and basis vectors as $R(x,y,z)$=[[Al,0.0,0.0,0.0],[Al,0,0.5,0.5],[Al,0.5,0.0,0.5],[Al,0.5,0.5,0.0]]. A wide list of crystalline materials database can be found in Inorganic Crystal Structure Database (ICSD) [14], American Mineralogist Crystal Structure Database [15], Materials project [16], Crystallography Open Database [17] and Inorganic Material Database (AtomWork) [18]. These databases are created generally based on experimental x-ray diffraction data. All the materials used here except in Chapters 3 and 5 are perfectly crystalline materials.

### 2.2.2 Random Structure Materials

Random structure materials can be generated in various ways such as a) increasing temperature of the material computationally and then quenching temperature (discussed later), b) using coarse grain models such as finite-extensible non-linear elastic (FENE) model especially for amorphous polymers, c) using quasi random structures and/or genetic algorithm for predicting materials. The bead-spring model and genetic algorithm are discussed below:

**Bead spring model.** It is generally used for investigating atomistic phenomenon of macromolecules/polymer chains. First, random beads inside a specific box size are created with the constraint that they should not overlap. Then, the amorphous structure can be relaxed using the bead-spring model (such as the finite-extensible non-linear elastic (FENE) model ) [19]. After relaxation, the monomers are placed between the beads to obtain the amorphous polymeric structure and are further relaxed using the empirical potentials. This model is used in Chapter 3.

**Genetic algorithm.** Genetic algorithm is used to predict global minimum energy structures. Other similar algorithms are simulated annealing, minima hopping, and Monte-Carlo methods. The evolutionary approach to structure prediction is modeled after the natural process. Each crystal structure is considered as an organism. In nature, the fitness of an organism is based on how well its phenotype is suited to its environment and, in particular, how successful it is in reproducing. We assign fitness value to the organisms based on their energy values and allow them to reproduce probabilistically based on those fitness. Fitness is defined as:

$$f = \frac{value - worstValue}{bestValue - worstValue} \tag{2-4}$$

The organisms are organized into generations. The algorithm proceeds by creating successive generations. The methods by which an offspring generation is made from parents are called variation operations or variations. They include operations which are analogous to genetic mutation and crossover. Each time the algorithm attempts to create a new organism using a variation, it must select parents (structures) using a selection method. We also try to improve on the biological analogy when possible. In particular, we would rather not let the most optimal solution worsen from one algorithmic iteration to the next. Therefore, a promotion operation is employed which promotes some number of the best organisms from one generation directly to the next. Also, mutations in nature are usually detrimental, hence information for mutating structures are also passed on during the progress of algorithm. This algorithm has been used in Chapter 4.

## 2.3 Energy Calculation Methods

The energy calculation methods used in thesis are: quantum mechanics based density functional theory method and classical mechanics based empirical potential method.

### 2.3.1 Density Functional Theory

Density functional theory (DFT) is a quantum mechanical modeling method used to investigate the electronic structure (principally the lowest energy state) of many-body systems, such as molecules and condensed phases. Density functional theory uses density (scalar) with 3 variables as compared to wavefunction (vector) for energy calculation leading to great computational efficiency. The governing equation for DFT is Kohn-Sham equation which can be derived from the many-body Schrodinger equation. It can be written as follows:

$$\left[-\frac{\hbar^2}{2m}\nabla^2 + V_{eff}\left[n(\vec{r})\right]\right]\varphi_i(\vec{r}_i) = \varepsilon_i\varphi_i(\vec{r}_i) \tag{2-5}$$

where $V_{eff}[n(\vec{r})] = V_{ext}[n(\vec{r})] + V_{e-e}[n(\vec{r})] + V_{xc}[n(\vec{r})]$, $V_{e-e}[n(\vec{r})]$ is the Hartree potential given by $V_{e-e}[n(\vec{r})] = -e\int \frac{n(\vec{r})}{\vec{r}-\vec{r}\,'} d\vec{r}\,'$ and $V_{xc}[n(\vec{r})] = \frac{\delta E_{xc}[n(\vec{r})]}{\delta[n(\vec{r})]}$

It is to be noted that the $\varphi_i$ and $\varepsilon_i$ are then the solution of Kohn-Sham orbitals rather than actual electronic wavefunction and energy. The interaction of each electron is described in the net field of all electrons, which is known as the mean-field approximation. The physical significance of exchange term is related to the same-spin electrons (Pauli's exclusion principle) and the correlation term is related to the fact that the interactions between electrons are interdependent/correlated in the mean-field approximation.

The Kohn-Sham equations are solved iteratively starting from an initial guess for charge density $n(\vec{r})$, calculate $V_{xc}[n(\vec{r})]$ and then solve Kohn-Sham equation for $\varphi_i$. Here, exchange is related to Pauli's exclusion principle. From this, the Kohn-Sham wavefunction charge density is calculated using

$$n(\vec{r}) = \sum_i^{occ} |\varphi_i(\vec{r})|^2$$

and then fed back to the KS equation to solve the equation iteratively until desired convergence criteria is reached. This process is called a self-consistent loop.

**Ionic and electronic minimization.** Ionic and electronic self-consistent loops are carried out during a typical DFT calculation for finding the minimum energy of a bulk structure. Forces acting on atoms are calculated using the Hellmann-Feynman method and the positions of the atoms are moved to minimize the forces on them until a particular force convergence criteria is reached. For each ionic minimization step, a number of electronic minimization is carried out using charge density $n(\vec{r})$ until a defined energy convergence criteria is reached. After the bulk structure has been minimized using the above procedure, defective and surface structures, which are discussed below, can be created and electronic minimization is carried out to calculate relevant properties of materials.

**Exchange correlation functional.** As the exact form of exchange-correlation functional is not known, approximations are made using already established theory. A simple approximation to $E_{xc}$ is the local density approximation (LDA) proposed by Kohn and Sham [4]. In LDA, exchange correlation energy per electron is approximated to the exchange correlation energy per electron of a homogeneous electron gas with the same density. Thus, LDA can work well for a system with a slowly varying electron density. The next approximation is generalized gradient approximation GGA is taken in which gradient of density is also taken into account for slowly varying density. Some of the common GGA functionals are Perdew –Wang (PW91), Perdew-Burke-Ernzerhof (PBE) [20].

One of the popular functionals are hybrid functionals, which are used to incorporate exact exchange term by combining Hartree-Fock and the density functional treatments of exchange. Correlation effects are still treated only within the density-functional scheme. Hybrid functionals have historically provided some of the most accurate energies and structures relative to LDA and GGA. Common hybrid functionals are Becke, Lee,Yang and Par ( B3LYP), PBE0, Hyde, Scuseria and Ernzerhof (HSE) [38, 39, 40, 41]. The inclusion of 25% of exact exchange for short distances in the HSE06 functional improves the band gap by recovering the derivative discontinuity of the Kohn-Sham potential for integer electron numbers.

**Basis set.** As mentioned above quantum mechanics is based on uncertainty and the associated probability theory, which requires a sample space consisting of all possible collection of events. The total sample space can be approximated by taking linear combinations of either analytical known solutions (such as for the hydrogen atom) or taking a series expansion and truncating it after a certain defined criteria. Basis sets can also be considered as a collection of vectors, which spans (defines) a space in which a problem is solved. Basis sets are crucial in determining the computational performance of materials. Two most common basis sets used are: delocalized (plane wave) basis and localized (Gaussian type) basis. Plane wave basis sets are generally employed for extended solids while for molecules Gaussian type orbitals give a better description of the system. The Gaussian basis sets are characterized by the number of Gaussian type orbitals taken into account to represent them and the plane waves are characterized by an energy cut-off. In this thesis only plane wave basis is used

**Pseudopotentials.** The pseudopotential is an effective potential constructed to replace the atomic all-electron potential (full-potential) such that core states (with very high binding energies) are eliminated and the valence electrons are described by pseudo-wavefunctions with significantly fewer nodes. This allows the pseudo-wavefunctions to be described with far more tractable, thus making plane-wave basis sets practical to use. Some of the common pseudopotentials are norm-conserving, ultrasoft, linearized augmented plane wave (LAPW) and projected augmented wave (PAW) pseudopotentials. In this thesis only PAW is used. A detailed description for DFT can be found in Ref. [7]. Some of the common softwares for DFT are: VASP, Quantum Espresso, Abinit and WIEN2k. DFT has been used for materials in Chapters 4 and 6.

### 2.3.2 Classical Empirical Potentials

While clearly successful at providing insights into the properties of materials, DFT calculations are typically limited to small system sizes (<1000 atoms) [21] and hence their ability to investigate some properties, such as energetics of interfaces due to lattice mismatch or mechanics of nanostructures. Classical empirical potentials/force-fields have been proven to describe materials properties at reasonable computational cost. In this method, mathematical models in terms of empirical potentials are developed to calculate energy as function of charges, interatomic distances, charge densities, bond angles, types of neighbor, and/or other cases. After formulating the empirical potential, either static minimization of structure or Newton's equations based dynamic simulation can be carried out. A detailed description of the molecular dynamics (MD) simulation procedure can be found in Ref. [11].

**Static energy minimization.** Energy minimization of the system with empirical potentials can be carried out by iteratively adjusting atom coordinates. Numerical Iterations are terminated when the stopping criteria for minimization (generally energy difference) is satisfied. At that point the configuration is supposed to reach to a local potential energy minimum. Some of the protocols for static minimization are steepest descent, Hessian-free truncated Newton algorithm and conjugate gradient methods. Interface structures in Chapter 4 were minimized with conjugate gradient algorithm.

**Dynamic simulations and time integration.** Time integration is needed to numerically evolve the system under the constraints of empirical potential. The most computationally expensive part of classical empirical potential based MD is evaluation of forces. Any method that requires more than one force evaluation per time step of integration (multistep) is considered inefficient. Some of the lower order methods [11] with easy implementation and stability are: leapfrog, Verlet, velocity Verlet method while predictor-corrector (multistep ) method is a much more accurate for large time steps but computationally much expensive. The velocity-Verlet algorithm can be given as:

$$v(t+\Delta t) = v(t) + a(t)\Delta t + O\left((\Delta t)^2\right) \tag{2-6}$$

$$r(t+\Delta t) = r(t) + v(t)\Delta t + \frac{1}{2}(\Delta t)^2 + O\left((\Delta t)^3\right)$$

In this dissertation, mainly the velocity-Verlet algorithm is used. Here, the time step for integration $\Delta t$ is of crucial importance to capture atomistic phenomenon such as lattice vibrations and typically used in the range of femtoseconds. Generally a typical MD simulations is characterized with the number of total time steps it was carried out.

**Statistical ensemble.** As mentioned above, MD simulations generate information at the microscopic level, including atomic positions and velocities. Hence, the conversion of this microscopic information to macroscopic observables such as pressure, energy, heat capacities, requires implementation of statistical methods. In order to connect the macroscopic system to the microscopic system, statistical averages are often introduced. A single point in phase space, denoted by *G*, describes the state of the system. An ensemble is a collection of points in phase space satisfying the conditions of a particular thermodynamic state. Three common ensembles are:

(1) *Microcanonical Ensemble* (NVE): used for an isolated system with N atoms, which keeps a constant volume (V) and a conserved total energy (E).

(2) *Canonical Ensemble* (NVT): used for a system with N atoms in a temperature bath. The volume (V) and the temperature (T) of the system are kept constant. In NVT, the energy of endothermic and exothermic processes is exchanged with a thermostat. Popular techniques to control temperature include velocity rescaling, the Nosé-Hoover thermostat, Nosé-Hoover chains, the Berendsen thermostat, the Andersen thermostat and Langevin thermostat.

(3) *Isobaric Isothermal Ensemble* (NPT): used for an isolated system with N atoms in a temperature and pressure bath. The pressure (P) and the temperature (T) of the system are kept constant.

**Periodic boundary condition.** Periodic boundary conditions (PBC) are necessary to relate the simulation to actual experiments. Bulk materials are simulated as being periodic in all the directions, while surfaces are modelled by breaking the PBC in a particular direction for a particular orientation of plane. Using PBC when a particle enters or leaves the simulation region, an image particle leaves or enters this region, such that the number of particles from the simulation region is always conserved. Molecules are represented having no periodicity any of the spatial (*x*,*y*,*z*) dimension.

Some of the common software for empirical potential based MD simulations are LAMMPS and GULP. A brief description of the empirical potentials used in this thesis is given below:

**Lennard-Jones potential.** The LJ potential is particularly suitable for studying inert gases. The LJ potential can be written as:

$$U(r_{ij}) = 4\varepsilon\left[\left(\frac{\sigma}{r}\right)^{12} - \left(\frac{\sigma}{r}\right)^{6}\right] \tag{2-7}$$

where, $\alpha$ and $\epsilon$ are parameters that are tuned/fitted for a particular material. Some other flavors of classical empirical potentials are: embedded atom potential (used mainly for metals), Buckingham potential (used mainly for covalent bonded materials), Tersoff potential (mainly used for covalent bonded materials), reactive empirical bond-order potential (REBO) (for covalently bonded materials especially hydrocarbons enabling bond-formation/breakage), charge-optimized reactive potentials such as ReaxFF, COMB (for heterogeneous bonding environments) and so on. In this thesis REBO and COMB potentials are mainly used.

**Embedded atom method (EAM).** The EAM method accounts for the behavior of an atom placed in a particular defined electron density. The method is proven to work well for describing systems consisting of metallic bonding. In EAM method, the potential energy of an atom *i* is given by:

$$E_i = F_\alpha\left(\sum_{i\neq j}\rho_\beta\left(r_{ij}\right)\right)+\frac{1}{2}\sum_{i\neq j}\phi_{\alpha\beta}\left(r_{ij}\right) \tag{2-8}$$

where $r_{ij}$ is the distance between atoms *i* and *j*, $\phi_{\alpha\beta}$ is a pair-wise potential function, $\rho_\beta$ is the contribution to the electron charge density from atom *j* of type $\beta$ at the location of atom *i*, and *F* is an embedding function that represents the energy required to place atom *i* of type $\alpha$ into the electron cloud.

A cutoff radius is generally needed in the EAM method as the electron cloud density is a summation over many atoms. For a single element system (A) of atoms, three scalar functions must be specified: the embedding function, a pair-wise interaction, and an electron cloud contribution function. For binary systems (AB), the EAM method requires seven functions: three pair-wise interactions (A-A, A-B, B-B), two embedding functions, and two electron cloud contribution functions. This method has been used in Chapter 4.

**Reactive empirical bond-order potential (REBO).** REBO is a non-charge based reactive (enabling bond-formation/breakage) empirical potential. For the simulations described in detail in Chapters 3 and 4, the short-range interactions were modeled using the second-generation reactive empirical bond order (REBO) potential for hydrocarbon systems developed by Brenner et al. [22]. REBO potential was further improved to accurately predict bond energies, bond lengths, and force constants. REBO potential was developed by Kemper et al. for C-H-S [23] system and has been used in this work.

The chemical binding energy $E_b$ for short-range interactions is determined as follows using the second generation reactive empirical bond-order (REBO) formalism:

$$E_b = \sum_i \sum_{j>i} \left[ V^R\left(r_{ij}\right) - b_{ij} V^A\left(r_{ij}\right) \right] \tag{2-9}$$

The functions $V^R(r)$ and $V^A(r)$ are pair-additive repulsive and attractive terms that represents all interatomic repulsions (such as core-core) and attraction from valence electrons, respectively. The quantity $r_{ij}$ is the distance between pairs of nearest-neighbor atoms *i* and *j* , and $b_{ij}$ is a bond-order between atoms *i* and *j* that is derivable from Huckel or similar level electronic theory [23].

**Charge-optimized many body potential (COMB).** COMB is a dynamic charge, reactive empirical potential. In this section the formalism of COMB potential is described. Under the third-generation COMB potential formalism, the total energy $U^{tot}(\{q\},\{r\})$ of the system consists of electrostatic energy $U^{es}[\{q\},\{r\}]$, short-range interaction energy $U^{short}[\{q\},\{r\}]$, van der Waals interaction energy $U^{vdW}[\{q\},\{r\}]$ and correction term $U^{corrt}[\{q\},\{r\}]$. Here, $\{q\},\{r\}$ are the charge and atomic coordinate arrays.

$$U^{tot}[\{q\},\{r\}] = U^{es}[\{q\},\{r\}] + U^{short}[\{q\},\{r\}] + U^{vdW}[\{q\},\{r\}] + U^{corrt}[\{q\},\{r\}] \quad (2\text{-}10)$$

The electrostatic part takes into account the charged nature of the atoms in the system.

The self-energy $U^{self}[\{q\},\{r\}]$ can be interpreted as the energy required to form a charge on an atom representing the ionization or affinity energy, and a correction function, termed as field effect representing change of electronegativity and atomic hardness of the atom within its environment. Short range energy is calculated with the bond energy comprising of pairwise attractive $V_{ij}^{A}(r_{ij}, q_i, q_j)$, an pairwise repulsive $V_{ij}^{R}(r_{ij}, q_i, q_j)$ terms based on Yasukawa and Tersoff potentials similar to the REBO potential. The long-range van der Waals (vdW) interactions, $U^{vdW}[\{r\}]$ are denoted by the classical Lennard-Jones.The correction terms, $U^{corrt}[\{q\},\{r\}]$ are generally used to modify the energy contribution from specific bond angles.

**Dynamic charge equilibration.** One of the advantages of COMB potential is that it can describe systems with different bonding environments (metallic/ionic/covalent) due to its charge equilibration methodology. In COMB, the charge on atoms are treated as a dynamical variable which can equilibrated/minimized based on electronegativity-equilibration (EE) principle. This principle was proposed by Sanderson [24] which states that "in a closed system of interacting ions at chemical equilibrium, the electron density is distributed so that electrochemical potential is equal at all atomic sites". COMB has been used in Chapters 4 and 5.

## 2.4 Properties of Materials

### 2.4.1 Bulk Properties

Bulk properties of materials can be described with properties such as lattice constants, radial distribution function for atoms, heat of formation and elastic constants [11]. DFT calculates lattice constant within ~ 2.5 % error and cohesive energy around ~20 % of experimental value, while classical empirical potentials are generally fit to these data. Heat of formation is defined as the energy required making formula unit of material from its constituents.

$$xA + y\frac{1}{2}B \rightarrow A_x B_y, \quad \Delta H_f = (x+y)E_{atom} - x\mu_A - y\mu_B \tag{2-11}$$

where, $x$ and $y$ are stoichiometric coefficients, $E_{atom}$ is energy per atom of a system. After equilibrium lattice constants are obtained, perturbation/stress-strain approach gives the elastic constants and modulus of a material. All of these quantities are directly comparable to experiments and act as tools for verifying an atomistic method. For empirical potentials, these values are generally fit, so they reproduces the experimental data as well. These properties are calculated for Al-O-N system in Chapter 4.

### 2.4.2 Defect Properties

Defects are very commonly encountered during material fabrication. These can be: 1) point defects (vacancies, interstitials) 2) line defects (dislocations), planar dislocations (grain boundaries, stacking faults).

Point defect energies are calculated as :

$$E_f\left[X^q\right] = E_{tot}\left[X^q\right] - E_{tot}[Perfect] - \sum_i n_i \mu_i + q[E_F + E_v + \Delta V] \tag{2-12}$$

where $E_f\left[X^q\right]$ is the total energy derived from a supercell calculation with one impurity or defect *X* in the cell, $E_{tot}\left[Perfect\right]$ is the total energy for the equivalent supercell containing only bulk perfect crystal. $n_i$ indicates the number of atoms of type *i* (host atoms or impurity atoms) that have been added to ($n_i$>0) or removed from ($n_i$<0) the supercell when the defect or impurity is created, and the $\mu_i$ s are the corresponding chemical potentials of these species. $E_F$ is the Fermi level, referenced to the valence-band maximum ($E_V$) in the bulk. Generally for empirical potential methods charged defect energetics are crucial to match with DFT data, hence for MD null-charge (i.e. $q$ = 0) defects are calculated, while for DFT methods charged defects can also be calculated. Line and planar defect energetics is also calculated in similar formalism. This property has been calculated in Chapter 4.

**2.4.3 Phonon Properties**

Phonons are important to describe dynamical properties such as infrared, Raman, and neutron scattering spectra; specific heat, thermal expansion, and heat conduction; electron-phonon interactions, resistivity and so on. The properties of phonons are described under a harmonic approximation based on a Taylor expansion of total energy about structural equilibrium co-ordinates

$$E = E_0 + \sum_{\kappa,\alpha} \frac{\partial E}{\partial u_{\kappa,\alpha}} u_{\kappa,\alpha} + \frac{1}{2} \sum_{\kappa,\alpha,\kappa',\alpha'} u_{\kappa,\alpha} \Phi^{\kappa,\kappa'}_{\alpha,\alpha'}(a) u_{\kappa',\alpha'} + \ldots \tag{2-13}$$

Here, $\kappa$ denotes a unit cell with N atoms, $R_{\kappa,\alpha}$ is the Cartesian coordinates for crystal in mechanical equilibrium, $\alpha = 1,2,3$ for *x*, *y*, *z*, $u_{\kappa,\alpha} = x_{\kappa,\alpha} - R_{\kappa,\alpha}$ denotes the displacement of an atom from its equilibrium position and $\Phi^{\kappa,\kappa'}_{\alpha,\alpha'}$ is the matrix of force constants. The force constants can be calculated from the both the empirical potential and DFT methods. From force constants data phonon density of states and dispersion relations for phonons can be calculated. This property has been evaluated in Chapter 4 for Al-O-N system.

### 2.4.4 Density of States (DOS) and Band-structure (BS)

The density DOS and BS are direct consequence of quantized propertied of quantum particles such as electrons, phonons. As these particles can occupy only certain allowed quantum-states, DOS represents the number of states per interval of energy at each energy level that are available to be occupied. DOS as a function of Energy, $E$ can be calculated by doing a sum of all bands over all energies of $k$-vectors in Brillouin zone , using the following equation

$$DOS(E) = \frac{1}{N_k} \sum_k \delta(E - E_k) \tag{2-14}$$

where $N_k$ is the total number of $k$-vectors covering the first Brillouin zone and $\delta$ is the delta-function . Here, it is important to note that simply using the delta function gives a histogram-like count for the DOS, instead of yielding a function of $E$. Since all electrical/thermal properties of solids depend on DOS as a function of $E$, a simple histogram is not as useful. Rather, the delta function is replaced with a Gaussian distribution. To reduce calculation time, and due to the fact that data points in the tail of the Gaussian distribution is much less important than the ones in the middle, any point that falls outside of 6$\sigma$ (can be defined to any value as desired by user) of the Gaussian function are neglected from the calculation. BS are generally used to represent availability of energy states at particular high-symmetry k-points (chosen for BS plot) in Brillouin zone. A high DOS at a specific energy level means that there are many states available for occupation. A DOS of zero means that no states can be occupied at that energy level. Band-gap value and nature of bands (steep/shallow) can be obtained from these plots. This property has been calculated for $Y_3Al_5O_{12}$ and $CH_3NH_3PbI_3$ in Chapter 6.

**2.4.5 Surface/Interface Properties**

To simulate the surfaces vacuum of few nm is added to the bulk structure in the direction normal to the surface. These surface planes are characterized based on the Miller indices for the plane. A generalized method for making surface structures exist in atomic simulation environment (ASE) and pymatgen tool [16].The surface energies are calculated as:

$$\sigma = \frac{E_{slab} - E_{bulk}}{2A} \tag{2-15}$$

where $E_{slab}$ is the total energy of the slab, $E_{bulk}$ is the energy of the bulk structure, and $A$ is the surface area.

Interface properties are measured in terms of the energy required to put them together, known as work of adhesion given by:

$$W_{adh} = (E_{Intf}^{Tot} + E_{slab-1}^{Tot} - E_{slab-2}^{Tot}) / A \tag{2-16}$$

where, $W_{adh}, E_{Intf}^{Tot}, E_{slab-1}^{Tot}, E_{slab-2}^{Tot}, A$ are the work of adhesion for the interface, total energy of the interface, total energy of first surface structure, total energy second surface structure, and surface area of the interface, respectively. Often straining of lattice is carried out to match the interface. A generalized algorithm to make interface with minimal mismatch was given by Zur et al. [25] . This property has been calculated in Chapter 4.

After the surfaces are made, they can wrapped in the direction perpendicular to the non-periodic direction of surface to make nanotubes. Nanostructured materials are discussed in Chapter 5.

### 2.4.6 Optical Properties

Optical properties of materials [26] can be obtained from energy level and wavefunction information obtained from DFT calculation. Probability of electron transition from the occupied energy level to the unoccupied energy level can be used to calculate the dielectric function of the materials. From the dielectric function reflectance, absorbance, transmittance, refractive index properties can be calculated. This property has been calculated in Chapter 6.

# CHAPTER 3
# ION-SPUTTERING INDUCED POLYMER SURFACE MODIFICATION

## 3.1 S, SH and SC Deposition on Polystyrene (PS)

### 3.1.1 General Introduction

Plasma induced chemical surface modification with ions, electrons, neutrals, photons and ions is used to deposit coatings, induce interfacial adhesion, introduce chemical functionalization, sterilize surfaces, ensure biocompatibility and in various other applications [27]. The interactions of sulfurous compounds with polymers include the vulcanization of rubber [28,29] and its reclamation, oil processing, [30] gas sensors [31,32], batteries [33,34], printing applications [35], organic electronics,[36,37] high strength materials,[38-41] and the etching of surfaces.[42] Recent advances in investigating polymer surface modification by ion beam deposition have revealed various mechanisms by which the surface chemistry and its properties are modified. This approach has been shown to be a good approach to experimentally isolate the role of mass-selected polyatomic ions on surface modification. These same ions are likely to be present in low-energy plasmas, but the complex environment of the plasma makes identifying their role difficult.

The motivation of the current work is to investigate the fundamental processes that occur during atomic and polyatomic particle-surface interactions, such as those that are likely to occur in low-energy plasmas and mass-selected ion-beam deposition. These processes include new product formation, quantifying the extent of particle penetration into, and chemical modification of, the surface, and determining and characterizing the molecules, radicals, and fragments that are sputtered from the surface. In particular, we consider incident S, SH, and SC onto a polymer surface of amorphous polystyrene (PS).

This work builds on previous computational and experimental work. For example, the deposition of polyatomic fluorocarbons and hydrocarbons on polymer surfaces was examined in classical molecular dynamics (MD) simulations by Hsu et al.[43-45] and Jang et al.[46] These prior studies noted that incident atoms tend to attach to PS chains through the replacement of native H or by capping the ends of broken chains. Complementary experimental work by Ada et al.[47] used mass-selected ion-beam deposition to examine the deposition of $SF_5^+$, $C_3F_5^+$, and $SO_3^+$ on PS. The results indicated the mechanism of fluorination of PS through spectroscopic analysis, and sulfur was detected on the surface for deposition at energies above 50 eV. In addition, at higher kinetic energies higher projectile dissociation and increased depth penetration of the incident ions was found to occur. The energy threshold for grafting sulfur onto the PS surface and dissociation of deposited ions was identified to be 3.38 eV.

In related work, Hanley et al.[48] examined surface morphology modification due to fluorocarbon and thiophene deposition at energies of 5-200 eV on silicon and PS surfaces with X-ray photoelectron spectroscopy (XPS) and atomic force microscopy (AFM). The root-mean square roughness of the PS was found to increase as the ion-beam energy increased. In particular, surface roughness varied from ~0.1 nm for the unmodified surface to ~0.2 nm after deposition of $C_3F_5^+$ with 25 eV energy and ~2 nm following $C_3F_5^+$ deposition with 100 eV of energy. This study also found that higher beam energies lead to greater penetration of the incident molecules and fragmentation of the ion beams. In related work, Karade et al.[49] used stopping range of ions in matter (SRIM) simulations to investigate how ion beams with energies of several keV can be used to tailor the mechanical properties of PS.

Here, classical MD simulations are carried out to examine the deposition of S, SH and SC beams on amorphous PS with external kinetic energies of 25, 50 and 100 eV. The forces on the atoms in the simulations are determined using the second-generation reactive empirical bond-order (REBO) potential for hydrocarbons[50] that has recently been extended to include sulfur[51] for the short-ranged interactions, and a Lennard-Jones (LJ) potential[52] for long-ranged interactions. This potential predicts the breaking and formation of chemical bonds through analysis of the instantaneous local environment of the atoms within the system. However, electronic charge is neglected, so the incident atomic and polyatomic particles are treated as reactive neutrals. Therefore, the simulations implicitly assume both that the ions are neutralized upon impacting the surface and that the cumulative effect of deposition has a negligible effect on charge accumulation on the PS. Additionally, electronic stopping[53] is not included in the simulations, which is known to be less important at low kinetic energies, such as are considered here.

The MD simulations are performed by integrating Newton's equation of motion with a third-order Nordsieck predictor-corrector algorithm[45]. It first calculates the forces acting on each atom, and then, subsequently, the position, velocity, acceleration and acceleration derivative of the atoms at future times. The chemical binding energy $E_b$ for short-range interactions is determined as follows using the second generation REBO formalism:

$$E_b = \sum_i \sum_{j>i} \left[ V^R(r_{ij}) - b_{ij} V^A(r_{ij}) \right] , \qquad (3\text{-}1)$$

where the repulsive pair term $V^R(r_{ij})$ is representative of the Pauli repulsion between electron clouds, and the attraction term $V^A(r_{ij})$ captures the atomic attraction due to valence electrons forming covalent bonds. The term $r_{ij}$ is the distance between atoms *i* and *j,* and $b_{ij}$ is the many-body, bond-order term. The long-range interaction is given by a standard LJ potential formalism:

$$V_{LJ} = 4\varepsilon\left[\left(\frac{\sigma}{r}\right)^{12} - \left(\frac{\sigma}{r}\right)^{6}\right], \qquad (3\text{-}2)$$

where $\varepsilon$ is the depth of the potential well, $\sigma$ is the finite distance at which the inter-particle potential is zero, and *r* is the interatomic distance.

The initial structure of the amorphous, syndiotactic PS surface slab considered is illustrated in Figure 3-1 A. First random beads inside a specific box size are created with the constraint that they should not overlap. Then, the amorphous structure is relaxed using the bead-spring model based on the finite-extensible non-linear elastic (FENE) model in the LAMMPS software package[19]. After relaxation, the monomers are placed between the beads to obtain the amorphous polymeric structure and are further relaxed using the REBO and LJ potentials. In the final step of placing the monomers between the beads, periodic boundary conditions are implemented in the direction of the chains. The FENE model was developed to study polymer melts, and it has been shown to compare well to experimental neutron spin-echo data[54]. It has the following energy function:

$$E = -0.5KR_0^2 \ln\left[1 - \left(\frac{r}{R_0}\right)^2\right] + 4\varepsilon\left[\left(\frac{\sigma}{r}\right)^{12} - \left(\frac{\sigma}{r}\right)^{6}\right] + \varepsilon, \qquad (3\text{-}3)$$

where the parameters have the following values: $K = 30.0$, $R_o = 1.5$, $\varepsilon = 1.0$ and $\sigma = 1.0$, in LJ units, are used, with $\varepsilon = 2^{1/6}\sigma$. An inter-chain LJ potential is also used, with a cutoff of $1.12$, and parameters $\varepsilon = 1.0$, and $\sigma = 0.9$. The following bond angle cosine potential was also used,

$$E = K[1 + \cos(\theta)], \tag{3-4}$$

with $K = 10.0$, to avoid small bond angles, which are found to be problematic during atomic monomer addition.

The PS substrate has dimensions of 7.5 nm x 7.5 nm x 6.5 nm with 27250 atoms. Periodic boundary conditions[45] are applied within the plane of the surface, while there is only free vacuum normal to the surface slab. Each polymer chain terminates at the pre-determined boundary and then effectively wraps upon itself such that there is no surface slab edge effect. A Langevin thermostat[55,56] is applied to 10% of the atoms in the outer portions of the surface substrate edges within the plane of the surface, and to 15% of the atoms in the bottommost portion of the substrate to effectively dissipate the excess heat that is generated during deposition and to maintain the temperature of the system at 300K, as shown in Figs. 3-1B and 3-1C. All other atoms are active, i.e., they evolve in time without any additional constraints.

Surface atoms were determined based on the highest coordinate values of the atomic centers in the direction normal to the surface after specifying a suitable grid throughout the surface slab. Surface roughness analysis was done in a manner that is similar to that carried out in Ref. [57]. In particular, the root-mean-square (RMS) roughness was calculated using the surface atoms' coordinates in the direction normal to the substrate surface ($Z_i$-for the $i^{th}$ atom) as follows:

$$R = \sqrt{\frac{\sum_{i=1}^{N}\left(Z_i - \bar{Z}\right)^2}{N}} \tag{3-5}$$

The RMS value of the pristine surface was found to be 0.48 nm, which is close to the experimentally measured value of 0.4 nm using AFM by Meyers et al.[58] A contour plot of the unmodified surface profile prior to deposition is provided in Fig. 3-1D.

In each deposition case, 100 atoms or dimers are deposited onto spatially randomized locations on the active portion of the PS substrate surface with normal incidence and kinetic energies of 25, 50, and 100 eV. The temperature of the substrate is monitored and sufficient time for dissipation of excess energy in the substrate is provided between deposition events. For S, SC, SH depositions at 25 eV and 50 eV of energy, the substrate is allowed to relax for 400 fs and 600 fs, respectively. For deposition at 100 eV the relaxation time is 1200 fs, 1000 fs and 500 fs respectively, for S, SC, SH. The fluence is 17 x $10^{14}$ ions/cm$^2$, which is comparable to experimental values.[47] The flux is in the range of 0.4 x $10^{24}$ ions/cm$^2$s.

### 3.1.2 S Deposition Results

The depth profiles for 25, 50 and 100 eV atomic sulfur beams are provided in Figure 3-2. The maximum depth attained for all cases is about 5 nm (i.e., around 77 % of the depth of the surface slab), resulting from only a few, random particles being able to penetrate deeply through the amorphous polymer. The incident atoms penetrate to an average depth of 2.0, 3.3, and 3.8 nm when their external kinetic energies are equal to 25, 50 and 100 eV, respectively. The maximum density of 3.32x $10^{18}$/cm$^3$ of deposited atoms is attained for incident energies of 50 eV. Not surprisingly, the incident atoms with 100 eV of energy penetrate to an overall larger extent than do those with 50 eV of energy, which, in turn, penetrate to a greater extent than at 25 eV. However, in the case of 50 eV a significant number of deposited atoms accumulate at a specific depth at about 1.9 nm, giving rise to a higher density of new products for this case.

Interestingly, the depth profile is skewed for the 25 eV deposition case while it is almost perfectly bell-shaped for the other deposition energies. This is because at 25 eV the deposited atoms mainly remain much nearer the surface than in the case of the higher beam-energies. The slight negative depth (on the order of a few angstroms) value reported in Figure 3-2 is due to the swelling of the PS surface as a result of the deposition process.

As expected, as the beam energy increases, the penetration depth increases, as illustrated in Figure 3-2. Not surprisingly, the amount of surface sputtering also increases as the kinetic energy of the beam goes up. The most common sputtered products were $C_2H$, $C_2HS$ and $C_3H_2S$ as shown in Figure 3-3. The most plentiful products formed during S-deposition regardless of incident energy are $CH_2$, $C_2H$, $C_4H_4S$, $C_6H_4S$, and the highest molecular weight product formed, $C_{41}H_{36}S_2$, occurs during the 50 eV deposition process. In general, the newly formed products as a function of beam energy can be summarized as $C_xH_yS_z$,: for 25 eV , $x < 48$, $y < 44$, $z < 4$; for 50 eV $x < 47$, $y < 44$, $z < 3$; for 100 eV , $x < 41$, $y < 34$ , $z < 3$.

In general, as the beam energy increases, the incident atoms tend to bond to a larger variety of sites on the polymer chains compared to the case of low beam energy, where the incident atoms bond to only a few specific sites on the polymer chains, as illustrated in Figure 3-4A. In particular, at low energies the incident S atoms tend to attach to the least crowded atom on the styrene phenyl group ($C_3$ ,$C_7$ in Figure 3-4B). The phenyl groups make up a greater fraction of the surface area of the PS substrate, which is what makes them a good target for attack despite their lower overall reactivity. An example of sulfur attaching to a PS chain is illustrated in Figure 3-4E. Therefore, at low beam energies, one can more readily predict the final location of the deposited atoms and the chemical modification of the substrate is more controlled. This variation in beam-attachment sites is due to the fact that the higher energies allow the reactants to more easily overcome reaction barriers leading to a larger variety of chemical products.

### 3.1.3 SH-Deposition Results

The depth profiles for SH-depositions are provided in Figure 3-5. As was the case for deposited S atoms, most of the deposited dimers remain near the surface during low-energy deposition and penetrate more deeply during high-energy deposition. In particular, the incident atoms generally remain within about 3 nm of the surface except in a few instances, as was also predicted in the case of atomic sulfur deposition. As the beam energy increases, the dimers tend to dissociate and the sulfur atoms penetrate more deeply into the substrate than do the hydrogen atoms, because the greater momenta of the heavier sulfur atoms. The maximum density of total deposited atoms (both S and H) is 1.65x $10^{18}$ /$cm^3$ and is achieved for 50 eV, which is comparable to the case for S-deposition. The depth profile of 50 eV is mostly bell-shaped implying a homogenized distribution of deposited atoms throughout the surface slab.

The molecular weights of the chemical products produced as a result of these reactions are shown in Figure 3-6A. The products formed are generally of the form $C_xH_yS_z$, where the values of x, y, and z depend on the incident energy as follows: x <

48, y < 48 , z < 2 for 25 eV; x < 21, y < 40, z < 3 for 50 eV; and x < 41, y < 40, z < 3 for 100 eV. Representative example products formed include $C_2H_2$, $C_5H_4S_2$, $CH_2S$, as indicated in Figure 3-6. Common sputtered products are $C_2H_2$, $C_8H_8$ and $C_5H_4S_2$, $CH_2S$, $C_5H_5S$, $C_6H_6S$ as shown in Figure 3-6B.

Similar to the atomic sulfur case, the incident atoms bond to a larger variety of atom sites on the PS chains as the beam energy increases (see Figure 3-4C). Specifically, at low incident energies the deposited atoms tend to attach to the least crowded atom on the styrene phenyl group ($C_1$, $C_6$, $C_7$ in Figure 3-4C). Hence, the deposited atoms not only attach to the phenyl groups, but also to the backbone carbons atoms. As the beam energy changes, the atoms are predicted to attach to a greater variety of sites, especially $C_2$ and $C_8$ in Figure 3-4C.

**3.1.4 SC-Deposition Results**

There are some distinct similarities between the results of the deposition of the SC and SH dimers. In particular, as the beam energy increases the dimers are more likely to dissociate and the sulfur atoms penetrate more deeply into the substrate than do the incident carbon atoms, as indicated in Figure 3-7. The maximum total atomic density of the incident beams (considering both S and C atoms) throughout the substrate obtained is 3.08x $10^{18}$ /$cm^3$ for 50 eV at about 1.85 nm, which is the highest of all the deposition densities predicted in this study. The density obtained here for 50 eV is highest amongst all depositions for reasons that are analogous to those given in the case of SH deposition. The dissociation of SC-dimers occurs less frequently than the SH-dimers, because of the nature of the chemical bond strength in these two incident molecules. Specifically, the bond strength of SC is 3.38 eV, while the bond strength of SH is 1.9 eV as predicted by REBO.

Few products are formed following deposition at 25 eV and the number of products and their molecular weights increase as the energy of the incident dimers increases. Once again, the products have the form of $C_xH_yS_z$, where x < 42, y < 50, z < 4 for 25 eV; x  <44, y < 45, z < 4 for 50 eV,; and x < 51, y < 45, z < 3 for 100 eV with representative example products of $CH_2S$ and $C_3H_2S$,$C_6H_4S$, as illustrated in Figure 3.8A. Here, the molecular weights of the chemical products are smallest compared to the other beam types because the SC dimer is the most saturated and hence least reactive incident particle considered. The most common sputtered products are $C_2H$, $C_2H_2$, $C_7H_5S$, and $C_4H_4S$ as shown in Figure 3-8B.

In the case of SC-depositions at 25 eV, the deposited atoms are predicted to preferentially bond to the carbon atoms within the phenyl group (i.e. $C_4$-$C_7$ in Figure 3-4A), with the majority bonding to the $C_6$ site, as illustrated in Figure 3-4D. The SC-dimers also bond to the $C_2$ site, which means they can separate the phenyl group from the polymer chain because the $C_2$ carbon atoms are already properly coordinated. Again, as the beam energy increases, the beam atoms attach to a greater variety of sites on the PS chain monomers, with the $C_5$ site the overall most preferred site for chemical reactions.

### 3.1.5 Discussions

The deposition cases considered in these simulations produce several interesting trends. With increasing ion-beam energies, the incident beam atoms or dimers chemically modify the PS surface in ways that strongly depend on the bond-saturation of the incident beams. However, based on depth profile and product yield analyses, 50 eV deposition with beams of atomic S (Figure 3-2B) is predicted to be the most efficient at producing PS surfaces with high total atomic densities. The greatest number of new products inside the substrate was formed for S deposition with 100 eV, while the smallest number was formed for SC with 25 eV of energy. On the other hand, the greatest amount of sputtering occurred for SC with 100 eV of energy, and the smallest amount occurred for S with 25 eV of energy. It should be noted that not all of the deposited beam atoms or molecules reacted with the surface. The probability of deposition for S was 95%, 95%, 100% for 25, 50 and 100 eV, respectively; 48%, 66%, 92% for SH for 25, 50 and 100 eV, respectively; and 57%, 82%, 95% for SC deposition for 25, 50 and 100 eV respectively. The corresponding sputter yields[59] were 0.96, 1.11, 1.19 for S-deposition; 0.81, 0.94, 1.23 for SH-deposition; and 0.82, 1.09, 1.32 for SC-deposition for 25, 50 and 100 eV, respectively.

Various product formation analyses inside the substrate (in the form of $C_xH_yS_z$) and the sputtered products after deposition describe the change in the chemistry of the surface. The most prevalent products formed on and within the surface slab are CHS, $CH_2S$, and $C_2HS_2$ for all the depositions considered. The highest molecular weight chemical product formed is $C_{41}H_{36}S_2$, which is formed for S deposition with 50 eV (Figure 3-3A) and its formation is as a result of the highly unsaturated form of the incident S atoms.

The selectivity of the chemical reactions in which the deposited atoms or dimers participate are predicted to depend strongly on the incident beam energies. However, the phenyl group carbons are targeted most often because of their higher surface area despite their lower reactivity. The attachment of the incident beam atoms with the allylic bonds of the PS is in fair agreement with the predictions of the classical MD simulations of Hsu et al.[43] and Jang et al.[39] This is unsurprising, given that the second-generation REBO plus LJ potentials were used in these studies as well.

The maximum density of deposited atoms throughout the substrate, which is obtained for S deposition with 50 eV of incident kinetic energy, is 3.32x $10^{18}$ /$cm^3$. For comparison, the maximum S-densities obtained at 50 eV for S, SH and SC depositions are 3.32x $10^{18}$/$cm^3$, 1.65x $10^{18}$ /$cm^3$, 3.08x $10^{18}$ /$cm^3$, respectively. Interestingly in all the cases considered the change in RMS roughness following deposition is negligible. In particular, the roughness is predicted to change from a minimum of 0.465 nm for SH deposited with100 eV to a maximum of 0.481 nm for SC deposited with 25 eV.

In this work we investigated the chemical modification of amorphous polystyrene through the deposition of 100 S atoms, SC dimers, or SH dimers with 25, 50 and 100 eV each of incident kinetic energy using classical molecular dynamics simulations with the second-generation REBO potential for short-range interactions and a standard Lennard-Jones potential for long-range interactions. The results allow for the determination of the ways in which these processes chemically modify the surface and produce new chemical products. Although limited to only three incident particles, this work has implications for the deposition of other sulfur-containing species, such as thiophene[60], which may fragment into these types of components during deposition. Taken as a whole, the results illustrate the richness of chemical activity that occurs during the chemical modification of polymer surfaces by hyperthermal reactive atoms and molecules.

## 3.2 $C_2H$, $CH_3$ and $C_3H_5$ Deposition on Polymethyl Methacrylate (PMMA)

### 3.2.1 Introduction

Polymethyl-methacrylate (PMMA) is a widely used industrial material for manufacturing implants and can be used as a lightweight, shatter-resistant alternative to glass[61]. Plasma processing is an important industrial approach for the chemical functionalization of polymer surfaces[62] that can make them wear-resistant[63] or can be used to deposit thin films[64-67], ensure biological compatibility[63], or optimize the hardness of the substrate material[28,31,32,42,68,69]. More generally, plasmas are used to modify polymer surfaces by adding a desired functionality. This, in turn, allows for control of surface reactions and the tailoring of surface properties for a particular environment[70]. In addition, adhesion to the modified surface can be aided by providing a means to form new bonds[71].

It is well established that low-energy or 'cold' plasmas[72] are typically dominated by polyatomic ions and neutral species[73]. Mass-selected polyatomic-ion beam deposition[69] is a related approach that can readily isolate the influence of specific polyatomics on surface modification and thin-film growth. There are four possible outcomes[74,75] to the deposition process, which can occur simultaneously to various extents: the polyatomic beams can be implanted beneath the surface, they can form new chemical bonds at the surface, they can bounce off the surface without ever forming a chemical bond, or they can sputter[76-79] the surface.

Li et al.[80] found that as a result of deposition of diamond-like carbon prepared using a multi-functional ion beam assisted deposition system, consisting of mainly three broad-beam Kaufman ion sources with 200 to1000 eV onto a PMMA substrate, the polymer's hardness and wear resistance significantly improved. Wijesundara et al.[81]

showed with X-ray photoelectron spectroscopy (XPS) that a chemical gradient on PMMA surfaces can be produced using $C_3F_5^+$ deposited at hyperthermal incident energies with varying fluences. Using XPS, Fuoco et al.[82] demonstrated that the nature of the incident species, $Ar^+$ versus $SF_5^+$, can have a significant influence on the sputtering of PMMA over a range of hyperthermal incident energies.

Insights into the atomic-scale mechanisms by which polyatomic particles modify polymer surfaces have been obtained from simulations.[76-79] For example, using molecular dynamics (MD) simulations Su et al.[83] predicted that varying the temperature of the PMMA substrate to its glass-transition temperature can lead to higher fragmentation during $Ar^+$ deposition at 1 keV. Thermal and hyper-thermal surface modification of PMMA was also previously examined computationally[84-86] and experimentally[87,88] to investigate ablation and scission of PMMA during ion and photon deposition induced surface modification. The laser-ablation behavior of PMMA was further investigated by Confronti et al.[89] using a hybrid of MD and Monte Carlo simulations to separate the thermal and chemical pathways of decomposition. Among a wide range of  plasmas, acetylene[90] is an attractive choice because of its highly reactive carbon-carbon triple bonds[91]. In addition, the effects of radiation on polyethylene and cellulose was examined using MD simulations by Polvi et al.[92], while Bringa et al.[93] investigated thermal spikes and sputter yields in Lennard-Jones solids. MD simulations were further used by Reinhold et al.[78] to investigate hydrogen reflection at amorphous carbon surfaces following low-energy deposition (0.1-50 eV); the reflection coefficient was predicted to be larger at  lower energies.

Previous MD simulations of hydrocarbon (HC), such as $C_3H_5$ and $CH_3$, and fluorocarbon (FC), such as $C_3F_5$ and $CF_3$, polyatomic deposition on polystyrene (PS) at 50 eV incident energy[43] predicted that backbone chains are modified significantly more than the phenyl groups. These simulations further predicted that smaller HC polyatomics chemically modify the PS to a greater extent than larger HC or FC polyatomics. Additional simulations[94] examined surface modification and etching by Ar, and predicted that small changes in the structure of the substrate surface, namely amorphous PS, poly ($\alpha$-methylstyrene) PαMS, or poly(4-methylstyrene) (P4MS), can drastically change the sputtering outcomes of deposition.

In the present work the chemical modification of PMMA with beams of $CH_3$, $C_2H$, $C_3H_5$ or hydrogen deposited with 4, 10, 25 and 50 eV of kinetic energy is investigated with classical MD simulations and a many-body, reactive potential. The objective is to quantify the associated chemical modification of the surface as a function of the chemical structure, size, and molecular complexity of the incident polyatomics in addition to their kinetic energies.

The MD simulations are carried out by integrating Newton's equation of motion with a third-order Nordsieck predictor-corrector algorithm[95]. It first calculates the forces acting on each atom, and then, subsequently, the position, velocity, acceleration and acceleration derivative of the atoms at future times. The chemical binding energy $E_b$ for short-range interactions is determined as follows using the second generation reactive empirical bond-order (REBO) formalism.

The initial structure of the amorphous, syndiotactic PMMA surface slab considered is illustrated in Figure 3-9. It is constructed using the following procedure.

First random beads inside a specific box size are created with the constraint that they should not overlap. Next, the amorphous structure is relaxed using the bead-spring model based on the finite-extensible non-linear elastic (FENE) model that is present in the LAMMPS software package[19]. Following relaxation, the monomers are placed between the beads to obtain the amorphous polymeric structure and are further relaxed using the REBO and LJ potentials. Lastly, periodic boundary conditions are implemented in the direction of the chains. The bond length and bond angle after bead-spring FENE relaxation are taken into account. Thereafter, there is a slight change in terms of chain orientation after the all-atom relaxation.

The PMMA substrate has dimensions of 7.5 nm x 7.5 nm x 7.5 nm with 28803 atoms. Periodic boundary conditions[95] are applied within the plane of the surface, while there is only free vacuum normal to the surface slab. Each polymer chain terminates at the pre-determined boundary and then effectively wraps upon itself such that there is no surface slab edge effect. A Langevin thermostat[55,56] is applied to 10% (i.e. 0.75 nm) of the atoms in the outer portions of the surface substrate edges within the plane of the surface, and to 15% (i.e. 1.125 nm)of the atoms in the bottommost portion of the substrate to effectively dissipate the excess heat that is generated during deposition and to maintain the temperature of the system at 300K, as shown in Figure 3-9. All other atoms are active, i.e., they evolve in time without any additional constraints.

Neither the REBO nor the LJ potentials used here include explicit charge. Therefore, rather than deposit actual ions the simulations consider the deposition of radicals. For each H or polyatomic considered, a beam of uniform composition and incident energy is deposited normal to the polymer surface slab. The rotational orientations of the beams are randomized. The beams are deposited only in the active region and are randomly oriented spatially within this region. The fluence for each beam is 17 x $10^{14}$ ions/$cm^2$, which is comparable to experimental values.[47] The flux is in the range of 0.4 x $10^{24}$ ions/$cm^2$s.

**3.2.2 50eV Deposition**

The first set of simulations considered the deposition of H, $CH_3$, $C_2H$, or $C_3H_5$ on PMMA with 50 eV of incident energy. The depth profile is determined once the entire series of impacts is complete. The depth profiles for each beam are given in Figure 3-10. Due to the number of atoms present in each beam, the maximum density of deposited atoms on the surface slab is 2.184x$10^{19}$/$cm^3$ for $C_3H_5$ at a depth of about 1 nm. Interestingly, the majority of the atoms in the polyatomic hydrocarbon beams remain within 1.5 nm of the surface after deposition regardless of the structural details of the polyatomic. However, only the depth penetration curve for $C_3H_5$ is relatively symmetric and bell shaped. This implies that $C_3H_5$ can lead to a homogenized distribution of polyatomic species within the polymer, while the other polyatomics lead to a gradually decreasing concentration profile with increasing depth. Unsurprisingly, the smaller H penetrates more deeply to an average depth of 4 nm.

A significant amount of sputtered material and new chemical products in and within the PMMA surface slab is produced for each beam, as illustrated in Figure 3-11. The products, given in the form $C_xH_yO_z$, can be generalized as follows: for H: $x < 12$, $y < 19$, $z < 5$; for $CH_3$: $x<30$, $y<43$, $z<13$; for $C_2H$: $x < 19$, $y < 23$, $z < 7$ and for $C_3H_5$: $x<6$, $y<9$, $z<3$. Very few products are produced as a result of H deposition. Products formed from deposited H are $CH_2$, $CH_3$, $CH_4$, and $C_{11}H_{18}O_4$; the corresponding products from deposited $CH_3$ are $C_5H_8O_3$ and $C_{17}H_{27}O_6$; the products from deposited $C_2H$ are $C_6H_9O_2$ and $C_{17}H_{26}O_6$, and the products from $C_3H_5$ are $C_5H_8$ and $C_2H_3$. Analysis of the surface after deposition and equilibration indicates that $C_2H$ forms the largest number of products inside the surface slab, followed by $CH_3$ and then $C_3H_5$. Very few new products are formed following H-deposition. This behavior can be explained based on the number of unsaturated bonds present and the bulkiness of the deposited atoms or polyatomics in the beam. In general, the greater the number of unsaturated bonds and less bulky the deposited particle is, the more reactive it is predicted to be.

The deposition process also leads to surface sputtering, and some of the species sputtered include $C_5H_8O_2$ and $C_7H_9O_2$ for deposited $C_2H$; $C_{15}H_{25}O_6$, $C_{29}H_{42}O_{12}$ and $C_4H_8$ for deposited $CH_3$; and $C_2H_3O_2$ for deposited $C_3H_5$. This trend can be explained by noting that the more reactive $C_2H$ forms a greater variety of products that are sputtered by subsequent deposition, while the less reactive polyatomics form a narrower number of products that can be sputtered.

Surprisingly, at 50 eV, the incident atoms or polyatomics are predicted to attach to several different sites on the PMMA monomer, as illustrated in Figure 3-12A, along the lengths of the polymer chains on the amorphous surface, as shown in Figure 3-12B. For example, while $C_3H_5$ is predicted to attach mostly to the carbonyl carbon and oxygen, all other incident beam polyatomics and atoms are predicted to attach to almost all other sites. Nonetheless, the carbonyl atoms are the preferred chemisorption site for the deposited atoms and polyatomics. The attachment of incident atoms to different sites along the chains gives rise to various new products, as is also evident from the product formation analysis.

### 3.2.3 25eV Deposition

The depth profiles for the 25 eV depositions are given in Figure 3-13. The atomic and polyatomic beams penetrate to a shallower depth compared to the 50 eV depositions. Here, the maximum atomic density is achieved for $C_2H$ at 8.8 $x10^{18}$ /$cm^3$ at a depth of about 1 nm. Similar to the 50 eV case, the maximum depth of penetration of the polyatomics is limited by their bulky nature to within 3 nm of the polymer surface, while the H penetrates up to 4 nm.

The molecular weights of the chemical products formed as a result of deposition can be denoted in the form of $C_xH_yO_z$, and can be summarized as follows: for H: x < 3, y < 5, z <3; for $CH_3$: x<4, y<6; for $C_2H$: x<3, y<6, or, in other words, no new compound with oxygen is predicted to form under the conditions of the simulations; for $C_3H_5$: x<8, y<11, z<3. Additionally, very little sputtering is predicted to occur. Similar to the 50 eV deposition case, the atoms from the beams are dispersed in their attachment to the PMMA monomers, but comparatively less so, as indicated in Figure 3-14A. Still carbonyl atoms namely $C_3$ and $O_1$, are the most preferred attachment sites for the incident atoms and polyatomics. A snapshot of the attachment of $C_2H$ is given in Figure 3-14B as a representative illustration of the attachment behavior predicted at these kinetic energies; the red circle in the figure emphasizes the above mentioned point. Fewer products are sputtered at this incident energy; those that are include $C_2H_4$ for $CH_3$, and $CH_3$ and $C_4H_{10}$ for $C_3H_5$, while there is negligible sputtering in the case of the other deposited particles.

### 3.2.4 10eV Deposition

The depth-profiles for the atomic and polyatomic beams considered here are given in Figure 3-15. The maximum atomic density of $2.03x10^{20}$ /$cm^3$ is achieved for $C_2H$ at 1.5 nm. The beams stop within 3 nm for $C_3H_5$ and 4 nm for the other beams considered, which is similar to the results for the other incident energy cases discussed above.

Several products formed during 10 eV deposition can be summarized as follows, where the products are given in the form of $C_xH_yO_z$,: for $C_2H$: x<5, y<5; for $CH_3$: x<3, y<7; and for $C_3H_5$: x<10, y<16. The monomer-attachment analysis for the incident atoms and polyatomics at this energy is summarized in Figure 3-16. The results indicate that the attachment sites are less dispersed within the PMMA monomer on the polymer chains relative to the other energies discussed above. In particular, the majority of the incident atoms attach to the carbonyl oxygen and carbon of the acrylate group of the monomer ($C_3$, $O_1$ and $C_4$).

Sputtering events occur less relative to higher incident energies. However, some products are predicted to sputter, including $C_2H_5$ and $C_2H_6$ during $CH_3$ deposition; $C_4H_3$ during $C_2H$ deposition; and $C_9H_{15}$ during $C_3H_5$ deposition.

### 3.2.5 4eV Deposition

The depth-profile for 4 eV for all the atomic and polyatomic beams is given in Figure 3-17. The maximum atomic density of 2.82 x $10^{19}$ /$cm^3$ is achieved for $C_3H_5$ at a depth of 1 nm. Unsurprisingly, less surface sputtering takes place than occurred at higher incident energies. Similarly, significantly less chemistry takes places between the deposited atoms and polyatomics at this energy and so fewer products are formed on or within the PMMA surface. Representative new products that do form, given in the form of $C_xH_yO_z$, are x<3 and y<7 for $CH_3$; and x<5 and y<4 for $C_2H$. In the other cases, the atoms or polyatomics deposited from the beam remain on the surface and undergo few reactions on the time scales of these classical MD simulations.

The attachment of the deposited atoms or polyatomics to the PMMA monomer throughout the surface is given in Figure 3-18. Here, the attachment sites are even more narrowly distributed than in the higher energy cases. This may be attributed to the activation energy barrier which is difficult to traverse if sufficient kinetic energy is not provided to the reactants.

### 3.2.6 Discussions

Several interesting trends emerge from these simulation results. For example, with increasing beam energies the incident atoms or polyatomics form a variety of chemical products by reacting with the PMMA surface. At 50 eV, $C_3H_5$-beam deposition results in the highest atomic density of 2.18 x $10^{19}$ /$cm^3$ at a depth of about 1 nm inside the PMMA surface slab and, while the lowest atomic density is about an order of magnitude lower and is predicted to occur following H-beam deposition. This trend is explained by the fact that $C_3H_5$ is a polyatomic and thus contains multiple atoms, in addition to the fact that it is bulky and thus spatially constrained from embedding within the polymer, both of which lead to a high atomic density at a set point within the PMMA. However, in the case of the 25 eV and 10 eV impacts (Fig 3-13 and 3-15) the highest peak in the depth profile occurs with the ethynyl radical.

Some of the deposited polyatomics are predicted to bounce off the polymer without undergoing any chemical reactions or modifying the surface. This occurs to the greatest extent for $C_3H_5$ and occurs to the smallest extent for $CH_3$. Additionally, this occurs to a greater extent at 4 eV compared to 25 or 50 eV. This may again be attributed to the bulky nature of the polyatomics and the fact that at the higher incident energies they were more easily able to embed themselves within the polymer and/or undergo chemical reactions, while at lower energies it was significantly more difficult for them to do either of these things.

Analysis of these results illustrates that while higher incident hyperthermal beam energies lead to more chemical reactions, less bounce-off, and more embedding of the polyatomics, the amount of surface sputtering is also increased. Less sputtering is predicted to occur for the lower energies, but this is accompanied by fewer reactions and more possibilities for bounce off. For the cases considered in these simulations, the highest atomic density at the surface with the least amount of surface sputtering is predicted for beams of $C_2H$ deposited with 50 eV energy. The sputtered products include atoms from the surface polymer chains. To minimize the extent to which the sputtered products interact with the incident atoms or fragments during deposition, following each deposition and the next 2000 steps of relaxation, the sputtered products are moved to the opposite side of the surface slab. Significant bond-vibration is not predicted to take place in the sputtered products prior to their being moved, which indicates that they are likely stable over longer time scales than are considered here.

The largest variation in the nature of the products formed is predicted for $C_2H$ and $CH_3$ deposition with 50 eV, both of which do a good job of chemically functionalizing the PMMA. The most common new chemical products that are predicted to form are $CH_3$, $C_3H_3$, $C_4H_8$, $C_6H_9O_2$, $C_5H_8O_2$, $C_7H_9O_2$, and $C_2H_3O_2$. The most common sputtered products are $CH_4$, C2$H_4$, $C_4H_{10}$, $C_4H_3$, $C_5H_8O_2$, $C_7H_9O_2$, and $C_3H_3O_2$. The highest molecular weight chemical product formed among all depositions considered is $C_{29}H_{42}O_{12}$, which is formed following $CH_3$ deposition for 50 eV. This can be understood by considering both the high reactivity of the unsaturated carbon and relatively un-bulky nature of this particular polyatomic.

The selectivity of the chemical reactions in which the deposited atoms or polyatomics participate in terms of where beam atoms bond to the monomers of the PMMA chains are predicted to depend strongly on the incident beam energies. However, the carbon and oxygen atoms of the carbonyl group are consistently predicted to be the preferred sites for attachment. The increasing variation in attachment site within the PMMA momoner with increasing beam-energy may be explained on the basis of an increasing number of kinetic pathways being accessed and the enhanced conformation entropy associated with less bulky polyatomics.

The chemical modification of amorphous PMMA through the deposition of beams of H, $C_2H$, $CH_3$, or $C_3H_5$ with kinetic energies of 4, 10, 25, or 50 eV of energy is investigated using classical molecular dynamics simulations with reactive potentials. The results allow for the determination of the ways in which these processes produce new chemical products and the products vary with the beam and energy. The simulations further illustrate how the atomic density of the PMMA and the ways in which the attachment of the new products to the PMMA monomer along the chains of the polymer surface vary with changes in deposition conditions. The results identify likely mechanisms associated with the predicted chemical reactions and illustrate the ways in which the various aspects of the reaction conditions, such as the reactivity, bulkiness, and incident energy, of the deposited polyatomics balance one other to produce the final modified surface and new products.

### 3.3 Summary

In this section surface modification of amorphous polymers PS and PMMA with polyatomic ion beams was carried out with REBO potential. The primary goal of the work was to quantify the atomistic phenomenon based on the density profile analysis, mass-spec analysis of products formed and sputtered, surface roughness and specification of preferred sites on monomers. The results clearly indicated the preferred site on monomer changes with incident beam energy and number of saturated bonds in the ion beam. It also quantified the upper limits for constituents atoms in the computational mass-spec analysis of products formed and sputtered after deposition events. The depth profile showed that given a particular energy and chemistry of the ion beam what is the theoretical maximum depth that the ion beam could be modify the surface. While these predictions are primarily from the MD simulations, results from ab-initio calculation and experimental works can be carried out for validation of the model.

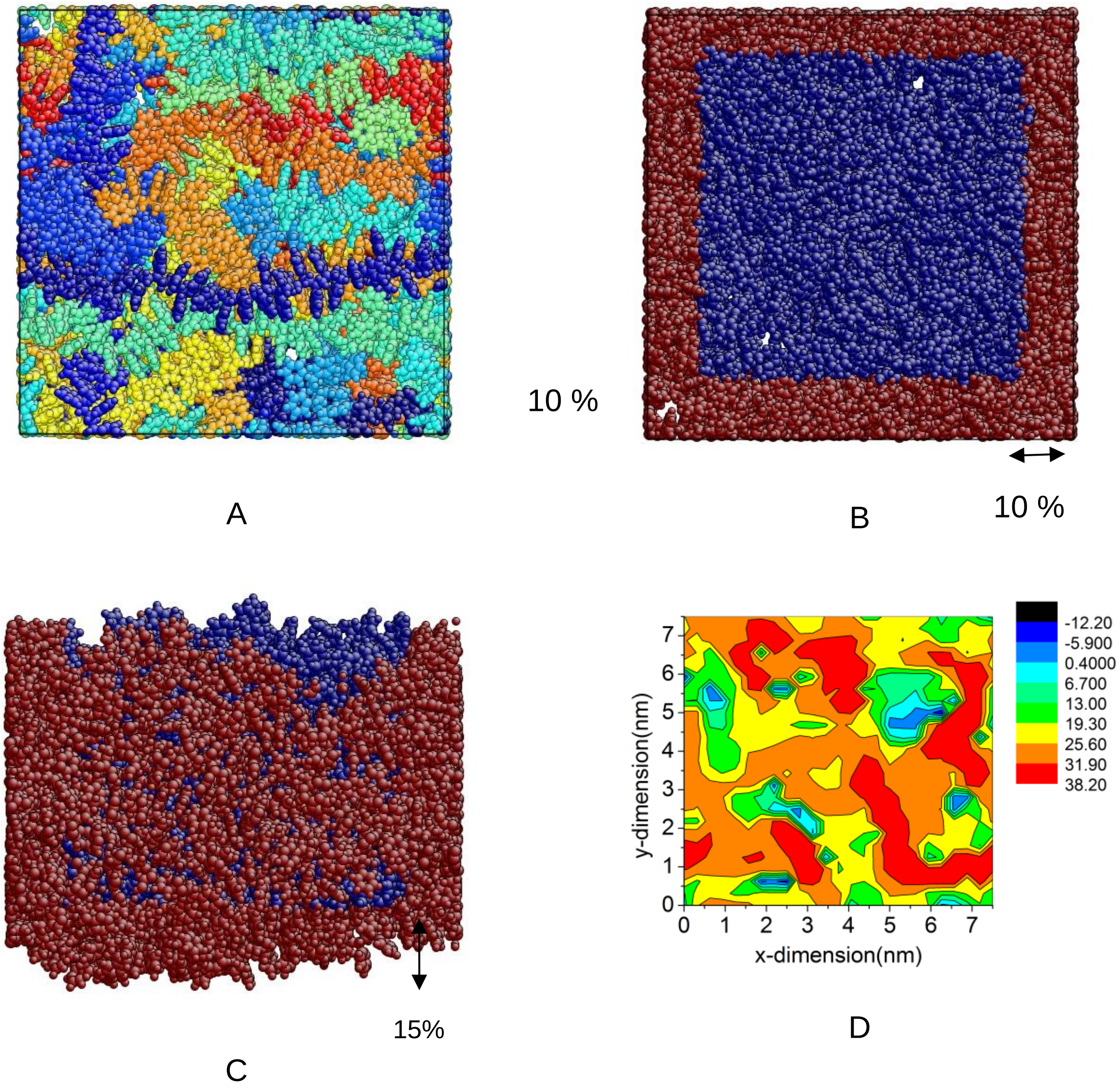

Figure 3-1. Initial set-up of amorphous polystyrene. A) Snapshot top view of the amorphous polystyrene surface slab made using the bead-spring model, where the different colors indicate different polymer chains, and top view B) and side view C) of the substrate that illustrates the thermostat and active regions (red and blue, respectively). D) Contour top view plot of the surface roughness of the as-constructed surface slab.

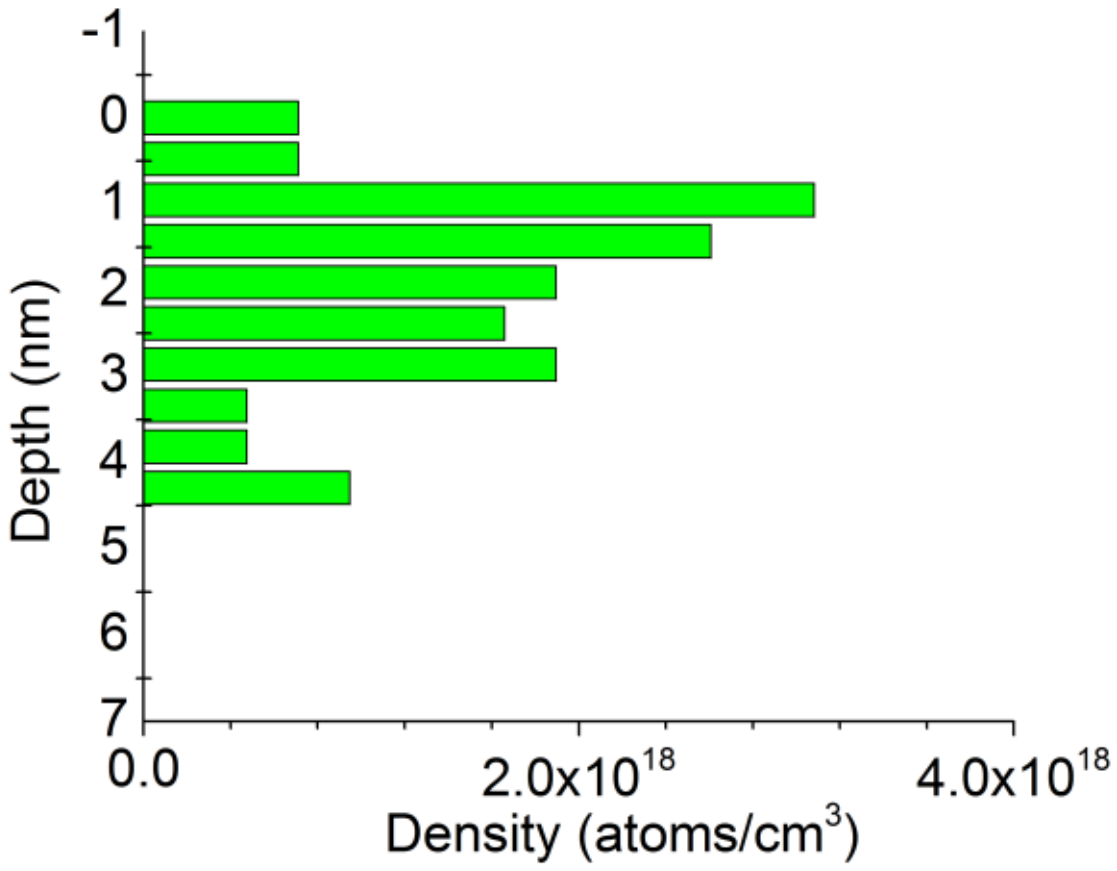

A

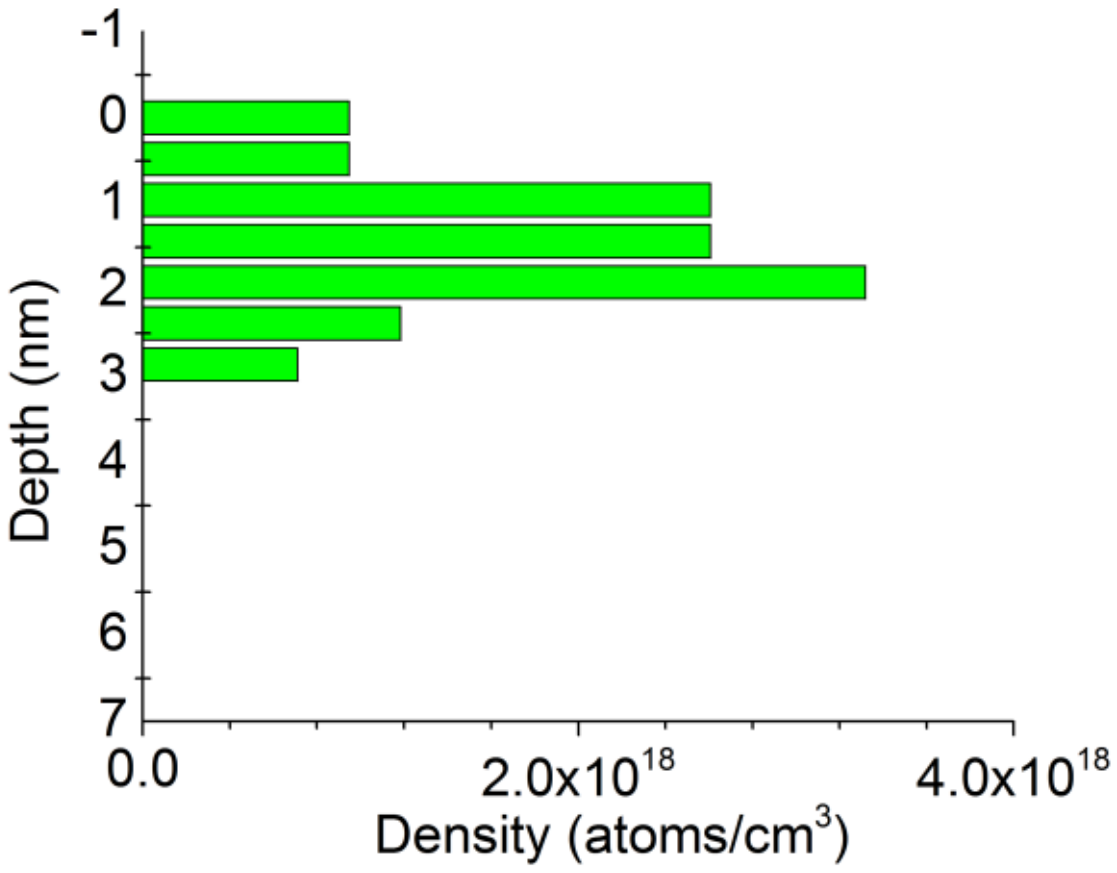

B

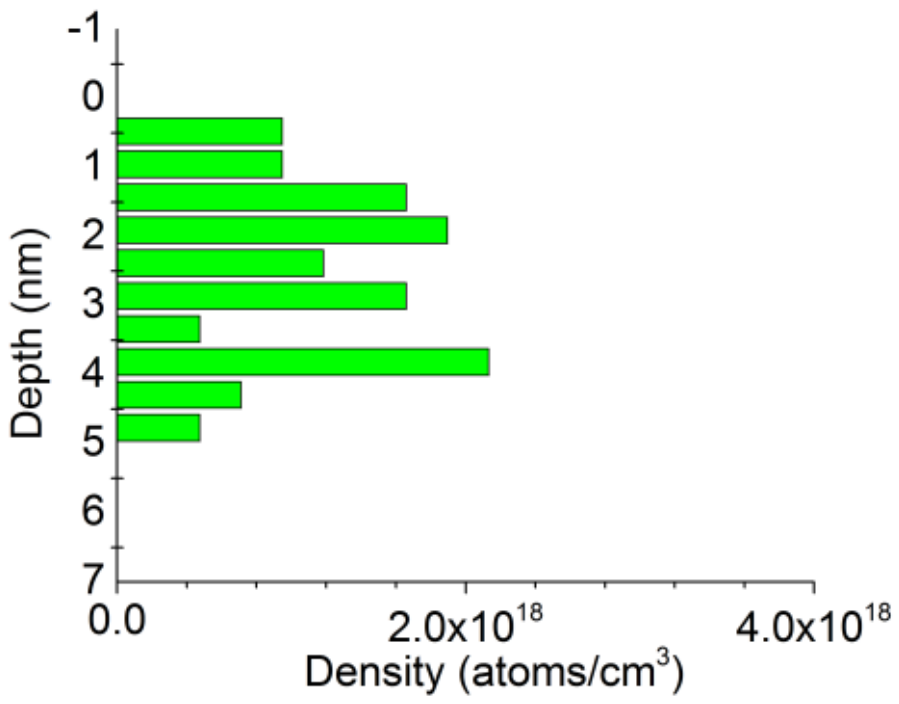

C

Figure 3-2. Depth profiles for atomic S deposition with different kinetic energies. A) 25 eV, B) 50 eV, and C) 100 eV.

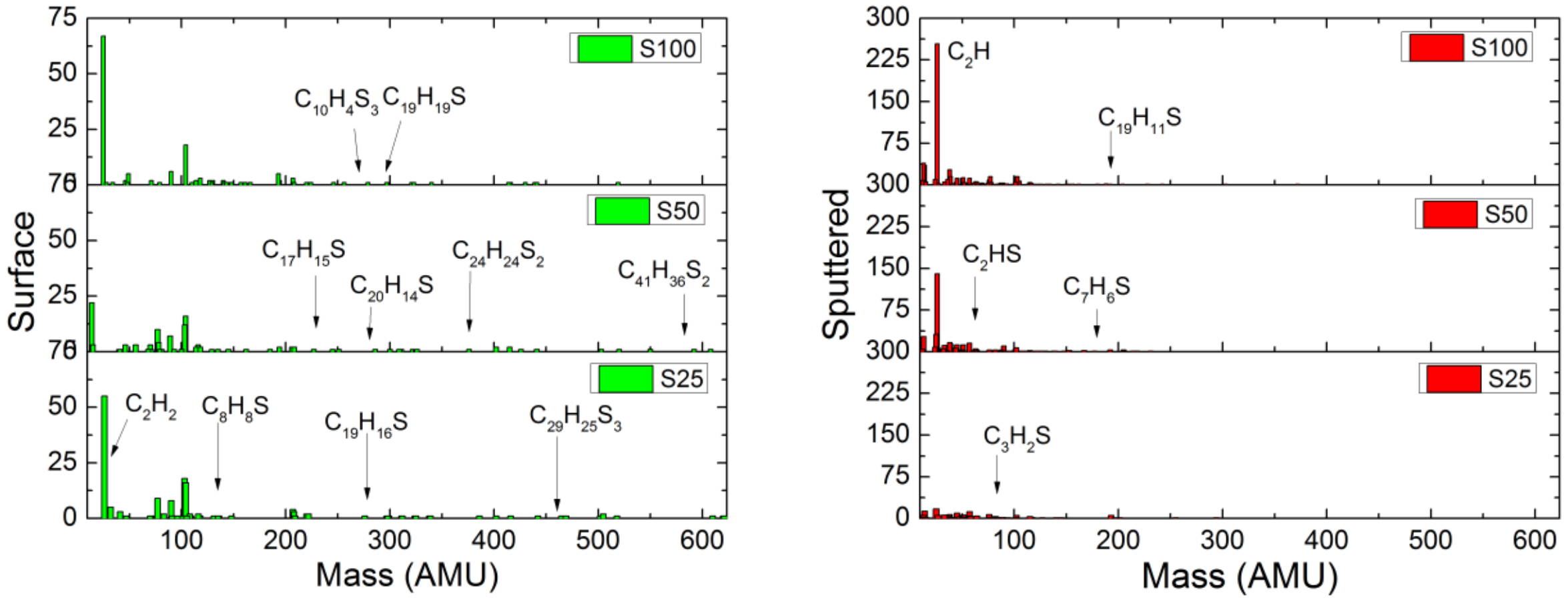

A B

Figure 3-3. Summary of the chemical products formed after the deposition of 100 S atoms: product formation A) within the surface slab and B) within the gas phase. The chemical products are analyzed in the form of $C_xH_yS_z$.

Figure 3-4. Atom bonding analysis after deposition. A) unmodified PS atoms and the PS following the attachment of B) S, C) SC, and D) SH. E) Snapshot of the attachment of S to a surface chain following the deposition of a beam of S with 50 eV of kinetic energy. Here, beam atoms represent the atoms in the incident beam that attach to monomers within the PS surface slab. This was measured by counting the number of incident atoms that attach to a particular atomic site on the PS monomer.

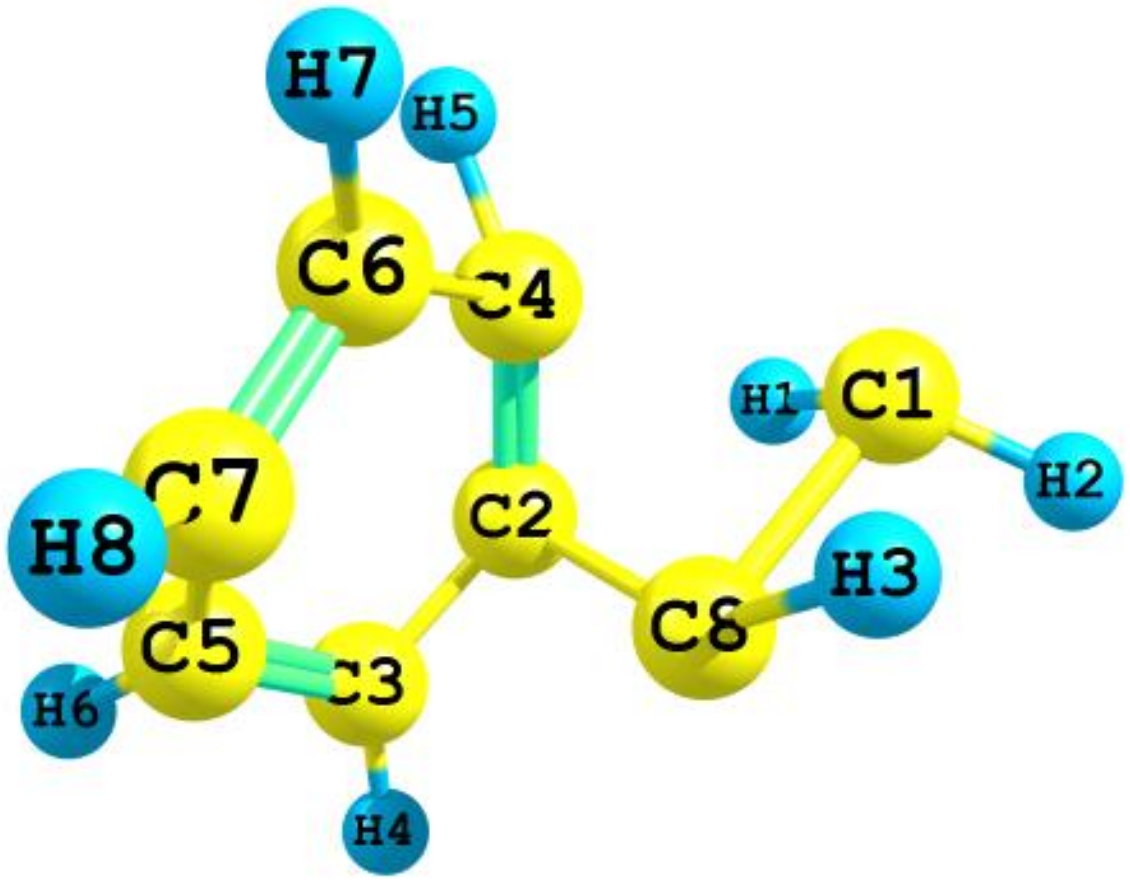

A

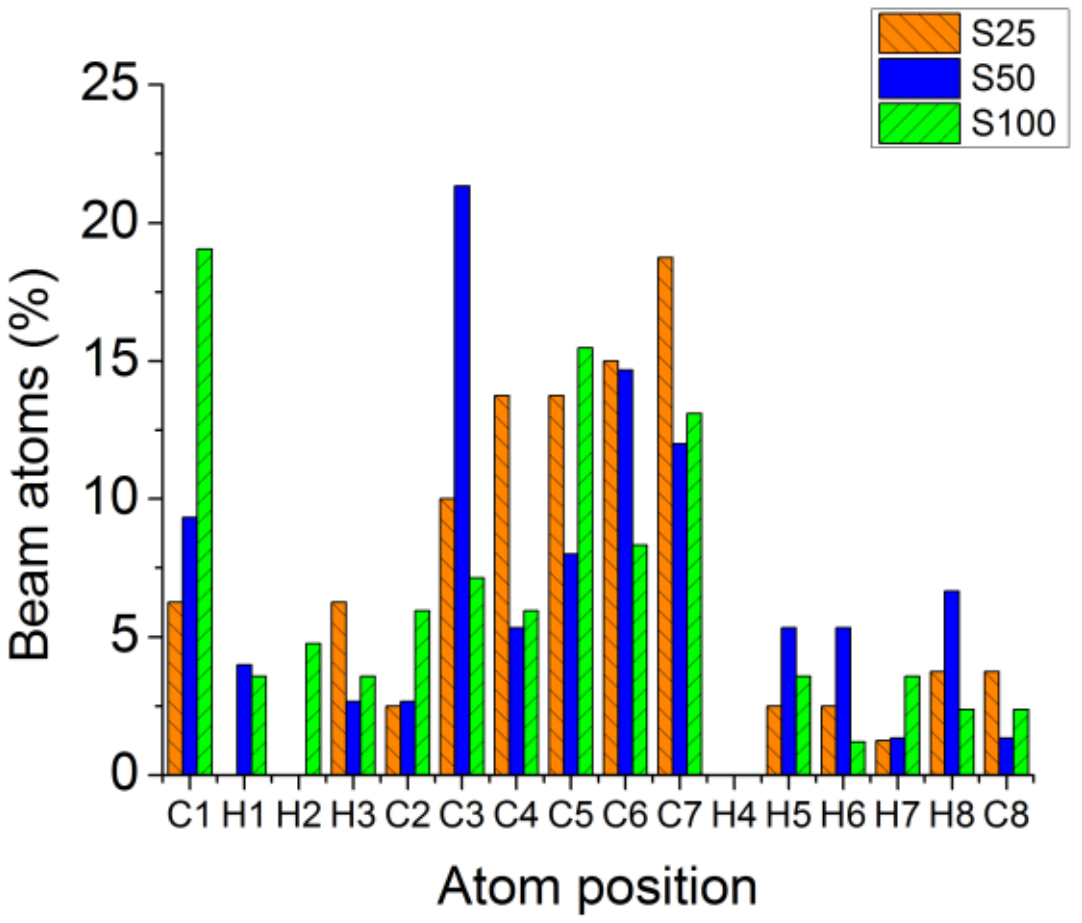
S25
S50
S100
Beam atoms (%)
Atom position
0
5
10
15
20
25
C1 H1 H2 H3 C2 C3 C4 C5 C6 C7 H4 H5 H6 H7 H8 C8

B

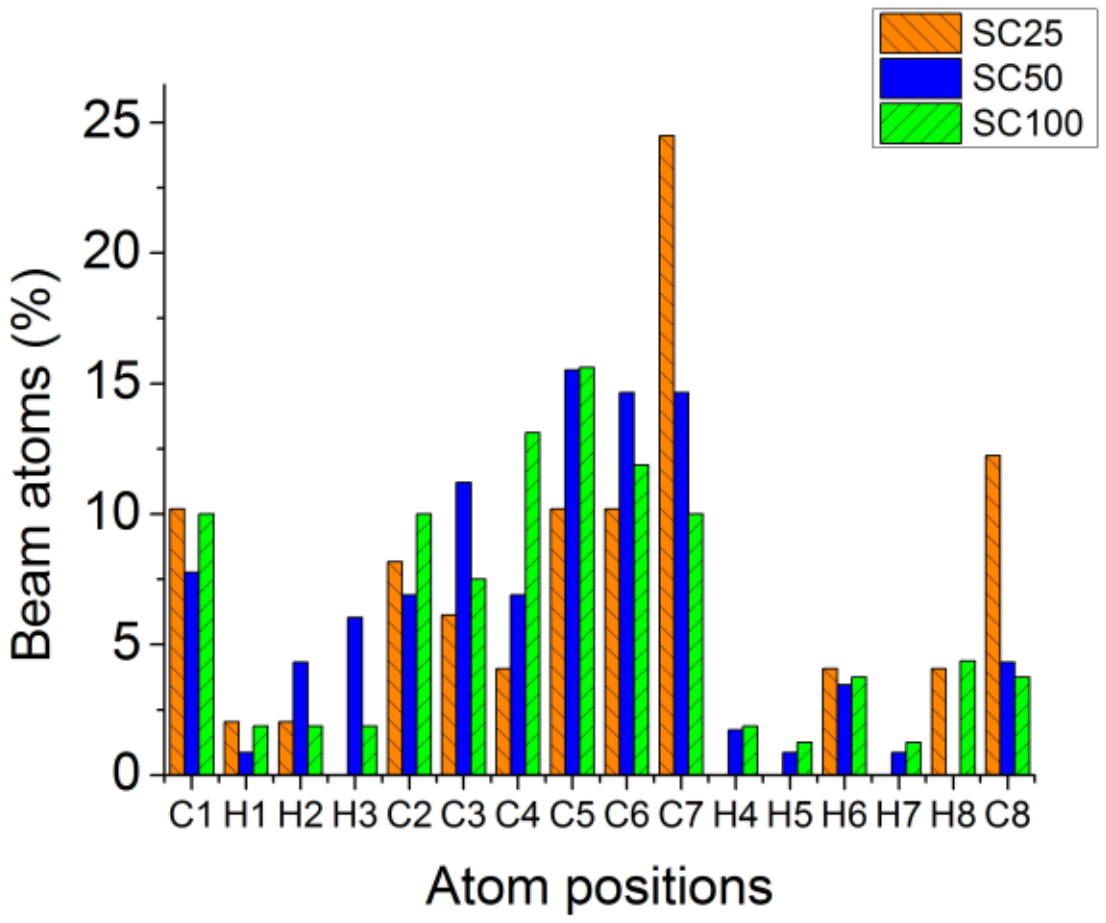
SC25
SC50
SC100
Beam atoms (%)
Atom positions
0
5
10
15
20
25
C1 H1 H2 H3 C2 C3 C4 C5 C6 C7 H4 H5 H6 H7 H8 C8

C

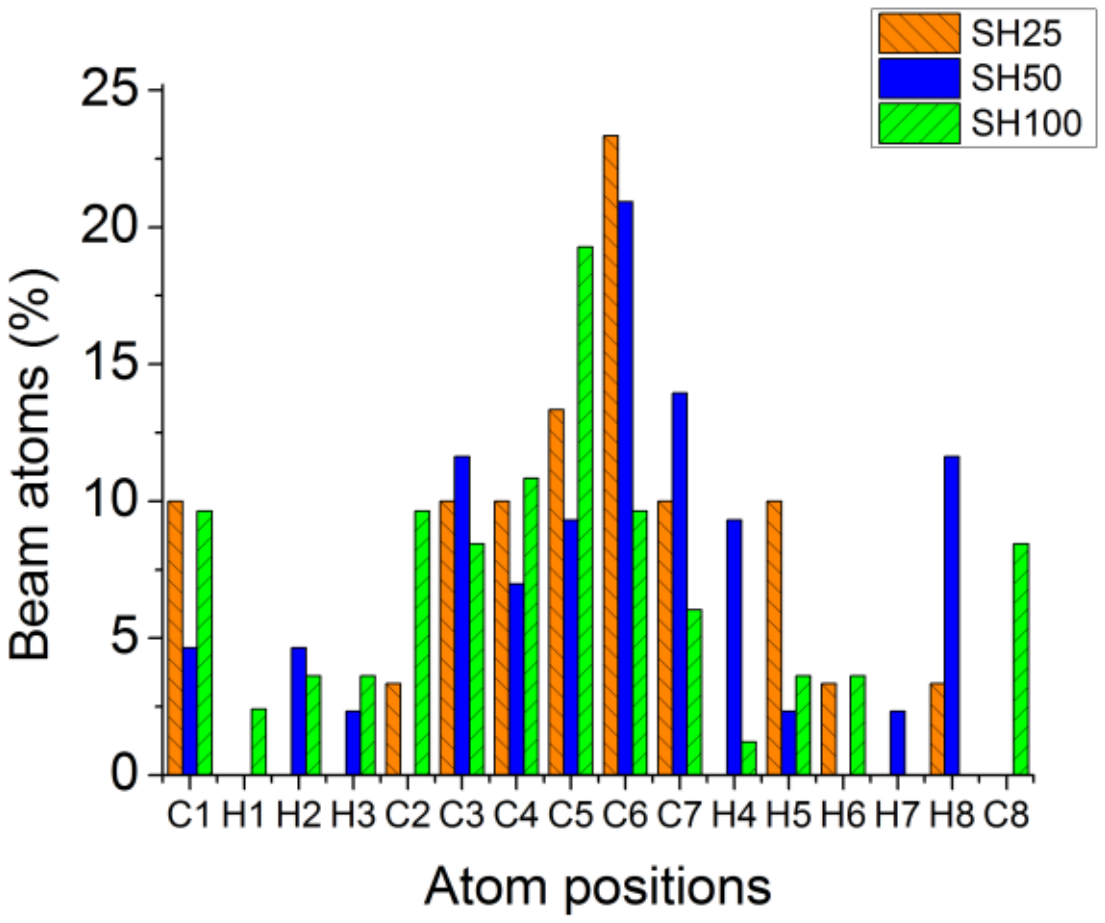
SH25
SH50
SH100
Beam atoms (%)
Atom positions
0
5
10
15
20
25
C1 H1 H2 H3 C2 C3 C4 C5 C6 C7 H4 H5 H6 H7 H8 C8

D

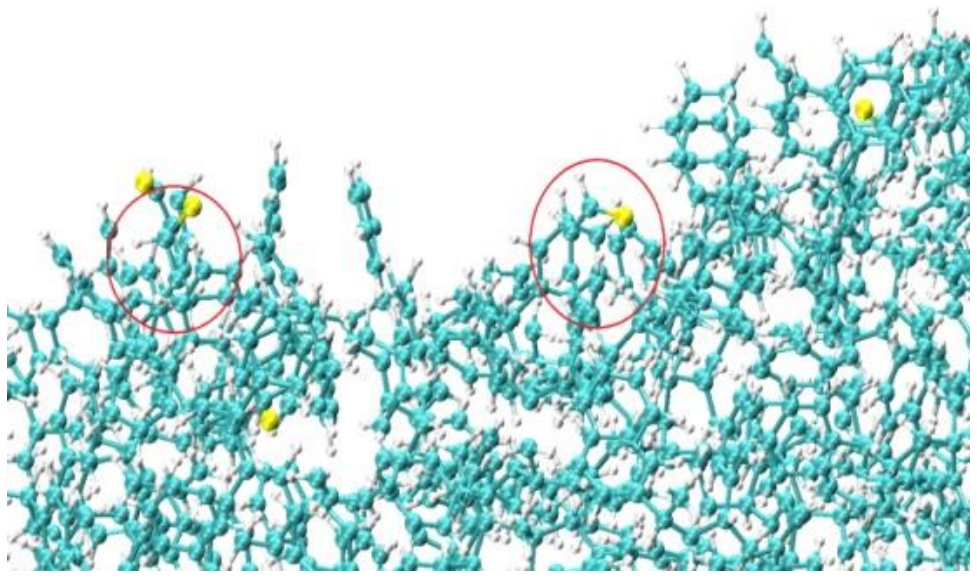

E

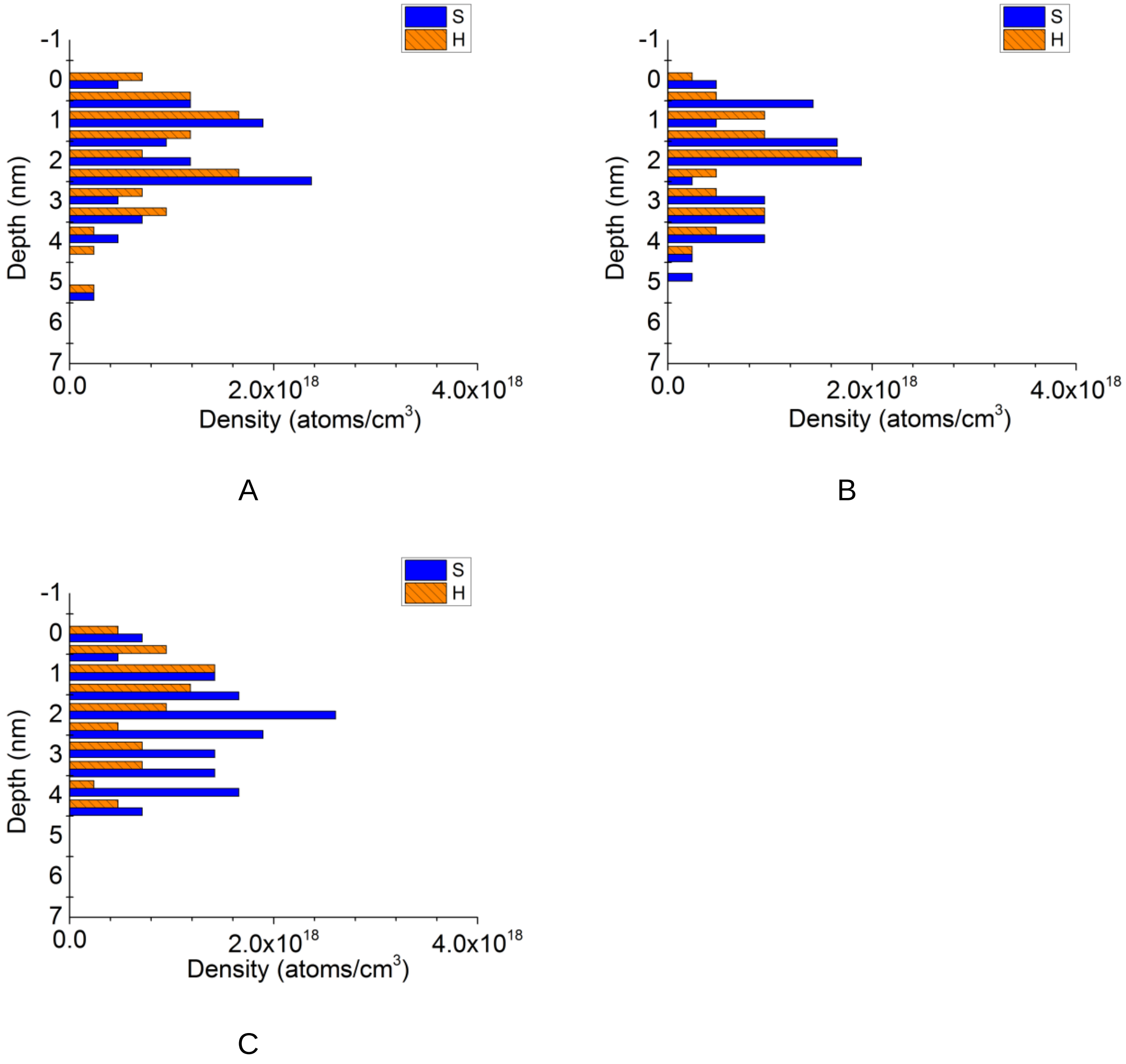

Figure 3-5. Depth profiles for SH deposition with different kinetic energies. A) 25 eV, B) 50 eV, and C) 100 eV.

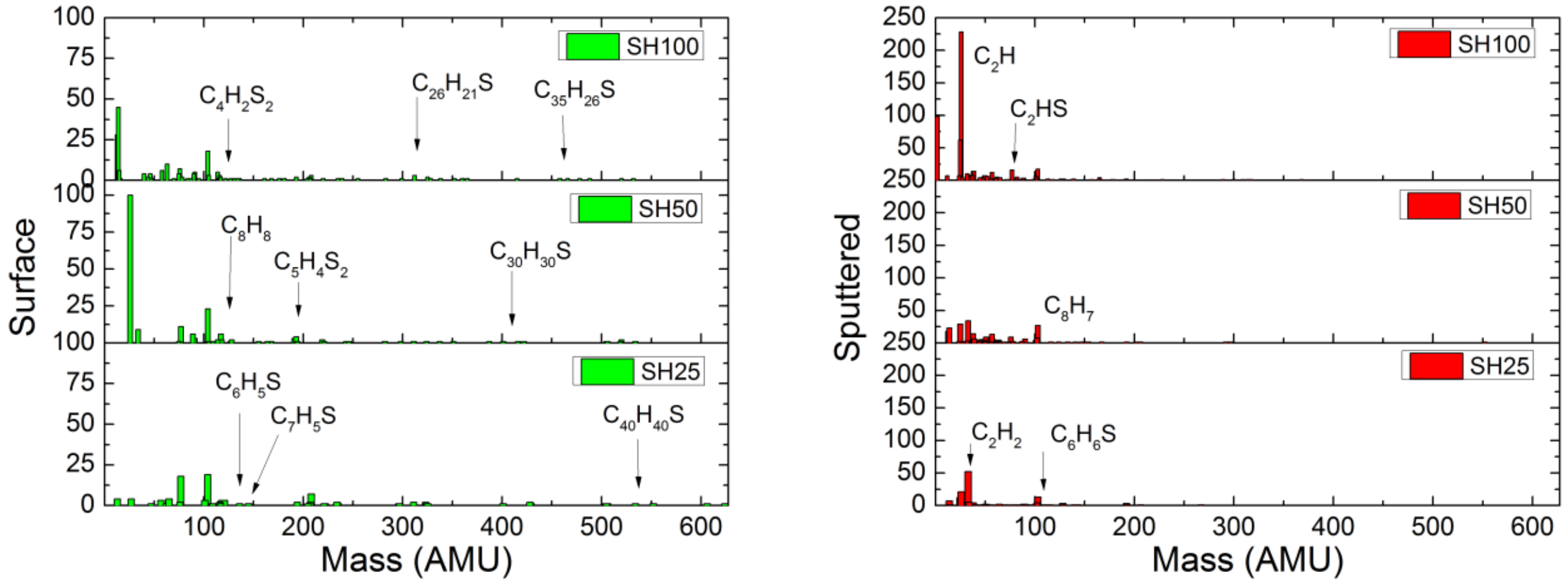

A B

Figure 3-6. Summary of chemical products formed after the deposition of 100 SH dimers. Product formation A) within the surface slab and B) within the gas phase.

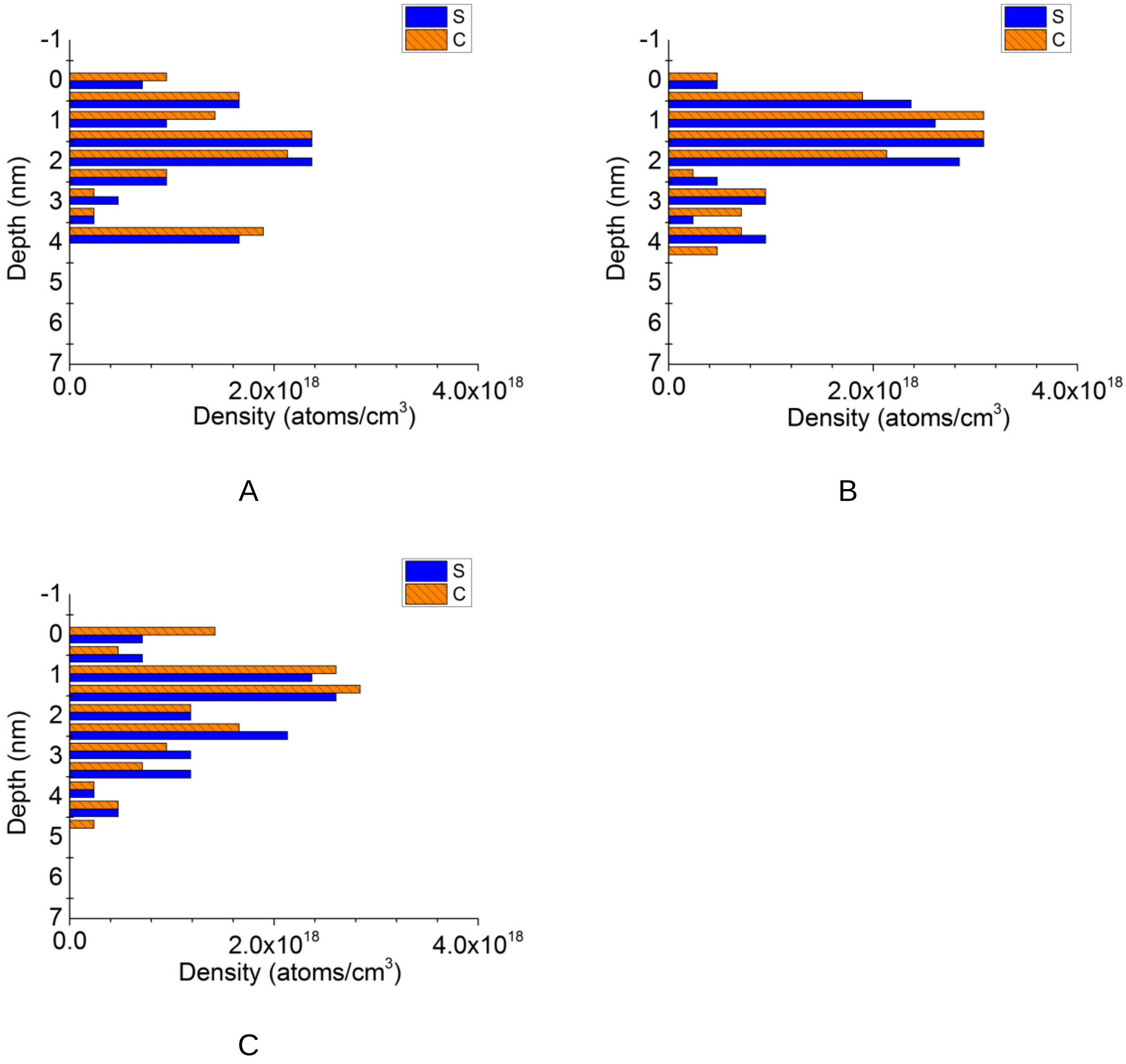

Figure 3-7. Depth profiles for SC deposition with different kinetic energies. A) 25 eV, B) 50 eV, and C) 100 eV.

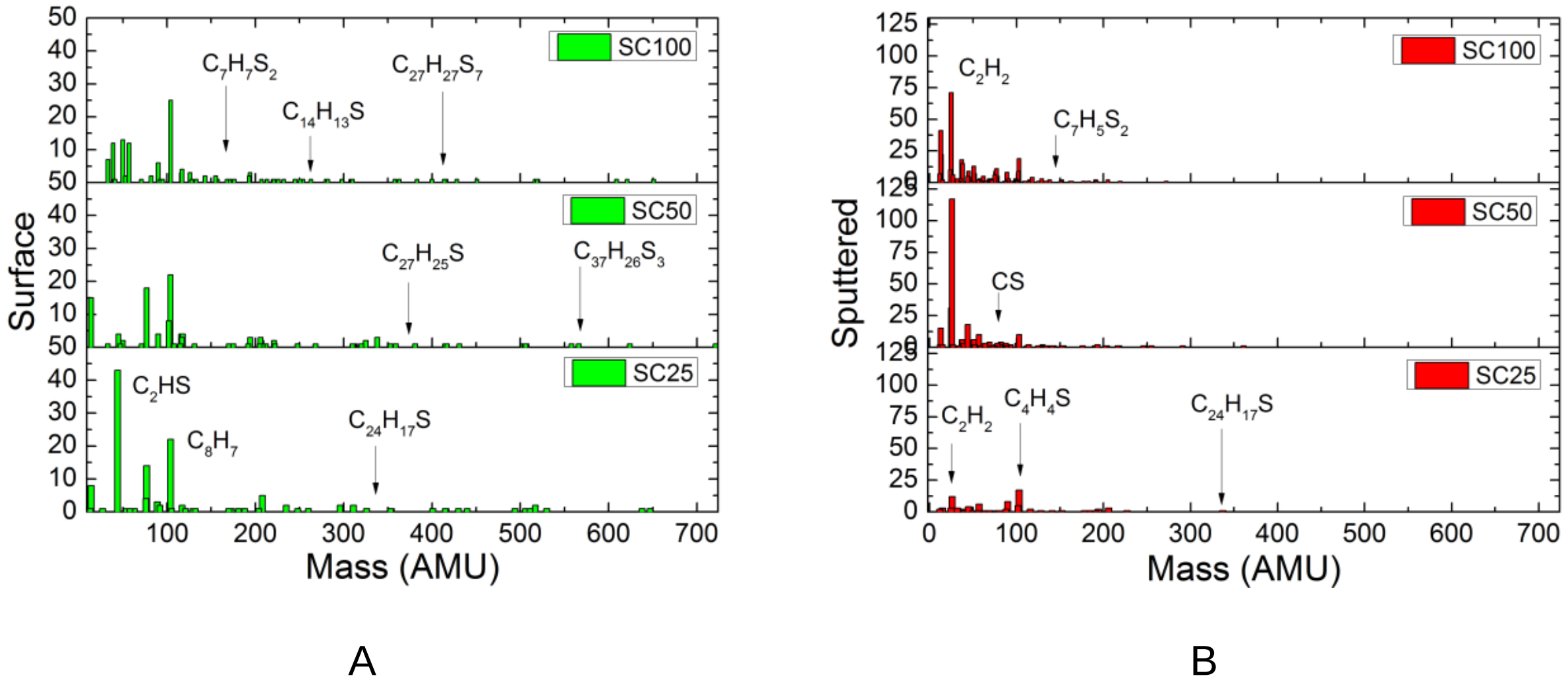

Figure 3-8. Summary of chemical products formed after the deposition of 100 SC dimers. Product formation A) within the surface slab and B) within the gas phase.

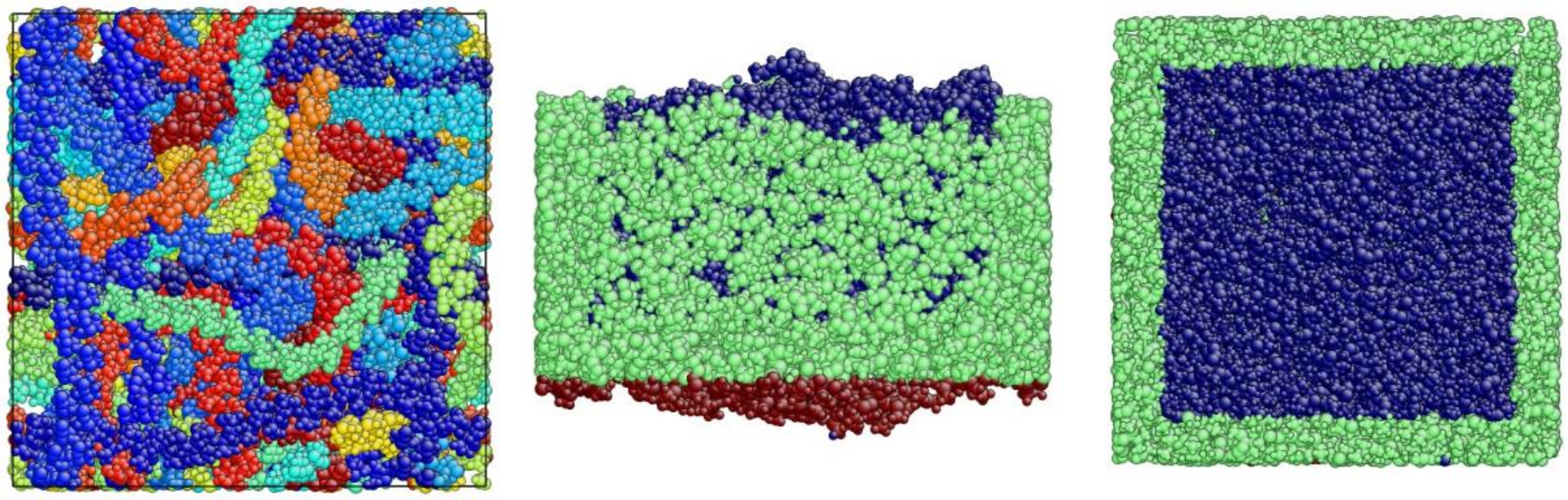

A B C

Figure 3-9. Initial set-up of PMMA system. A) Snapshot of the amorphous PMMA surface slab rendered in bead-spring mode, where the different colors indicate different polymer chains. B) Side and C) top view of the atomic scale surface slab, where blue indicates active region, green indicates thermostat region, and red indicates fixed atoms.

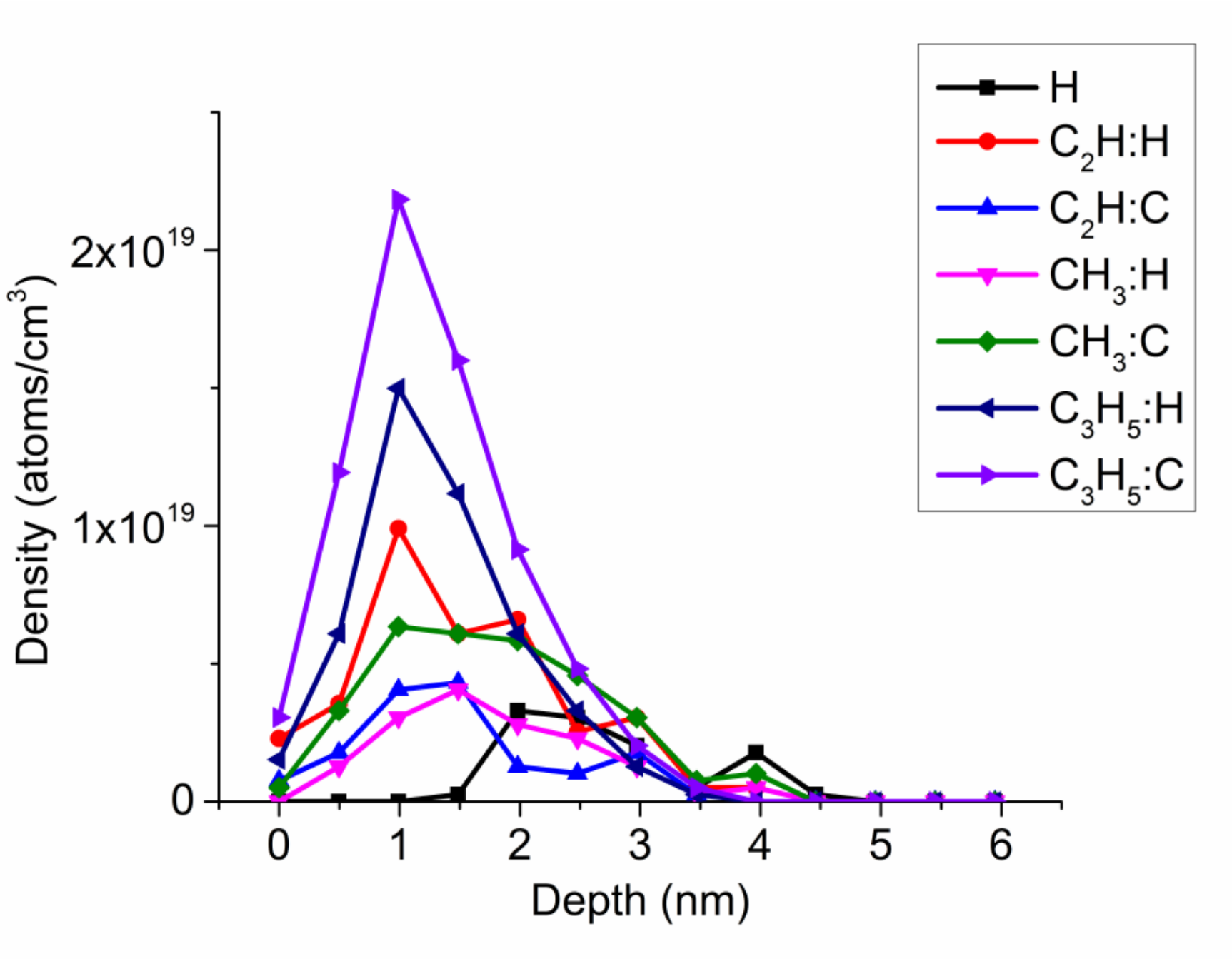

Figure 3-10. Depth profiles for H, $C_2H$, $CH_3$, and $C_3H_5$ deposition with 50 eV of energy. The highest atomic density in the PMMA surface slab is 2.184x$10^{19}$/$cm^3$ which occurs at a depth of 1 nm following $C_3H_5$ deposition.

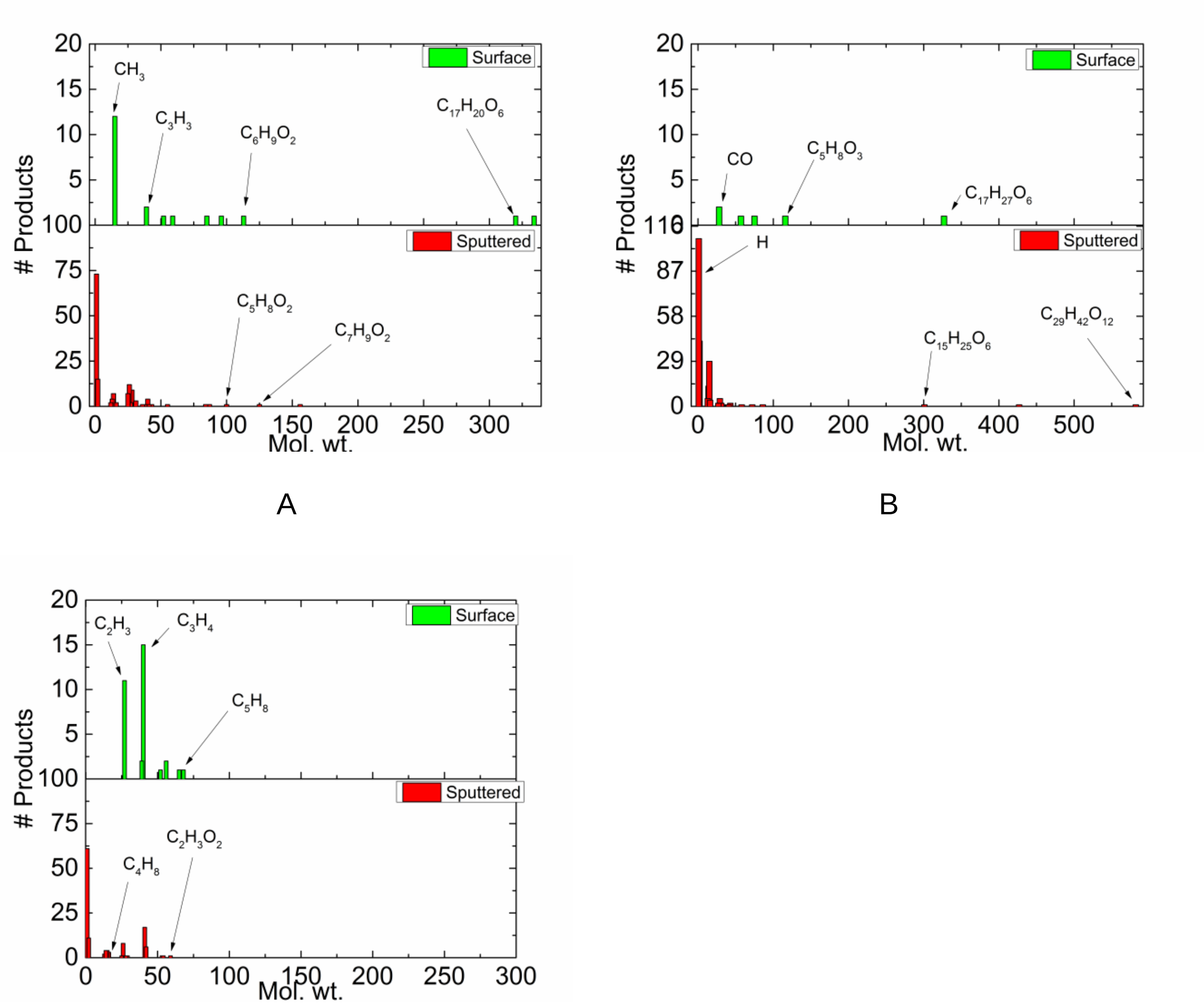

C

Figure 3-11. Product-formation and sputtering analysis for different ion-beams. A) $C_2H$, B) $CH_3$, and C) $C_3H_5$ beams deposited with 50 eV of energy. The most prevalent new chemical products that either remain on the PMMA surface or are sputtered are illustrated.

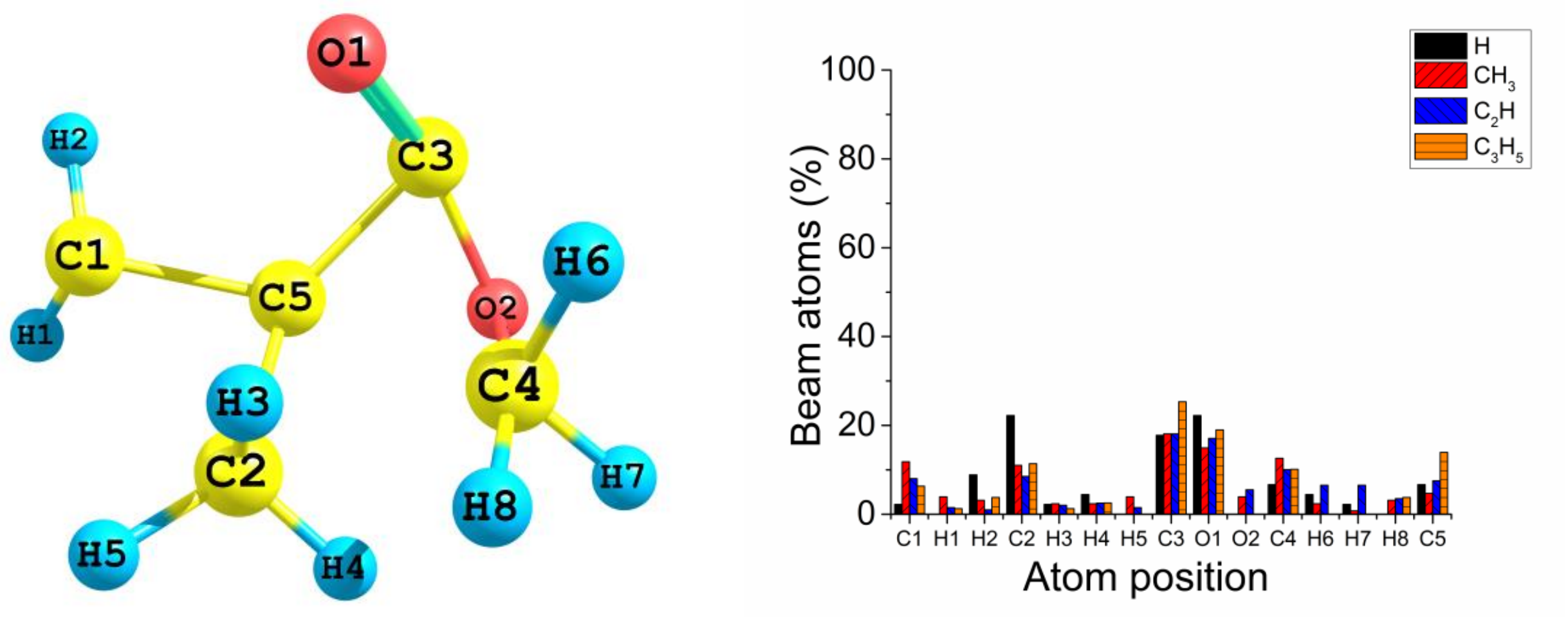

A B

Figure 3-12. Monomer attachment analysis after deposition with 50 eV. A) Labeled PMMA monomer. B) Chemical bonding analysis following 50 eV deposition of beams of H, $CH_3$, $C_2H$, or $C_3H_5$. The results are given relative to the sites in Figure 4A and indicate that the new chemical products formed as a result of deposition bond to a dispersed range of sites on the PMMA monomer.

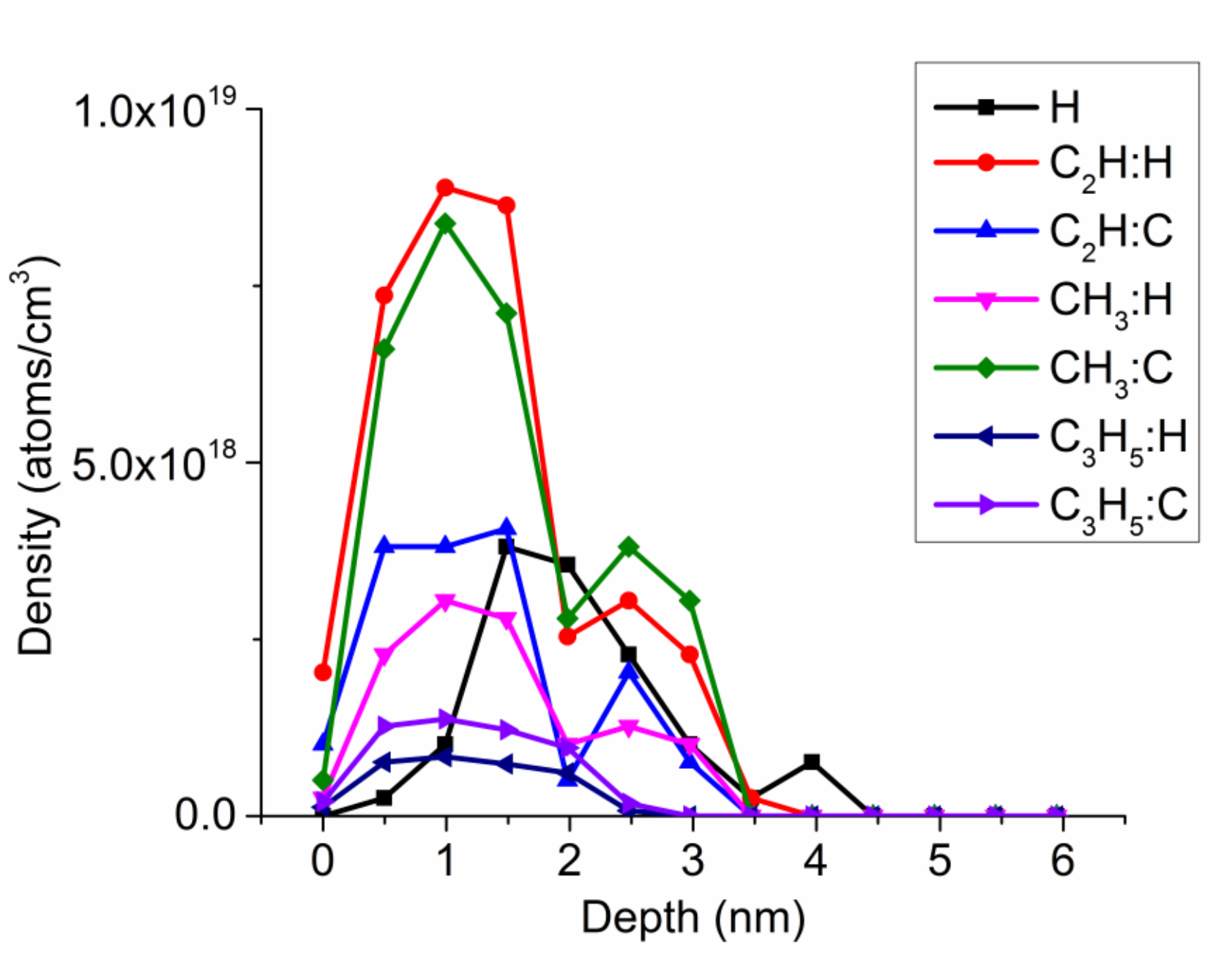

Figure 3-13. Depth profiles for H, $C_2H$, $CH_3$, and $C_3H_5$ with 25 eV of energy. The highest atomic density in the PMMA surface slab is 8.8 x$10^{18}$ /$cm^3$ which occurs at a depth of 1 nm following $C_2H$ deposition.

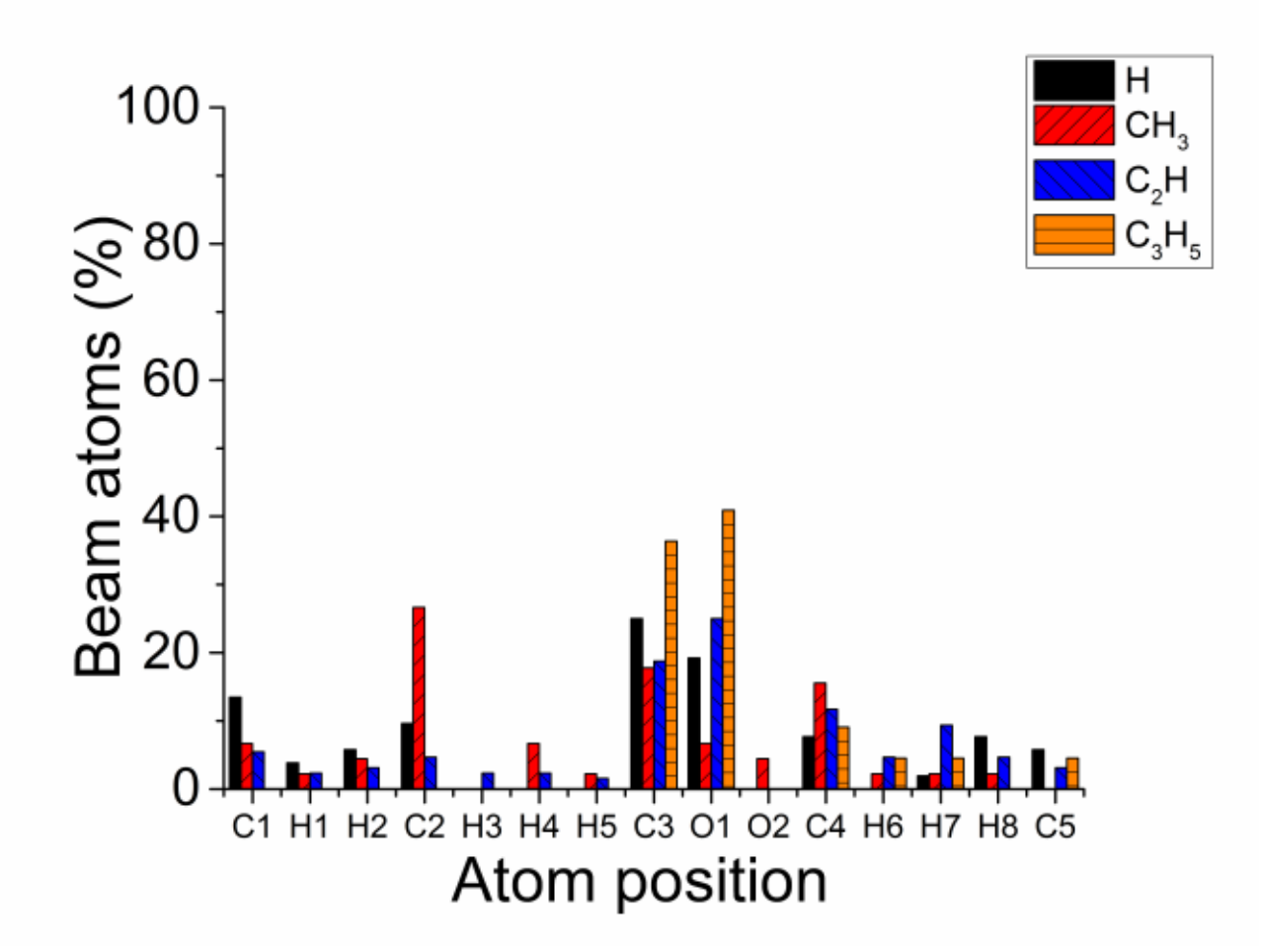

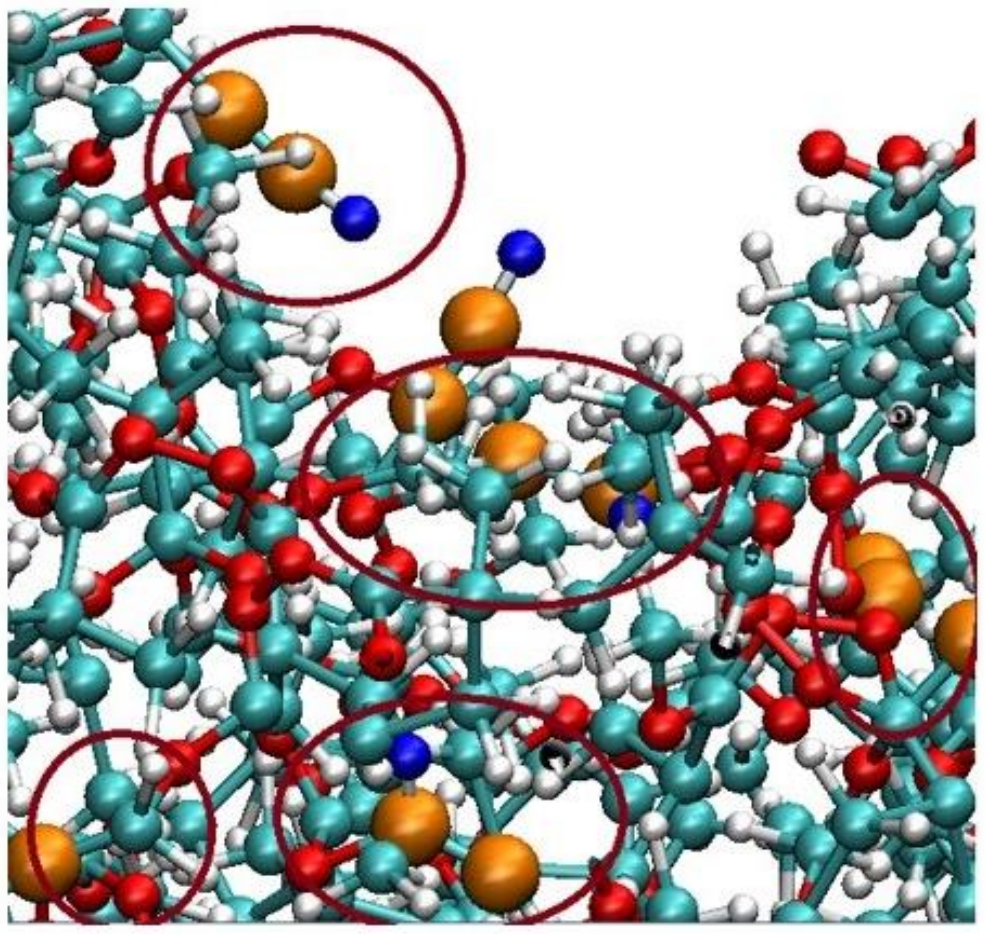

A B

Figure 3-14. Monomer attachment analysis for 25 eV. A) Chemical bonding analysis following 25 eV deposition of beams of H, $CH_3$, $C_2H$, or $C_3H_5$. The results are given relative to the sites in Figure 4A and indicate that the new chemical products formed as a result of deposition bond to a narrower range of sites than at 50 eV. B) An illustrative snapshot of the chemically modified PMMA surface following $C_2H$-beam deposition. The red and indigo spheres represents C and H of substrate, while orange and dark blue spheres represent the C and H of $C_2H$. The C and H of $C_2H$ has been exaggerated to clearly distinguish them and highlight the site-specificity of attachment.

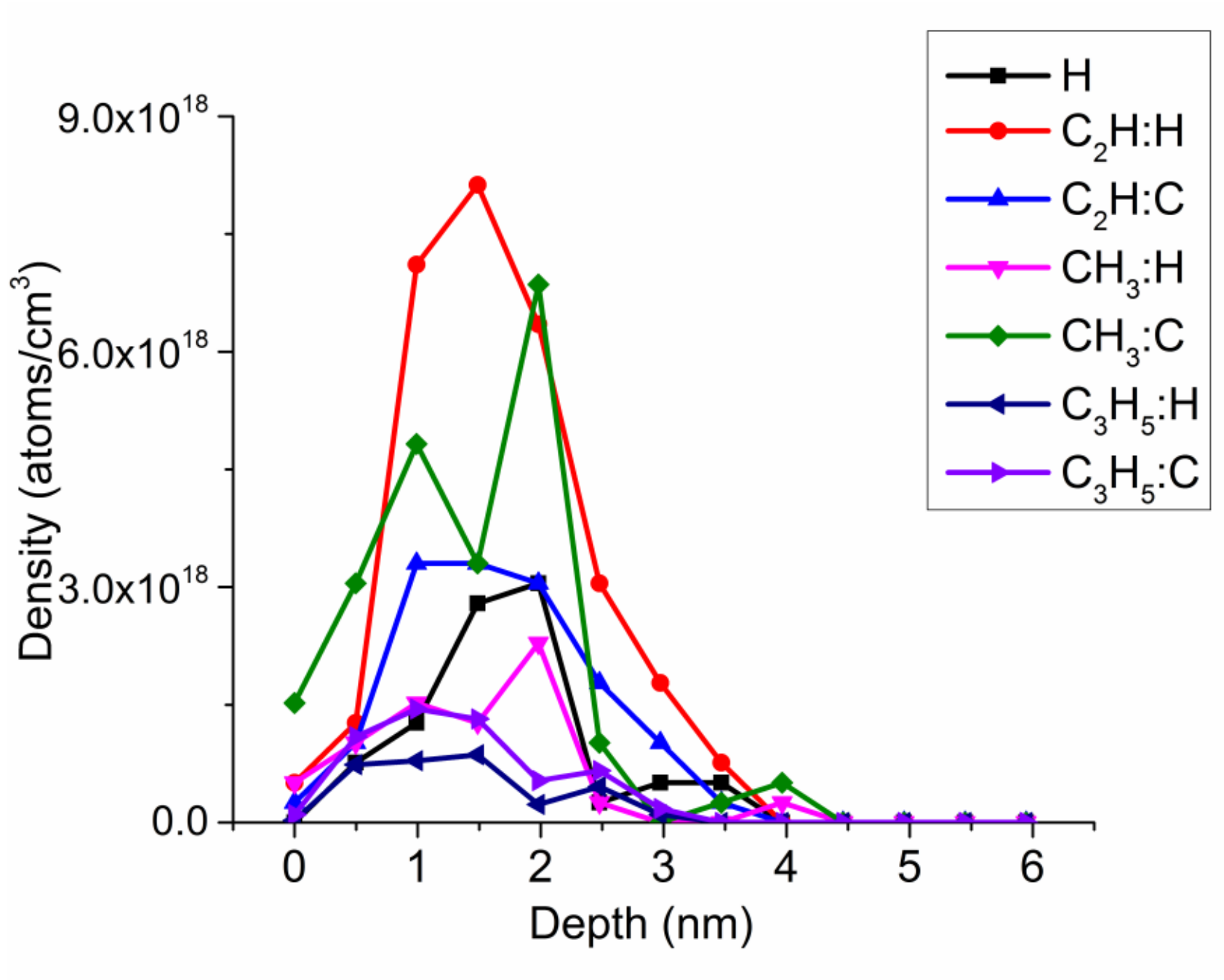

Figure 3-15. Depth profiles for H, $C_2H$, $CH_3$, and $C_3H_5$ with 10 eV of energy. The highest atomic density in the PMMA surface slab is $2.03x10^{20}/cm^3$ which occurs at a depth of 1.5 nm following $C_2H$ deposition.

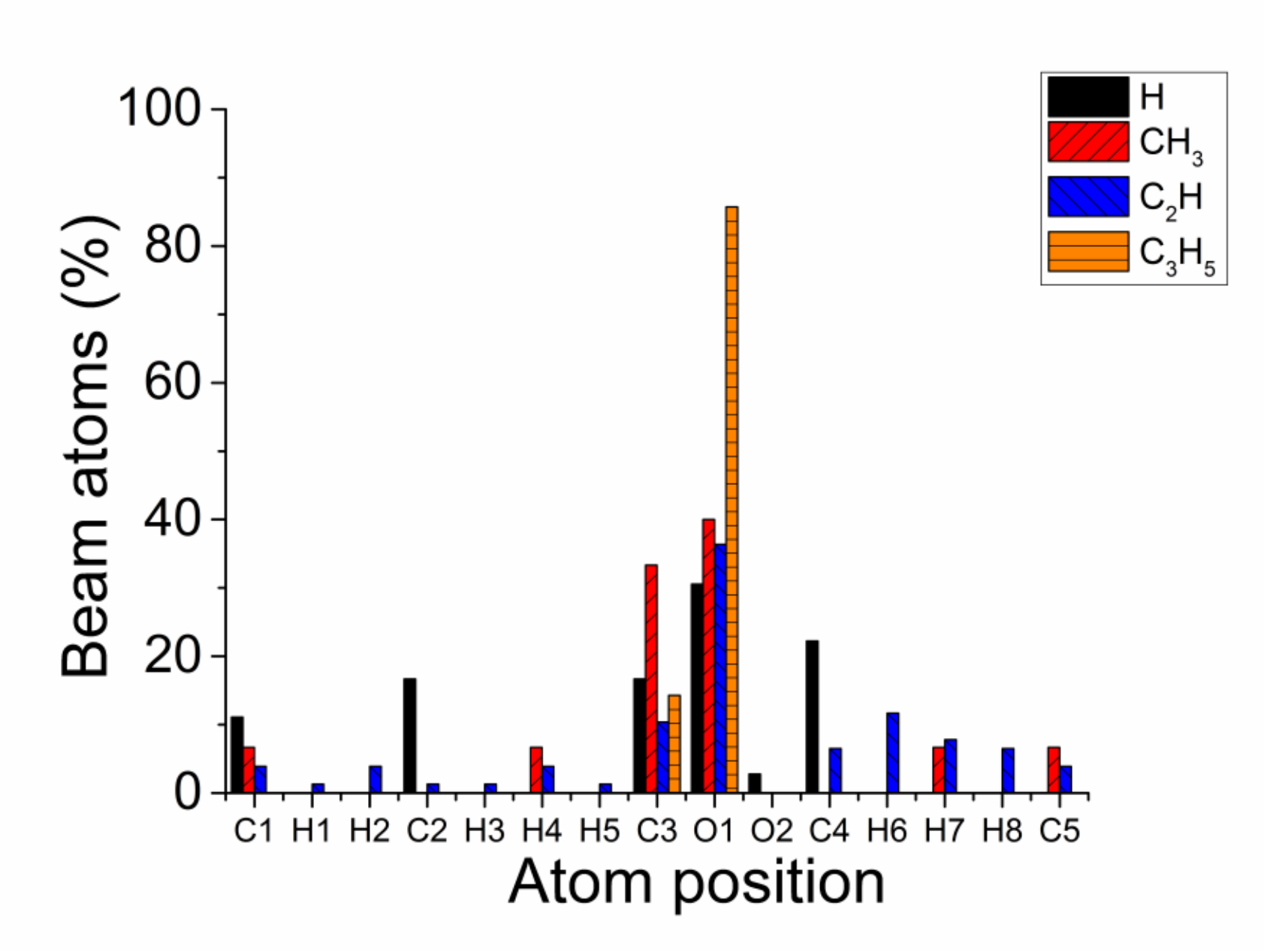

Figure 3-16. Chemical bonding analysis following 10 eV deposition of beams of H, $CH_3$, $C_2H$, or $C_3H_5$. The results are given relative to the sites in Figure 4A and indicate that the new chemical products formed as a result of deposition bond to a narrow range of sites.

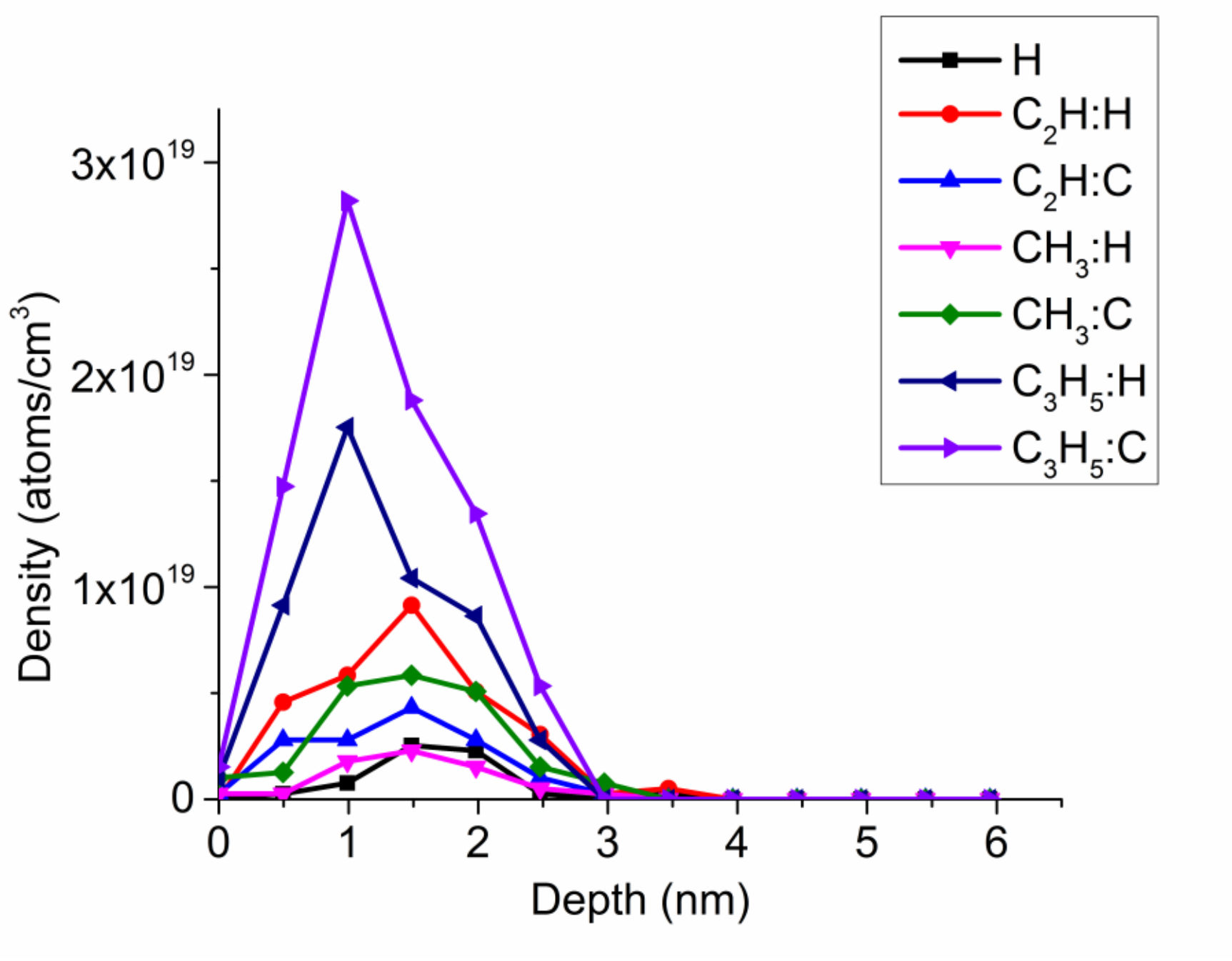

Figure 3-17. Depth profiles for H, $C_2H$, $CH_3$, and $C_3H_5$ with 4 eV of energy. The highest atomic density in the PMMA surface slab is 2.82 x $10^{19}$/$cm^3$ which occurs at a depth of 1 nm following $C_3H_5$ deposition.

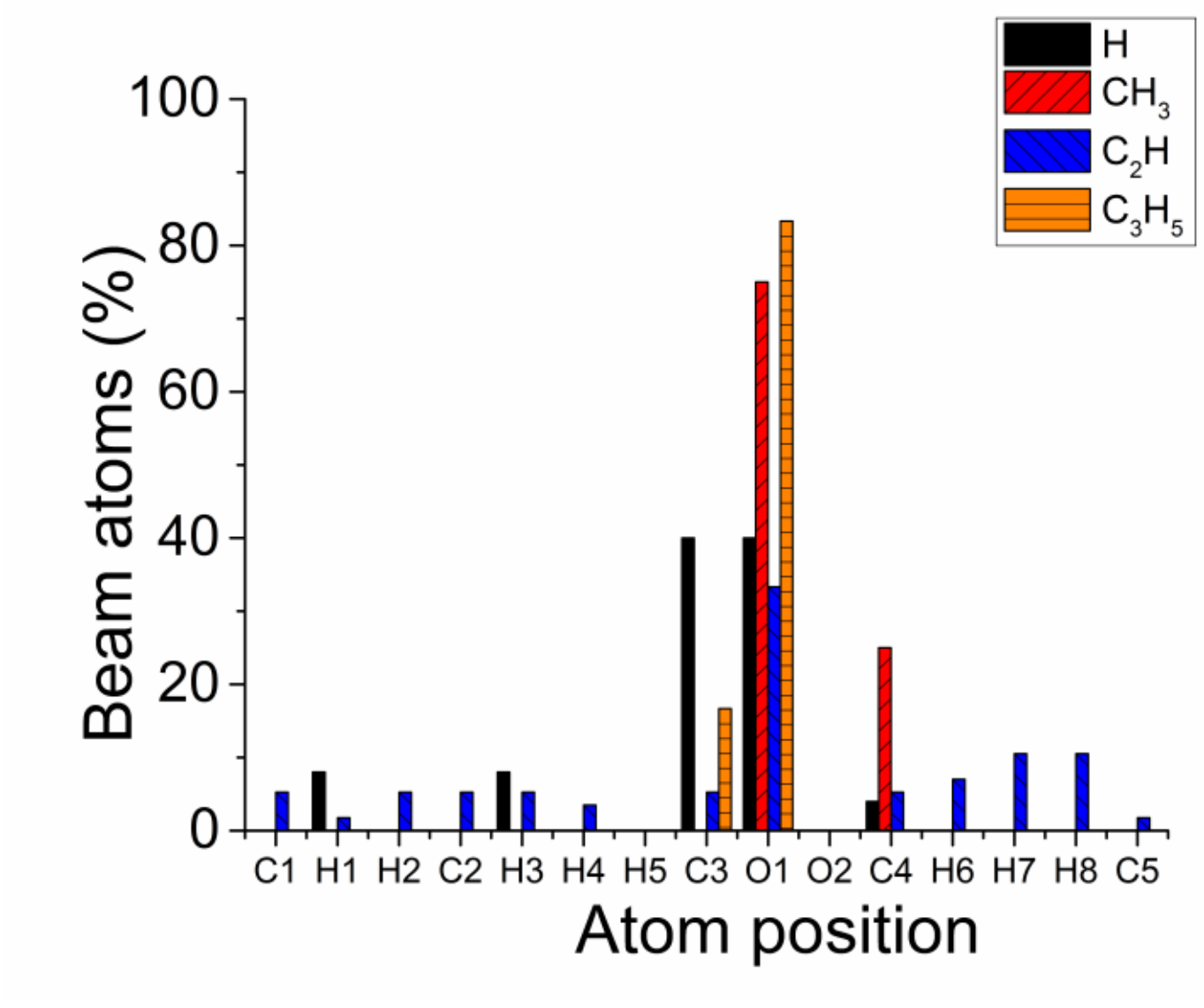

Figure 3-18. Chemical bonding analysis following 4 eV deposition of beams of H, $CH_3$, $C_2H$, or $C_3H_5$. The results are given relative to the sites in Figure 4A and indicate that the new chemical products formed as a result of deposition bond to the narrowest of sites of the cases considered here.

# CHAPTER 4
# METAL CERAMIC INTERFACE

## 4.1 COMB3 Potential for Al

### 4.1.1 Introduction

Aluminum is a widely used non-ferrous metal in numerous applications, including transportation, packaging, construction, and electrical transmission [96]. Several high-quality interatomic potentials for Al have been developed over the last three decades. These include multiple parameterizations of the embedded atom potential (EAM) [97-100]; each has been optimized to capture specific properties of Al. For example, the EAM potential parameterized by Voter and Chen (VC) shows especially good agreement with experimental data [101] in terms of cohesive energy, lattice constant, and elastic constants. In addition, Winey et al. used the atomic volume combined with the second and third order elastic constants in the fitting database to accurately capture the thermo-elastic response [102]. In contrast, Mendelev et al. developed an EAM potential to accurately simulate the crystalline and liquid states [103]. In some instances, not all of the properties are described well. For example, Oh, Johnson, and co-workers [104] developed a short-ranged EAM potential that does a good job reproducing the cohesive energy and lattice constant of Al, but also predicts unsatisfactory elastic constant values; Rohrer et al. [105] improved on this parameterization to enhance the potential’s ability to predict elastic constants. Numerous EAM potentials were developed specifically to allow for the modeling of mechanical deformation of Al and its alloys. One of the best of these is the parameterization of Mishin et al. [106] that was able to capture elastic constants and stacking fault values better than most other parameterizations. Subsequently, Liu et al.

[107] developed a similar potential based on the force-matching method that has improved elastic constant and stacking fault values. More recently, Jelinek et al. [108] used a modified embedded atom potential (MEAM) to obtain even better generalized stacking fault energies, thermal expansion coefficient, elastic moduli and defect formation energies. In addition to these potentials for pure aluminum, several interatomic potentials for Al alloys and mixtures have been developed. These include EAM potentials for Al-Cu [109,110], Al-Fe [111], Al-Mg [112], Al-Mn-Pd [113], Al-Ti [114], and Ni-Al-H [115]. Additionally, MEAM potentials for Al-Si-Mg-Cu-Fe [108] systems have been parameterized. They do reasonably well at capturing effects such as migrant atom segregation in intermetallic alloys and solid-liquid interfacial properties [111,112], but some of them are prone to overestimating the binding energies for cluster of atoms. While EAM parameters exist for non-metallic materials, it is not the preferred approach for these systems because of its lack of terms to address bond directionality [116]. Additionally, while MEAM potentials have been recently developed for $Al_2O_3$ [117], their exclusion of explicit Coulombic interactions and atomic charge limits the problems to which they may be applied.

A few dynamic charge-based interatomic potentials have also been developed. For instance, the ES+EAM potential of Streitz and Mintmire [118] is able to capture variable charge transfer between ions that is critical to model processes such as oxidation; it was first applied to examine the oxidation of Al. More recently, a ReaxFF potential was used to model interfacial phenomena associated with the $Al/Al_2O_3$ system, including structure, energetics, and adhesion [119]. Unfortunately, the surface energies of Al were not especially well described and the stacking fault energies were not reported.

In this work, we develop a third-generation COMB, or COMB3, potential for Al and illustrate its performance in several applications. This is the key first step in our on-going development of a COMB3 potential for the Al-O-N system.

**4.1.2 Potential Development :**

The training database consists of fourteen structures: the Al dimer; five solid-state phases, namely the face centered cubic (FCC), hexagonal close packed (HCP), body centered cubic (BCC), simple cubic (SC), and diamond structure; the three low-index Miller index surfaces of (111), (110) and (100); and defective structures with a vacancy, an interstitial atom, or one of three different stacking faults. The solid-state phases in the fitting database that are not present in the Al phase diagram are included to enable the parameterization of Al atoms as a function of the number and angular orientation of their neighbors.

The properties of all of these structures is determined using density functional theory (DFT) with the PBE (Perdew, Burke, and Ernzerhof)-GGA (generalized gradient approximation) density functional using the Vienna Ab initio Simulation Package (VASP) software package [120-123]. The calculations uses an 8x8x8 k-point mesh with a 600 eV energy cutoff; the self-consistency energy convergence criterion is taken as $10^{-6}$ eV and with $10^{-2}$ eV/Å force convergence criterion. Ultimately, the targets for the lattice constant, elastic constants, and bulk modulus of bulk FCC Al is based on experimental values, which are also close to DFT (Table 4-1), while the target values for cohesive energy, point defects, surface energies, and stacking fault energies is taken from the DFT calculations because the lack of individual experimental values for some of these quantities. The most modern method-based DFT data available in literature are chosen for target values during fitting.

Optimization of COMB3 parameters was carried out using the POSMat (Potential Optimization Software for Materials) [124] code, employing a simplex algorithm [125]. The optimization uses the weighted least square method for absolute deviation of lattice parameters, elastic constants, surface energies, defect formation energies and stacking fault energies. The COMB3 predictions and the parameterization target values are given in Table 4-1. These results are compared to the predictions of other selected potentials. As described in the previous section, there are numerous parameterizations of the EAM potential available to model Al. Here, we compare only to the EAM potential results parameterized by Mishin and co-workers [106] in addition to the predictions of ES+EAM [118] and ReaxFF [119].

### 4.1.3 Results and Discussion:

The COMB3 potential reproduces the correct phase order of solid-state structures and bulk properties for FCC Al that is comparable to the predictions of DFT and other empirical potentials. Although the predicted energy differences for the various phases of Al are not identical to the differences among the target values, they show a similar level of agreement as the predictions of ReaxFF, ES+EAM, and EAM potentials. The difference between the cohesive energy of the FCC and HCP phases is especially important because these phases have the same packing fraction and play important roles in the stacking fault, and hence mechanical response behavior. Legendre polynomial terms were included in the fitting scheme to distinguish between FCC and HCP phases, as was done previously for an earlier generation of the COMB potential for Cu [141].

As expected, the relaxed ideal Al (111) surface energy is predicted to be lower than the (100) surface energy, which is, in turn, predicted to be lower than the (110) surface energy, as indicated in Table 4-1; this is not a rigorous test as this order is predicted by a simple bond-breaking argument. The values calculated with DFT in the literature span a surprisingly wide range. The Al (111) surface energy predicted by COMB3 lies below the lower range of these literature values, while the values predicted for the Al (100) and Al (110) surface energies lie in the middle of their reported ranges. Thus, COMB3 successfully captures the trends in surface energies for Al. It should be noted that the EAM potential can also differentiate between these surfaces, while surface energies for different ones are not reported for ReaxFF.

Table 4-1 also illustrates that the COMB3 potential predicts the Al vacancy and interstitial energies to within 8% of the target data. As a result, the point defect formation energies predicted by the COMB3 potential are close to the DFT, EAM and ReaxFF values. The table further shows that the DFT stable (SF) and unstable stacking fault (USF) energies for Al vary over a wide range of value. To generate the stacking fault structure for use in the calculations, the mid-plane atoms of the (111) surface structure were displaced with a 0.01 nm step and the energies were determined and plotted in Figure 4-1. The first crest and trough correspond to the USF and SF, respectively. The SF predicted by COMB3 lies at the lower end of the range of values (Table 4-1) while the predicted USF lies in the middle of the range. It is known that in FCC metals with sufficient energy edge dislocations split into two partial dislocations [142] in the usual manner:

$$\frac{1}{2}[\bar{1}10] \rightarrow \frac{1}{6}[\bar{2}11] + \frac{1}{6}[\bar{1}2\bar{1}] \quad . \qquad (4\text{-}1)$$

COMB3 predicts the $\langle 101 \rangle$ SF to be 552 mJ/m$^{2,}$ while the SF and USF along $\langle 12\bar{1} \rangle$ are 146 and 194 mJ/m$^2$, respectively. The difference in SF energy values along the $\langle 12\bar{1} \rangle$ and $\langle 101 \rangle$ directions is clearly illustrated in the two-dimensional and three-dimensional contour plots of the stacking fault energies with directionality in Figure 4-2. The symmetry in the shapes of the contours verifies the periodic nature of the system. Despite the fact that the EAM and COMB3 contour plots in Figure 4-2 are quantitatively different from each other, the two potentials qualitatively predict the same overall stacking fault behavior.

The mechanical response of an Al polycrystal with four grains was considered using the LAMMPS software [19]; the polycrystal contained 252,822 atoms and is illustrated in Figure 4-3. The system was relaxed at 300 K using the NPT ensemble with the Nose-Hoover thermostat and a 0.1 femtosecond time-step. The tensile test was carried out using a constant strain rate of 2x10$^9$ s$^{-1}$; such a high strain rate is typical for simulations of nanocrystalline materials [143]. Common neighbor analysis (CNA) [144] of the 13.8% strained structure after a constant strain rate test (along the z-direction in Figure 4-3) is also provided. A large number of extended dislocations are predicted to form along the (111) plane. In addition, twinning behavior is observed in the simulations, as illustrated in Figs. 4-4A and 4-4B. Similar results were obtained for deformation of Al work carried out using EAM potentials by, e.g., van Swygenhoven et al. [145], while comparable experimental findings were reported by Liao et al. [146]. As discussed by Yamakov et al. [143], the deformation twinning predicted can either accelerate the deformation through the generation of additional slip planes or by facilitating dislocation-twin reactions. By contrast, newly formed twins can repel certain types of gliding dislocations and give rise to pile-ups that lead to strain hardening. Hence, the above mentioned observations are consistent with previous work.

As a further test of the potential, the phonon density of states (PDOS) and phonon-band structure of pure Al are determined with COMB3; the results are provided in Fig. 4-5 and 4-6. Some empirical potentials, such as the EAM parameterization of Mishin, [129], were explicitly parameterized for phonon properties, which was not done for this COMB3 Al potential. Nonetheless, the reasonable elastic constant values predicted, which are related to the slopes of the acoustic modes at the gamma point, result in physically meaningful dispersion and PDOS curves. In particular, the two peaks in the PDOS in Figure 4-5 correspond to the transverse (first peak) and longitudinal phonons (second peak); these peaks are ubiquitous for FCC metals [147]. The figure further illustrates that the COMB3 peak positions are in reasonable agreement with EAM and DFT values but their magnitudes differ.

Neither EAM nor COMB3 reproduce the DFT DOS, though EAM does seem to do a better job as its DOS peaks are in reasonably fair positions, which represent better mechanical responses. Peak positions are more important than the peak heights to correctly describe the lattice dynamics of the system. The COMB3 phonon energy is lower at the X, K and L points of the dispersion curve compared to DFT and EAM values. This is clearly seen in Figure 4-6, where the phonon dispersions are presented along the high symmetry lines in the first Brillouin Zone. The phonon behavior around the $\Gamma$ point is primarily a result of the fit, as the slopes of the dispersion curves are related to the elastic constants.

It should be noted that the COMB3 is not as computationally efficient as the EAM potential because of the number of many-body terms [148]. Other many-body potentials with similar formalisms, including ReaxFF and ES+EAM, have a computational cost that is approximately the same as COMB3 [148]. However, inclusion of Al in COMB3 will not only enable the modeling of this metal by COMB3, it is also the foundation of aluminum oxides, nitrides, hydrides, carbides, and other compounds and composites by the suite of COMB3 potentials.

## 4.2 COMB3 Potential for $Al_2O_3$

### 4.2.1 Introduction

Alumina is a widely used protective thin film material [149,150]. Additionally, composites of metals with alumina, including Al-$Al_2O_3$, Cu-$Al_2O_3$, and Ni-$Al_2O_3$, are used as protective coatings, castings, and in smelting processes [151-153]. Alumina is also used in nanoparticle form for catalysis [154-156] for its surface properties, and in thin film form in gate microelectronic devices [157] to take advantage of its high dielectric constant and high electron tunneling barrier. Currently, alumina nanowires and nanotubes [158] are being investigated for a range of applications including catalysis and cancer therapy [159]. Despite voluminous experimental and theoretical investigations [1] for alumina-based systems, atomic-level descriptions are needed to provide mechanistic insights into the structure-property relationships for its use in commercial and academic applications [118,119,155,160,161].

A high fidelity approach for the examination of Al-O systems is at the level of electronic structure calculations, typically with density functional theory (DFT) [20,162,163]. These types of calculations have been successfully used to investigate most stable Al-$Al_2O_3$ interfaces and the structural properties of amorphous alumina [164]. For example, the DFT calculations of Siegel et al. [165,166] and Zhang et al. [167] showed that the calculated work of adhesion of Al (111) with Al-terminated $Al_2O_3$ [160] is in excellent agreement with experiments. Recently, thermodynamic stability of defective Al-$Al_2O_3$ interfaces has been investigated by Kang et al. using ab-initio-molecular dynamics [168]. While clearly successful at providing insights into the properties of $Al_2O_3$ interfaces, DFT calculations are typically limited to small system sizes (<1000 atoms) [21] and hence their ability to investigate some properties, such as dislocations at interfaces or nanostructures such as nanowires, are very limited using commonly available resources.

Therefore, empirical methods are desirable for the study of materials at higher length and time scales than those achievable with DFT calculations [169]. Fixed charge empirical potentials have been developed in the past and have been successfully used to investigate the alumina system. For example, the Buckingham-type potential of Gale et al. [170] described this system's lattice constants within 1% and elastic constants within 30% accuracy, which are reasonably well. Vashishtha et al. [161] developed a fixed charge bond-order based potential for modeling the amorphous and liquid phases of alumina that predicts the liquid state density to be comparable to experimental data obtained through neutron and x-ray diffraction. The obvious limitation with such a potential is that it uses fixed charge on the ions, which prevents its applications to systems with bonding environments that are substantially different from bulk alumina.

A dynamically charged empirical potential for $Al_2O_3$ was developed by Streitz and Mintmire, who coupled a variable charge electrostatic potential with a charge independent Finnis-Sinclair formalism [171] to develop the Electrostatics Plus (ES+) [118] potential. This potential was used to investigate the mechanisms associated with the oxidation of Al surfaces and associated processes [118,172,173]. A significant drawback of their parameterization is that it predicts bixbyite to be the most stable phase of $Al_2O_3$. As an extension to the ES+ potentials, Zhou et al. [174] developed the charge transfer ionic potential (CTIP) by taking the non-electrostatic part of the total energy of the system in the form of a pair potential. The CTIP predicts the lattice constants, elastic constants, and surface energies of alumina quite well. Lazic et al. [175] showed that angular forces within a charge transfer ionic potential formalism is necessary to accurately predict *α*-alumina as the lowest energy configuration and introduced the 'Reference Free'' version of the Modified Embedded Atom Method (RFMEAM) for $Al_2O_3$. It is possible to attribute this angular force dependence to the specific formalism and use of embedding functions by the RFMEAM potential. It should be noted that the training database for this potential included various $Al_xO_y$ configurations for alumina, and the charges on Al and O for several systems were determined. Pilania et al. [176] successfully used this RFMEAM potential to examine the dislocation structure at Al-$Al_2O_3$ interfaces

Zhang et al. [119] developed a ReaxFF potential for Al-$Al_2O_3$ systems that includes both bond-order and dynamic variable charge. In this case, the authors noted that neither bond bending nor torsion terms are required to predict the correct phase order of $\alpha$-$Al_2O_3$. Rather, the correct structure is predicted when dispersion interactions and a penalty function for over-coordination are taken into account. The ReaxFF potential has been successfully used to explore various phenomena such as the properties of Al-$Al_2O_3$ interfaces and liquid alumina [119], in addition to the oxidation induced softening of aluminum [177].

Here, a third-generation, charge optimized many-body (COMB3) potential [178] for $Al_2O_3$ is developed and applied to the study of Al-$Al_2O_3$ interfaces and the deformation of Al, $Al_2O_3$, $Al_2O_3$-coated Al nanowires subjected to tensile stress.

**4.2.2 Potential Development**

The training database used to develop parameters consisted of twelve structures as follows: $\alpha$-$Al_2O_3$ (space group-$R\bar{3}c$), bixbyite (space group Ia3) , $\theta$-$Al_2O_3$ (space group C2/m), $AlO_2$ ($CaF_2$ structure, space group $Fm\bar{3}m$), $AlO_2$ ($TiO_2$, space group P 42/mnm), $Al_2O_3$ ($Fe_2O_3$, space group P4132), AlO (NaCl, space group $Fm\bar{3}m$), point defects-$V_{Al}$, $V_O$, $Al_i$, $O_i$ in α-$Al_2O_3$ and α-$Al_2O_3$ (0001) surface. It is important to note that the above structures probed different coordination for the cations and anions – for example $CaF_2$ has 8-4 (cation-anion) coordination, NaCl has 6-6, $TiO_2$ has 6-3, while that for $\alpha$-$Al_2O_3$ is 4-6. Thus these choices not only show that there are no obvious phases with lower energy, but also sample the likely bonding environments in any physically reasonable structure. The $\alpha$-$Al_2O_3$ structure is analyzed in the orthorhombic, rather than rhombohedral, representation to simplify the calculations. The solid-state phases in the fitting database, that are not present in the $Al_2O_3$ phase diagram are included to enable the parameterization of $Al_2O_3$ atoms as a function of the number and angular orientation of their neighbors.

The properties of these structures are determined using DFT with the PBE (Perdew, Burke, and Ernzerhof)-GGA (generalized gradient approximation) density functional using the Vienna Ab initio Simulation Package (VASP) software package [120-123]. The calculations use a 700 eV energy cutoff; the self-consistency energy convergence criterion is taken as $10^{-6}$ eV and with $10^{-2}$ eV/Å force convergence criterion. Ultimately, the targets for the lattice constant, elastic constants, cohesive energy, point defects, surface energies, elastic constants and bulk modulus of α-alumina are used in the fitting database from the DFT calculations, as indicated in Table. 4-2. Here, lattice constants and heat of formation data are from experimental measurements, while other data are from DFT calculated values.

The relative stability of the phases is determined based on their heats of formation and heats of reaction needed to form them. Specifically, the heat of formation is the energy required to form the material from its constitutive elements. For example, in the case of $Al_2O_3$, face-centered cubic (FCC) Al and an isolated, gas-phase $O_2$ molecule, respectively, as illustrated in Eq. (4-1).

$$xAl + y\frac{1}{2}O_2 \rightarrow Al_xO_y, \quad \Delta H_f = (x+y)E_{atom} - x\mu_{Al} - y\mu_{\frac{1}{2}O_2} \tag{4-1}$$

Here, x and y are stoichiometric coefficient, $\Delta H_f$ is the heat of formation and $\mu$ is the chemical potential of chemical species.

After the heat of formation for each individual compounds is calculated, the heat of reaction is calculated as the energy difference between phases, such as, for example, the bulk α-$Al_2O_3$ phase (A) and a different phase (B):

$$Al_mO_n(A) \rightarrow Al_xO_y(B) \tag{4-2}$$

$$\Delta H_{rxn} = (m+n)E_{atom(A)} - (x+y)E_{atom(B)} - (x-m)\mu_{Al} - (y-n)\mu_{\frac{1}{2}O_2} \tag{4-3}$$

For the $\alpha$-$Al_2O_3$ phase to be the ground state, the heats of reaction for other phases should be positive. Optimization of COMB3 parameters was carried out using the POSMat (Potential Optimization Software for Materials) [124] software, which employs a simplex algorithm [125]. The optimization uses the weighted least squares method for absolute deviation of lattice parameters, elastic constants, surface energies and defect formation energies. The parameters were confined by attempting to keep the elastic constant, bulk modulus and other properties mentioned above to be at maximum 20% deviation from the target values. Appropriate choice of weighting factors during the fitting procedure played important role in getting the current parameters (see ref. [178] for details). The COMB3 predictions and the parameterization target values are given in Table 4-2. and the corresponding parameters developed for the $Al_2O_3$ COMB3 potential are given in the Appendix. The COMB3 parameters are also distributed in the LAMMPS open source MD software [179].

### 4.2.3 Results

The values of the bulk and surface properties of alumina provided by the COMB3 potential are provided in Table 4-2 and compared to experimental data, the results of DFT calculations, as well as published data obtained using the ReaxFF potential [180], the potential of Vashishtha et al. [161] and the ES+ potential [118]. The heats of reaction for all the phases considered besides α-alumina were predicted to be positive by COMB3, which correctly predicted the corundum α-$Al_2O_3$ phase to be the lowest energy state amongst the database structures studied here. In addition, the average charges on the Al and O in bulk alumina are +1.22e and -0.82e, respectively. These values are comparable with the Born-effective charges of +1.67e and -1.21e correspondingly obtained from the DFT calculations [181]. Table 4-2 shows the defect formation energy for interstitials, vacancies using

$$E^{f} = E^{def} - E^{perf} \pm \sum_{i} n_i \mu_i \quad , \tag{4-4}$$

where $E_{def}$ and $E_{perf}$ are defective and perfect structure energies of α-alumina, $\mu_i$ is the chemical potential of the atom removed or added, and $n_i$ is the number of defects. Although defect formation energies were included in the fitting, compromises had to be made to produce good values for the heats of formation and elastic constants.

Consequently, the Al vacancy and O interstitial defect formation energy values are underestimated by COMB3 compared to the DFT target values, but the non-negative nature of the defect formation energies was an important outcome that ensures phase stability. The phase order for Al-O systems studied here from DFT was $\alpha$-$Al_2O_3$ < $\theta$-$Al_2O_3$ < Bixbyite < $Fe_2O_3$ ($Al_2O_3$) < $TiO_2$ ($AlO_2$) < $CaF_2$ ($AlO_2$) < NaCl (AlO), while that from COMB3 was α-$Al_2O_3$ < Bixbyite < θ-$Al_2O_3$ < $Fe_2O_3$ ($Al_2O_3$) < $CaF_2$ ($AlO_2$) < $TiO_2$ ($AlO_2$) <NaCl (AlO). The charges on Al and O in various phases of $Al_2O_3$ are provided in Table 4-3.

In addition to the defect formation energies, the parameters were further refined through fitting to the surface energies. To simulate the surfaces 3 nm of vacuum was added to the bulk alumina (mentioned above) in the direction normal to the (0001) surface. The surface energies were calculated as

$$\sigma = \frac{E_{slab} - E_{bulk}}{2A} \tag{4-5}$$

where $E_{slab}$ is the total energy of the slab, $E_{bulk}$ is the energy of the bulk structure, and $A$ is the surface area of the (0001) surface. The surface energy was found to be overestimated compared to the DFT values and shows room for additional improvement in the parametrization.

Overall, COMB3 qualitatively predicts phase stability, lattice constants, elastic constants, defect formation energies and surface energies for α-alumina. Quantitative agreement is also generally good, but various underestimations in defect formation energies and deviations from target DFT surface energies prevent perfect quantitative agreement for all properties.

To test the dynamical stability, the phonon density of states (PDOS) for *α*-alumina was calculated using PhonTS software [194] with interatomic interactions described by DFT and by the COMB3 potential. The low energy acoustic modes in the PDOS (Figure 4-7) are the responses around the gamma point in the first Brillouin zone. The COMB3 acoustic modes are in good agreement with the DFT because we explicitly fit the elastic constants for the system. The energies of the van Hove PDOS peaks [195] at 54 and 98 meV are in fair agreement with the DFT values at 49 and 92 meV, respectively. The higher end optical modes are similar but slightly off compared to DFT (by about 5 meV). The differences might be because we didn't explicitly fit to all of these perturbed structures. However, the absence of imaginary phonon frequency signifies that the α-alumina structure should be dynamical stable.

## 4.3 COMB3 Potential for AlN

### 4.3.1 Introduction

Aluminum nitride is an important III-V semiconductor material for its wide band gap, high thermal conductivity, electrical insulation, thermal expansion and non-toxicity [196]. Nanostructures of AlN are used for field emitters in flat panel displays [197] and photo-detectors [198]. Additionally, Al-O-N is also an important engineering material for its use in thin-film growth [199-203] and fabrication of transparent bulletproof armor of aluminum oxinitride. Atomistic understanding for various thermodynamic and mechanical phenomenon in these systems can play critical role in optimizing the device performance [204,205]. While transmission electron microscopy (TEM), calorimetry, atomic force microscopy (AFM) are experimental approaches [13]for getting insights of the atomistic phenomenon, theoretically, density functional theory (DFT) and molecular dynamics (MD) [206] can provide relevant information based on quantum and classical mechanics respectively. To list some of the important DFT work for Al-O-N systems, Felice and Northrup [207] used DFT to understand the effect of chemical potential of Al on formation energies and structures of AlN on $Al_2O_3$. Ogata et al. [208] used DFT to demonstrate the relative behavior of mechanical strength of Al/AlN system compared to Al and AlN. Li et al. [209] used DFT for AlN nanowires to find relative efficiency of atomic and molecular hydrogen storage in AlN nanowires and nanotubes. While DFT is a compelling high-fidelity quantum mechanical approach for simulation, it generally suffers from the problem of system size limitation which are important in relating to practical applications due to limitations such as lattice mismatch of interfacial systems. Molecular dynamics (MD) is an important tool meant for relatively larger system, but it needs force-filed/empirical potential for classical dynamics simulation. Again, empirical potentials are also of various types, such as fixed charge, dynamic charge, non-reactive and reactive empirical potential. A short description for molecular dynamics approach can be given as follows. A Buckingham type of fixed charge potential was put forward by Chisholm et al. [210] for AlN. The potential had reasonable lattice constant, elastic constant and defect formation energies for AlN. Various Tersoff-based interatomic potential have also been proposed AlN [211-213] and have been successfully utilized for thermal expansion coefficient [211] and AlN nanotubes [214] . Vashishtha et al. [215] developed three-body interaction potential with modification of the Stillinger–Weber potential for AlN to study the mechanical and thermal properties of crystalline and amorphous AlN. The potential did well in predicting the lattice constant, elastic constant, vibrational density of states.

**4.3.2 Potential Development**

The training database used to develop parameters consisted of twelve structures as follows: wurtzite-AlN (space group- $P6_3/mmc$), zinc blende-AlN (space group $F\bar{4}3m$) , rock-salt (space group $Fm\bar{3}m$), AlN ($CaF_2$ structure, space group $Fm\bar{3}m$ ), CsCl ( space group Pm3m), point defects (vacancies and interstitials)-$V_{Al}$, $V_N$, $Al_i$, $n_i$ in w-AlN and the w-AlN (100) non-polar surface. The w-AlN structure analyzed in the orthorhombic, rather than rhombohedral, representation to simplify the calculations. The solid-state phases in the fitting database are included to enable the parameterization of AlN atoms as a function of the number and angular orientation of their neighbors.

The properties of these structures are determined using DFT with the PBE (Perdew, Burke, and Ernzerhof)-GGA (generalized gradient approximation) density functional using the Vienna Ab initio Simulation Package (VASP) software package [120-123]. The calculations use a 500 eV energy cutoff; the self-consistency energy convergence criterion is taken as $10^{-6}$ eV and with $10^{-2}$ eV/Å force convergence criterion. Ultimately, the targets for the lattice constant, elastic constants, cohesive energy, point defects, surface energies, elastic constants and bulk modulus of w-AlN are used in the fitting database from the DFT calculations, as indicated in Table. 4-4.

Here, lattice constants, heat of formation data are from experimental measurements, while other data from DFT calculated values.

The relative stability of the phases is determined based on their heats of formation and heats of reaction needed to form them. Specifically, the heat of formation is the energy required to form the material from its constitutive elements. For example, in the case of AlN, face-centered cubic (FCC) Al and an isolated, gas-phase $N_2$ molecule, respectively, as illustrated in Eq. (4-6).

$$xAl + y\frac{1}{2}N_2 \rightarrow Al_xN_y,\ \Delta H_f = (x+y)E_{atom} - x\mu_{Al} - y\mu_{\frac{1}{2}N_2} \tag{4-6}$$

After the heat of formation for each individual compounds is calculated, the heat of reaction is calculated as the energy difference between phases, such as, for example, the bulk w-AlN phase (A) and a different phase (B):

$$Al_mN_n(A) \rightarrow Al_xN_y(B) \tag{4-7}$$

$$\Delta H_{rxn} = (m+n)E_{atom(A)} - (x+y)E_{atom(B)} - (x-m)\mu_{Al} - (y-n)\mu_{\frac{1}{2}N_2} \tag{4-8}$$

For the w-AlN phase to be the ground state, the heats of reaction for other phases should be positive values. Optimization of COMB3 parameters was carried out using the POSMat (Potential Optimization Software for Materials) [124] software, which employs a simplex algorithm [125]. The optimization uses the weighted least squares method for absolute deviation of lattice parameters, elastic constants, surface energies and defect formation energies. The parameters were confined by keeping the elastic constant, bulk modulus and other properties mentioned above to be within 20% of the target values.

### 4.2.3 Results

The COMB3 predictions and the parameterization target values are given in Table 4-4 and the corresponding parameters developed for the AlN COMB3 potential are distributed in LAMMPS open source MD software [179]. As the fitting database for $Al_2O_3$ and AlN mentioned here consisted of limited structures, it was important to investigate the performance of the potential using random structure search algorithm such as genetic algorithm. Genetic algorithm using GASP software package [216] was employed to map the energy landscape of the AlN and $Al_2O_3$ COMB3 potentials by searching possible low energy structures. Genetic algorithm (GA) is a heuristic optimization algorithm based on biological process of evolution of survival of fittest species. Here, the species in GA search were the low energy structures and their fitness were calibrated based on the energies of the structures obtained from COMB3 potential. The structures used in fitting database were used as an initial guesses in parent generation of GA. New structures were then constructed using genetic algorithm using mutation and mating operations on the parent structures. After subsequent generations in genetic algorithm search, phase diagrams (or convex hull plot) were predicted in Fig. 4-8. The algorithm was constrained within some hard parameters: maximum and minimum lattice parameters, minimum bond-lengths of Al-O and Al-N system (about 80 % of equilibrium bond length), maximum number of generations (here 40), maximum number of structures in each generation (here 30). Hence, total 1200 structures were searched and their energies were calculated using COMB3. Only the structures with negative energies were shown in the phase diagram. The primitive structures were replicated in supercell of at least 21 Å size to meet long range coulomb cut off criteria for COMB.

The structures lying on the phase diagram curve (also known as convex hull plot) are generally stable. Any other structure not lying on the curve can be represented as the mixture of materials of the stable structures. For Al-O phase diagram, $Al_2O_3$ was found on the curve. Additionally, AlO structure was also found in the convex hull plot. Experimentally or computationally (using materials project data ) [217], AlO is not found in the phase diagram of Al-O systems. For Al-N phase diagram search, AlN phase (with composition fraction 0.5 in the Fig. 4-8B) was found to be ~0.05 eV energy/formula unit above the convex hull curve. Instead $AlN_4$ was found to be the stable phase in the Al-N phase diagram search. Similar to the Al-O case, $AlN_4$ is not an stable structure in experimental or computational phase diagram for Al-N system. Such behavior were also observed during ReaxFF parametrization for $LiS_2$ potential [218]. These results can be considered as artifacts of the limited fitting database in our parametrization procedure and show scope of further improvement of the parameters.

### 4.4 Al-$Al_2O_3$ Interfaces

To further test the reliability of the $Al_2O_3$ potential, the work of adhesion for Al(111)-$Al_2O_3$ interface was calculated for both Al- and O-terminated $Al_2O_3$ (0001), as these interfaces are of major technical importance [223]. DFT calculations [224] predict that the clean Al-$Al_2O_3$ interface may be either O- or Al-terminated depending on the ambient oxygen partial pressure. Recent theoretical [168,225] and experimental [226] studies further indicate that self-regulated Al vacancies at the Al-terminated interface lead to the most stable interfacial energies.

Both the aluminum and alumina slabs were 4 nm thick, with 1080 atoms in the Al slab and 1800 atoms in the $Al_2O_3$ slab. Due to the lattice constant mismatch between Al and $Al_2O_3$, the Al structure was strained to match the $Al_2O_3$ by compressing it as follows:

$$\delta_{Al} = \frac{a - a'}{a} = 0.041 \quad (4\text{-}9)$$

where $\delta_{Al}$ is the change in aluminum lattice constant from a to a'.

The work of adhesion was calculated as follows:

$$W_{adh} = (E^{Tot}_{Al_2O_3} + E^{Tot}_{Al-slab} - E^{Tot}_{Al_2O_3+Al-slab}) / A \quad (4\text{-}10)$$

where, $W_{adh}, E^{Tot}_{Al_2O_3+Al-slab}, E^{Tot}_{Al_2O_3}, E^{Tot}_{Al-slab}, A$ are the work of adhesion for the interface, total energy of the interface, total energy of the (0001) alumina surface structure, total energy for the strained (111) aluminum surface structure, and surface area of the interface, respectively. The work of adhesion for Al (111)-$Al_2O_3$ (0001) with Al termination (Al occupying the face-centered cubic, or FCC, sites of alumina) as indicated in Figure 4-9D was calculated to be 0.95 J/m$^2$, which is in excellent agreement with the DFT measured value of 1.13 J/m$^2$. Additionally, the O-terminated $Al_2O_3$ (0001) with Al (111) (Al on oxygen top, OT site) as indicated in Figure 4-9E yields a work of adhesion of 9.47 J/m$^2$, which is also in good agreement with the DFT-determined value of 8.75 J/m$^2$. The naming-convention used to describe these systems is similar to that used by Siegel et al. [166].

In the case of the Al (111)-Al-terminated $Al_2O_3$ (0001) interface, the charges on the topmost Al and O atoms within the $Al_2O_3$ surface are 0.75e, and -0.64e, respectively. On the other hand, these charges for the Al (111)-O-terminated $Al_2O_3$ (0001) system are 1.29e and -0.43e, respectively. These charges were consistent with the findings of DFT calculations by Li et al. [227] and suggests strong ionic-covalent bonds. It is important to note that the energy associated with straining Al was nullified using Eq. 4-10, so no further correction was needed for the extra strain introduced to Al.

Two additional cases were considered for the Al (111)-Al-terminated $Al_2O_3$ (0001) interfacial system. Specifically, Al on the hexagonal closed packed (HCP) and on the oxygen top (OT) sites were considered by shifting the Al (111) slab. The work of adhesion trends were compared with universal binding energy relation results [228] by Siegel et al. [166]. Using the labels shown in Table 4-5, the trend predicted by Sigel et al. [166] was A<B<C<D while the COMB3 results were A<C<B<D. The inconsistency associated with the B and C cases can again be explained by differences in the details of the potential energy surfaces predicted by DFT and COMB3. It is important to note that none of the coherent interfacial structures were part of the potential fitting database and the results obtained for the work of adhesion are pure predictions. However, the overall trends for the work of adhesion with different terminations are consistent with both DFT and experiment, as indicated by Table 4-5. One way to interpret the overall work of adhesion trend is that it is more energetically favorable to break Al-Al bonds, while it is more energetically favorable to form Al-O bonds. The fact that COMB3 consistently predicts this important behavior gives confidence in its application to systems with a heterogeneous bonding environment.

### 4.5 Al-AlN-$Al_2O_3$ Interfaces

To further test the reliability of the potential, the work of adhesion for Al (111)-AlN (0001) and AlN (0001)- α-$Al_2O_3$ (0001) interfaces were investigated as they are of major technical importance [229]. Furthermore, we used Al and N terminated AlN (0001) surface for making the interface structures. So the interface structures considered in this study are: A) Al-terminated AlN (0001)/Al (111), B) N-terminated AlN (0001)/Al (111), C)

Al-terminated AlN (0001)/Al-terminated $Al_2O_3$ (0001), D) N-terminated AlN (0001)/Al-terminated $Al_2O_3$ (0001). Furthermore, we used the interface building algorithm by Zur et al. [25] to build interfaces with minimal mismatch. The algorithm was designed to search for all possible matches of surfaces to make interface structures. The interface structures are shown in Fig. 4-10.

The Al(111)-AlN(0001) three dimensional periodic box consisted of 38808 Al (111) and 64680 AlN(0001) atoms respectively. The mismatch between the two surfaces was 0.04. Strain was applied on the metallic part of the interface, because ceramics are generally brittle in nature. Periodic boundary conditions were imposed in x and y dimensions. We calculated work of adhesion for Al(111)-AlN(0001) to be 2.70 and 2.05 J/m$^2$ for case A and B using eq. (4-11):

$$W_{adh} = (E^{Tot}_{slab-1} + E^{Tot}_{Slab-2} - E^{Tot}_{Intf}) / A \quad (4\text{-}11)$$

where $W_{adh}, E^{Tot}_{Intf}, E^{Tot}_{Slab-1}, E^{Tot}_{Slab-2}, A$ are the work of adhesion for the interface, energies of slab-1 and slab-2 and A is the area of the interface. The work of adhesion for this system from DFT is 1.8 J/m$^2$ [230] while the experimental value is 1.42 J/m2 [200]. The Al-N equilibrium bond length was found to be 0.22 nm at the interface. As the Al atoms near the AlN surface acquire a positive charge (0.11 e), partial ionic character was predicted for the interface Al atoms. The work of adhesion value indicates that the AlN COMB3 potential is in good agreement with experimental as well as previous first-principles calculation data.

Next, we calculated the work of adhesion for $Al_2O_3$ (0001)-AlN (0001) system consisting of 60840 $Al_2O_3$ and 47320 AlN atoms. As the system consisted of ceramic materials, mismatch was constrained within 0.01. It is to be noted that for the present system DFT calculations are too computationally expensive hence they were not carried out here. The work of adhesion was determined to be 0.37 and -0.41 J/$m^2$for case C and D respectively. These values are much lesser than the Al(111)-AlN(0001) work of adhesion energy. It indicates that the Al(111)-AlN(0001) interface should be more thermodynamically stable than the $Al_2O_3$ (0001)-AlN (0001) interface. Interface charge transfer was again clearly predicted to take place. The Al-Al bond length found in the case of $Al_2O_3$-AlN interface case was 0.31 nm. The results indicate Al-terminated AlN can make relatively stable interfaces compared to N-terminated cases for both Al(111) and $Al_2O_3$ (0001). Similar, results for Al-terminated α-$Al_2O_3$ has been predicted before that Al-termination makes the most stable interfaces for α-$Al_2O_3$ [224,231].

## 4.6 Summary

A newly developed COMB3 potential for Al-$Al_2O_3$-AlN was shown to capture the key physical properties of respective polymorphs; it can also be applied to Al/$Al_2O_3$ and Al/AlN interfacial systems. Most importantly, it can be seamlessly coupled with recently developed COMB3 potentials for other materials to enable MD simulation studies of a wide range of heterogeneous material systems. The new potential should therefore prove to be a useful new tool for the computational toolbox and an effective method for carrying out large-scale atomistic simulations of systems of technological importance.

Table 4-1. Properties of Al determined with COMB3 properties compared with experiments, DFT and other empirical potentials.

| | *Exp. or DFT* | *EAM*[d] | *ES+* | *ReaxFF* | *COMB3* |
|---|---|---|---|---|---|
| FCC Properties | | | | | |
| a (Ǻ) | 4.05[a] | 4.05 | 4.05 | 4.01 | 4.05 |
| $E_0$ (eV/atom) | -3.36[b] | -3.36 | -3.39 | -3.36 | -3.36 |
| B (GPa) | 79.0[c] | 77.0 | 83.0 | 79.0 | 83.0 |
| G (GPa) | 26.0[c] | 28.0 | | | 33.0 |
| $C_{11}$ (GPa) | 114.0[c] | 114.0 | 94.0 | 119.0 | 113.0 |
| $C_{12}$ (GPa) | 62.0[c] | 61.6 | 77.0 | 57.0 | 66.0 |
| $C_{44}$ (GPa) | 32.0[c] | 31.6 | 34.0 | 50.0 | 39.0 |
| Phase Transitions (eV/atom) | | | | | |
| ΔE(HCP-FCC) | 0.03[d] | 0.03 | | 0.0016 | 0.013 |
| ΔE(BCC-FCC) | 0.11[d] | 0.12 | | 0.05 | 0.05 |
| ΔE(SC-FCC) | 0.33[d] | 0.33 | | 0.37 | 0.7 |
| ΔE(DIAM-FCC) | 0.67[d] | 0.67 | | 0.70 | 1.2 |
| Planar defects ($mJ/m^2$) | | | | | |
| γ (111) | 980[e],855[f],750 | 870 | | 576 | 720 |
| γ (100) | 980[e],1209[f],840[g] | 943 | | 576 | 908 |
| γ (110) | 980[e],1286[f],910[g] | 1006 | | 576 | 1067 |
| $\gamma_{ISF}$ $\langle 12\bar{1} \rangle$ | 116,160[h],120-144[i],164[j] | 146 | | | 146 |
| $\gamma_{USF}$ $\langle 12\bar{1} \rangle$ | 154, 224[j], 291[k] | 168 | | | 194 |
| $\gamma_{SF}$ $\langle 101 \rangle$ | 250[j], 663[k] | - | | | 552 |
| $\Delta H_f$ Point defects (eV) | | | | | |
| $V_{Al}$ | 0.68[m] | 0.67 | | 0.85 | 0.69 |
| $Al_i$ | 2.8[n] | 2.79 | | | 2.6 |

Ref. [a] [126], Ref. [b] [127], Ref. [c] [128], Ref. [d] [129], Ref. [e] [130], Ref. [f] [131], Ref. [g] [132], Ref [h] [133], Ref. [i] [134,135], Ref. [j] [136], Ref. [k] [137], Ref [l] [138], Ref [m] [139], Ref [n] [140]

Table 4-2. Properties of $\alpha$-$Al_2O_3$ predicted by the indicated computational and experimental methods.

| $\alpha$-$Al_2O_3$ | Exp./ DFT | [118] ES+ | [180] ReaxFF | [161] Priya et al. | COMB3 |
|---|---|---|---|---|---|
| a (nm) | 0.476[a] | 0.476 | 0.481 | 0.476 | 0.476 |
| c (nm) | 1.30[a] | 1.30 | 1.31 | 1.30 | 1.31 |
| $\Delta H_f$ (eV) | -17.37[b] | -17.37 | -23.54 | -17.29 | -17.75 |
| B (GPa) | 254.0[c] | 250 | 248 | 255 | 249.0 |
| $C_{11}$ (GPa) | 497[c] | 537 | | 498 | 458 |
| $C_{12}$ (GPa) | 164[c] | 180.0 | | 163 | 141 |
| $C_{13}$ (GPa) | 111[c] | 106.0 | | 117 | 132 |
| $C_{14}$ (GPa) | -24[c] | -30.0 | | -23 | -2.5 |
| $C_{33}$ (GPa) | 498[c] | 509.0 | | 502 | 518 |
| $C_{44}$ (GPa) | 147[c] | 130.0 | | 147 | 136 |
| $C_{66}$ (GPa) | 168[c] | 179.0 | | 167 | 153 |
| $\Delta H_{rxn}$ (Bixbyite-$\alpha$) | 0.11 | | | | 2.9 |
| $\Delta H_{rxn}$ (Theta-$\alpha$) | 0.02 | | | | 4.02 |
| $\Delta H_{rxn}$ (NaCl-$\alpha$) | 14.25 | | | | 9.4 |
| $\Delta H_{rxn}$ ($TiO_2$-$\alpha$) | 9.56 | | | | 8.9 |
| $\Delta H_{rxn}$ ($CaF_2$-$\alpha$) | 10.54 | | | | 7.19 |
| $\Delta H_{rxn}$ ($Fe_2O_3$-$\alpha$) | 4.97 | | | | 5.92 |
| O vacancy | 6.09[d] | | | | 4.6 |
| Al vacancy | 8.44[d] | | | | 2.24 |
| O interstitial | 8.84[d] | | | | 6.39 |
| Al interstitial | 16.8[d] | | | | 11.32 |
| $\gamma$ (0001) ($J/m^2$) | 1.5-3.5[e] | | 1.0 | | 4.2 |

Ref. a [182], Ref. b [183], Ref. c [184], Ref. d [185], Ref. e [186-193].

Table 4-3. Average charges on Al and O for the phases of $Al_2O_3$ that are included in the fitting database.

| *Phases* | $q_{Al}$ *(e)* | $q_O$ *(e)* |
|---|---|---|
| α-$Al_2O_3$ | 1.22 | -0.82 |
| Bixbyite | 1.32 | -0.88 |
| Theta | 1.27 | -0.86 |
| NaCl | 1.00 | -1.00 |
| $TiO_2$ | 1.39 | -0.69 |
| $CaF_2$ | 1.39 | -0.69 |
| $Fe_2O_3$ | 1.32 | -0.95 |

Table 4-4. Properties of w-AlN predicted by the indicated computational methods and by experimental methods.

| AlN | Exp./ DFT | Chisholm et al. [210] | Vashishtha [215] | Tersoff [211] | COMB3 |
|---|---|---|---|---|---|
| a (nm) | 0.311[a] | 0.311 | - | 0.311 | 0.311 |
| c/a | 1.601[a] | 1.60 | - | 1.60 | 1.62 |
| $\Delta H_f$ (eV/atom) | -1.56 [a] | | | | -1.30 |
| B (GPa) | 228[b] | 248 | 211 | 210 | 218 |
| $C_{11}$ (GPa) | 464[b] | 417 | 435 | | 463 |
| $C_{12}$ (GPa) | 149[b] | 178 | 148 | | 92 |
| $C_{13}$ (GPa) | 116[b] | 152 | 107 | | 104 |
| $C_{33}$ (GPa) | 409[b] | 432 | 356 | | 437 |
| $C_{44}$ (GPa) | 128[b] | 125 | 81 | | 194 |
| $\Delta H_{rxn}$ (ZB-WZ) | 0.05 | | | | 0.25 |
| $\Delta H_{rxn}$ (NaCl-WZ) | 0.4 | | | | 2.2 |
| $\Delta H_{rxn}$ (CsCl-WZ) | 4.03 | | | | 1.37 |
| $\Delta H_{rxn}$ ($CaF_2$-WZ) | 4.64 | | | | 2.24 |
| $\Delta H_{rxn}$ ($P6_3$/mmc-WZ) | 0.27 | | | | 1.21 |
| N vacancy | 6.79,1.44-5.36 [c] | | | | 4.8 |
| Al vacancy | 7.97,2.36-6.31 [c] | | | | 0.30 |
| N interstitial | 3.81 | | | | 8.92 |
| Al interstitial | 11.1 | | | | 12.8 |
| γ (10-10) ($J/m^2$) | 2.3 | | | | 2.13 |
| γ (0001) ($J/m^2$) | 5.8 [d] | | | | 2.4 |

Ref. a [219], Ref. b [220], Ref. c [221], Ref. d [222]

Table 4-5. Interface properties of Al-$Al_2O_3$ predicted by DFT [166] and COMB3.

| | | | $d_0$ *(nm)* | | $W_{ad}$ *(Jm⁻²)* | |
|---|---|---|---|---|---|---|
| | Stacking | Termination | DFT | COMB3 | DFT | COMB3 |
| A | FCC | Al | 0.255 | 0.236 | 1.14 | 0.95 |
| B | HCP | Al | 0.226 | 0.270 | 1.33 | 1.07 |
| C | OT | Al | 0.209 | 0.271 | 1.55 | 1.06 |
| D | OT | O | 0.171 | 0.211 | 9.43 | 9.47 |

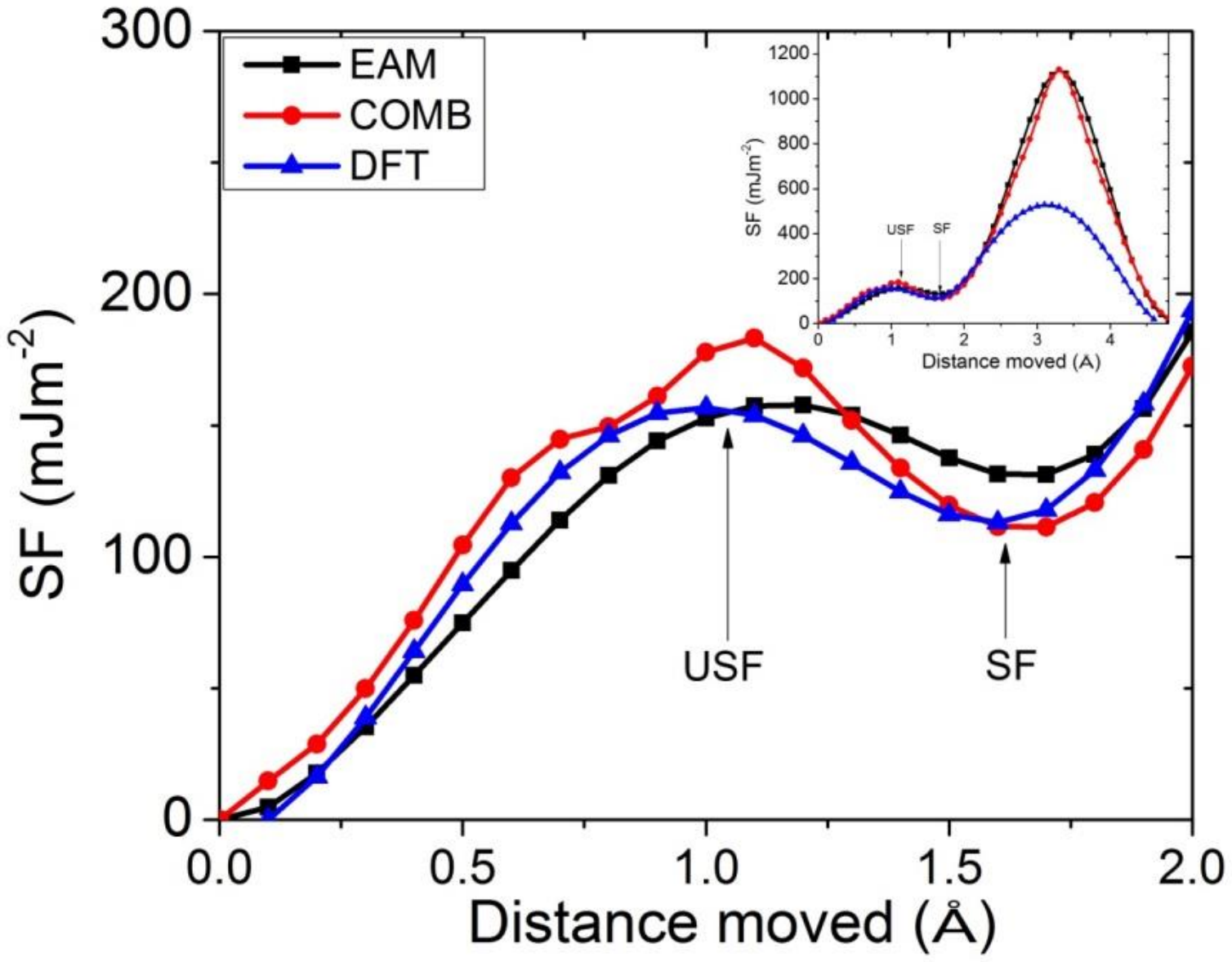

Figure 4-1. Comparison of Al stacking fault behavior as predicted by DFT and the COMB3 and EAM potentials.

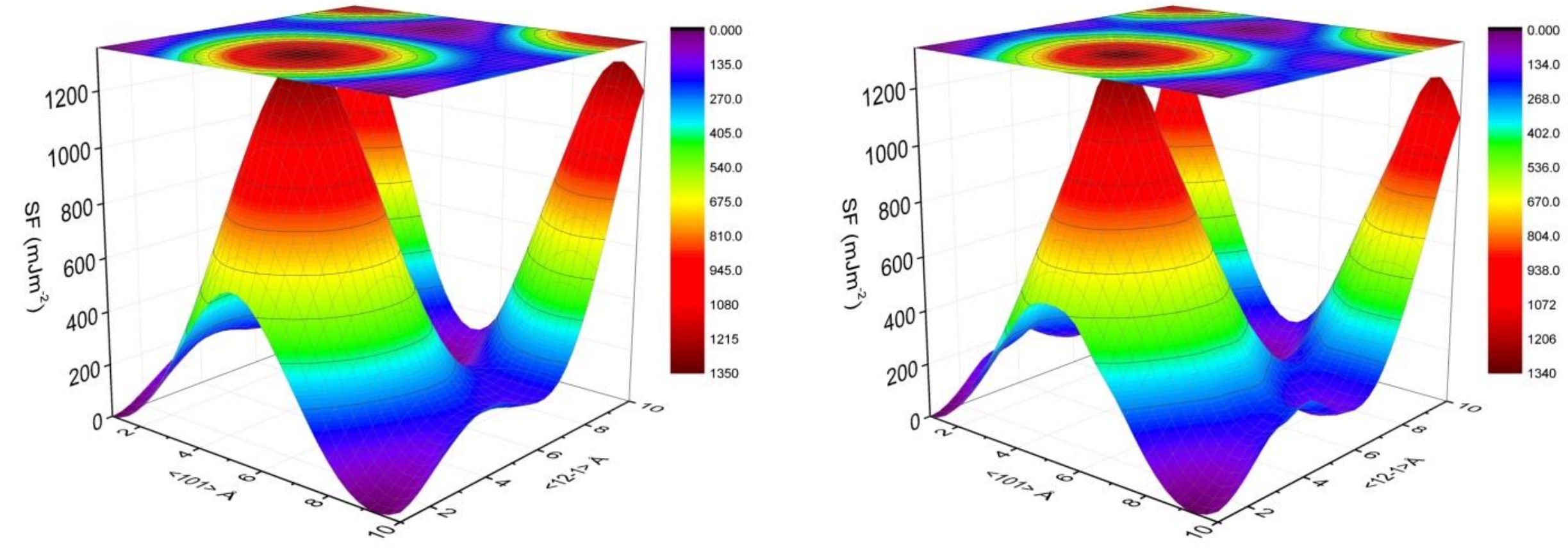

A) B)

Figure 4-2. Stacking fault energies illustrated in three-dimensions. A) EAM and B) COMB3, which correctly predict that the $\langle 12\bar{1} \rangle$ direction is preferred over $\langle 101 \rangle$ for dislocation propagation.

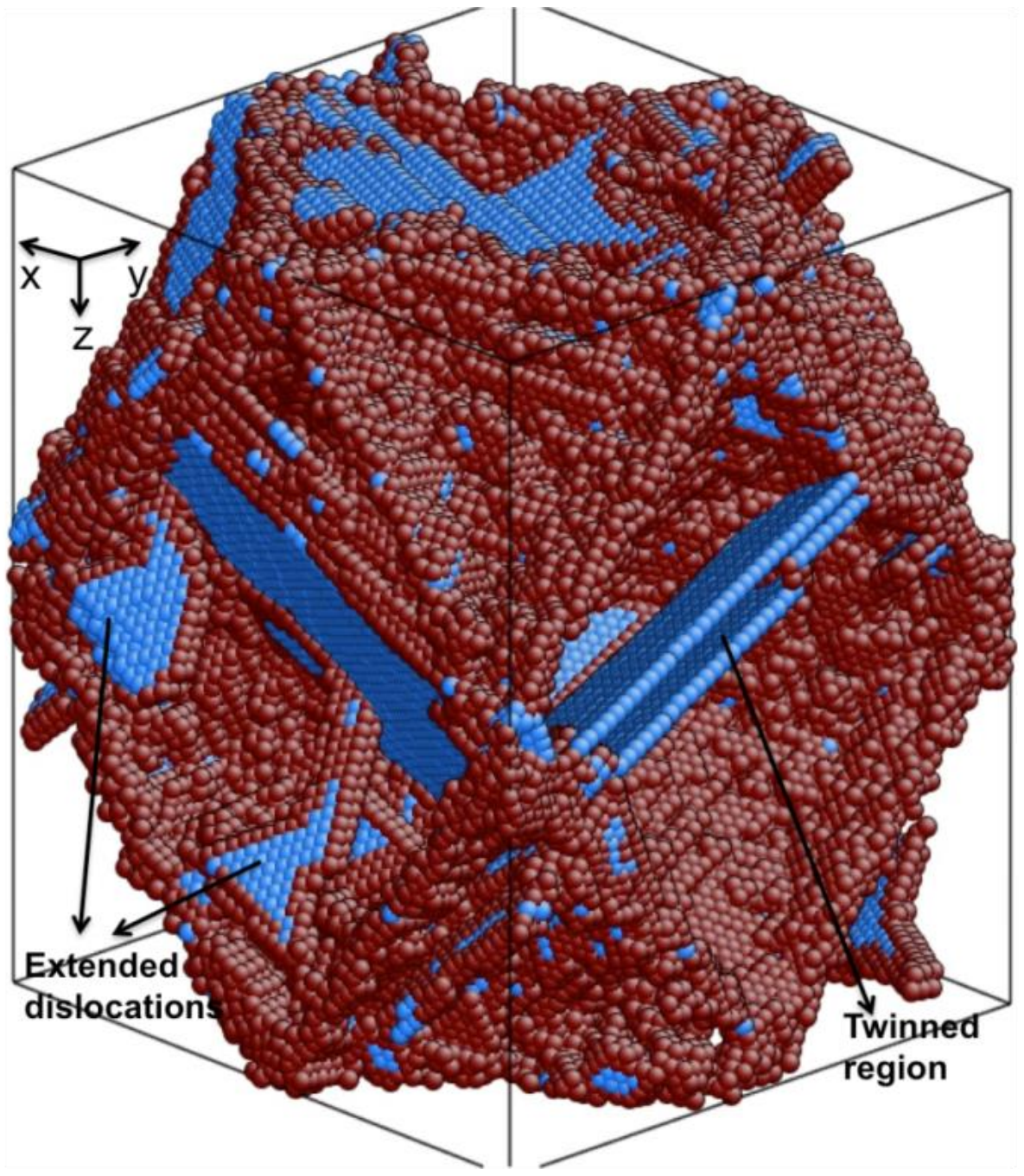

Figure 4-3. Snapshot of CNA analysis of a 13.8% strained, four-grain polycrystal of Al following constant strain rate along the z-direction using the COMB3 potential. Red spheres represent disordered atoms that lack 12-fold coordination. The light blue spheres represent atoms in an HCP environment, and the FCC atoms are not shown for clarity.

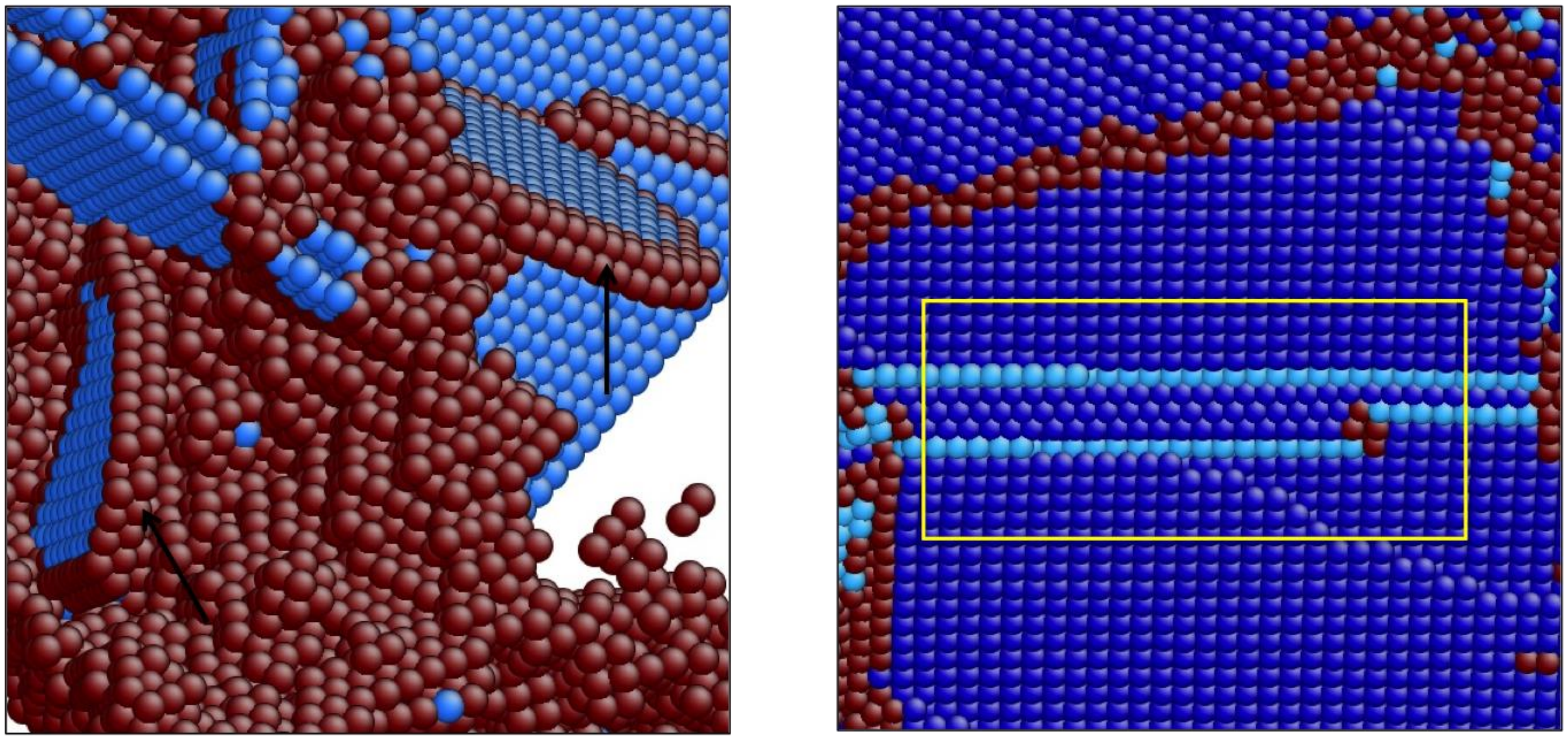

A B

Figure 4-4. Snapshpts of Al-polycrystal deformation simulation. A) Extended dislocations on [111] plane are shown. The black arrows point to the extended dislocations. B) Twinned region. The yellow square highlights twinned region. Red spheres represent disordered atoms that lack 12-fold coordination. The light blue spheres represent the atoms in HCP environment. The dark blue spheres represent the atoms in an FCC environment.

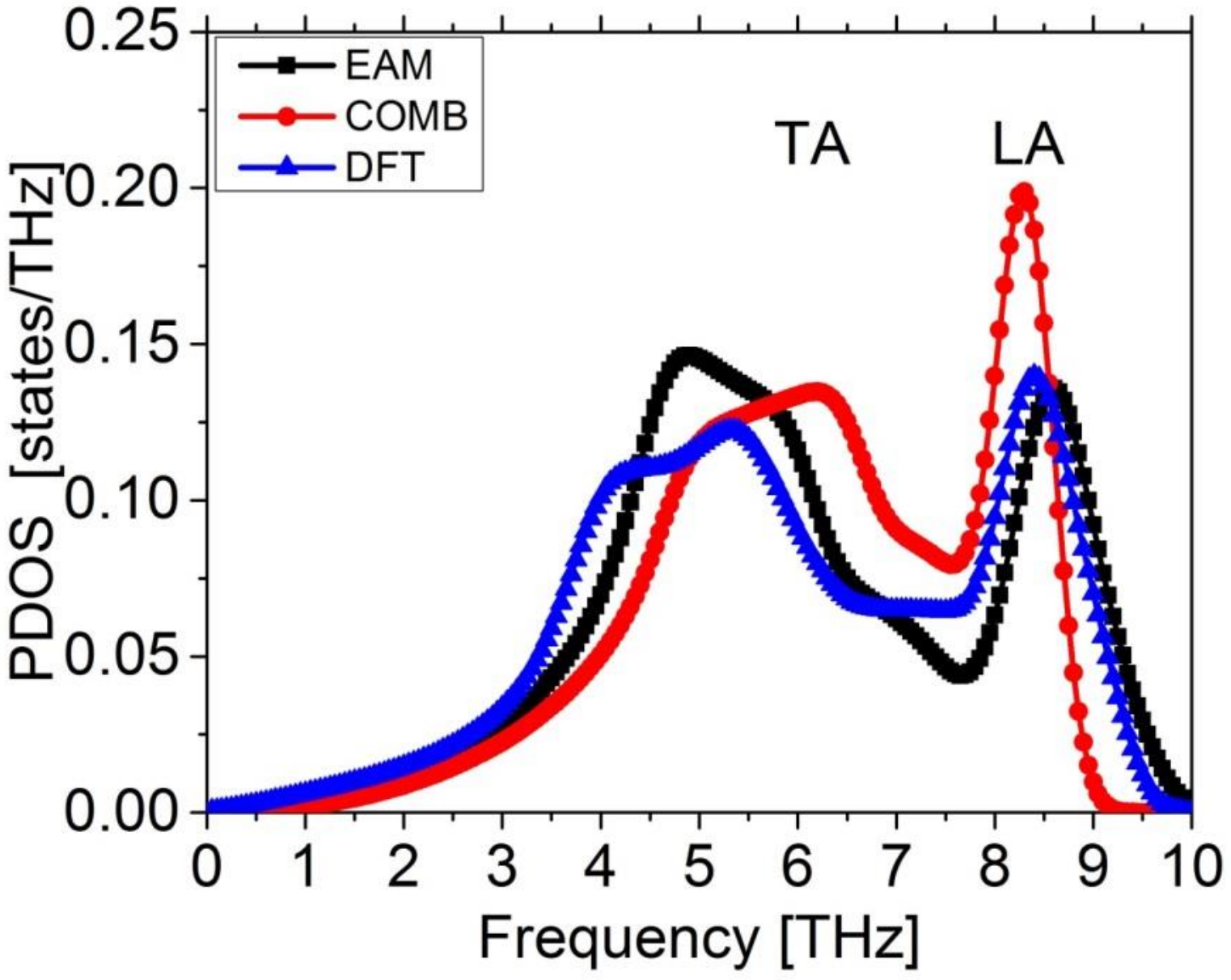

Figure 4-5. Phonon density of states (PDOS) as predicted by DFT, EAM and COMB3. The peaks corresponding transverse and optical modes are visible at almost the same position for all the methods considered.

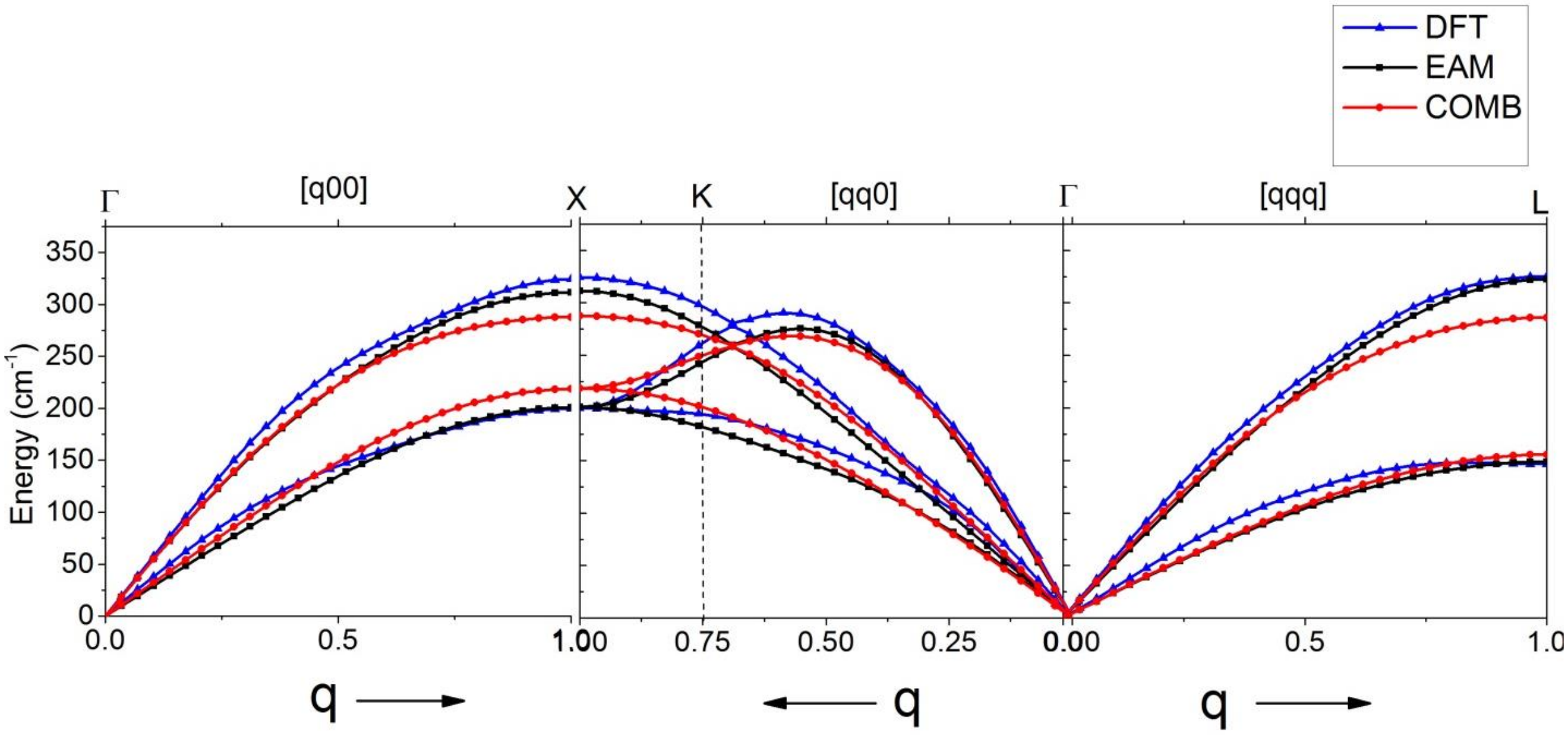

Figure 4-6. Phonon band-structure/dispersion curves calculated using EAM, DFT and COMB.

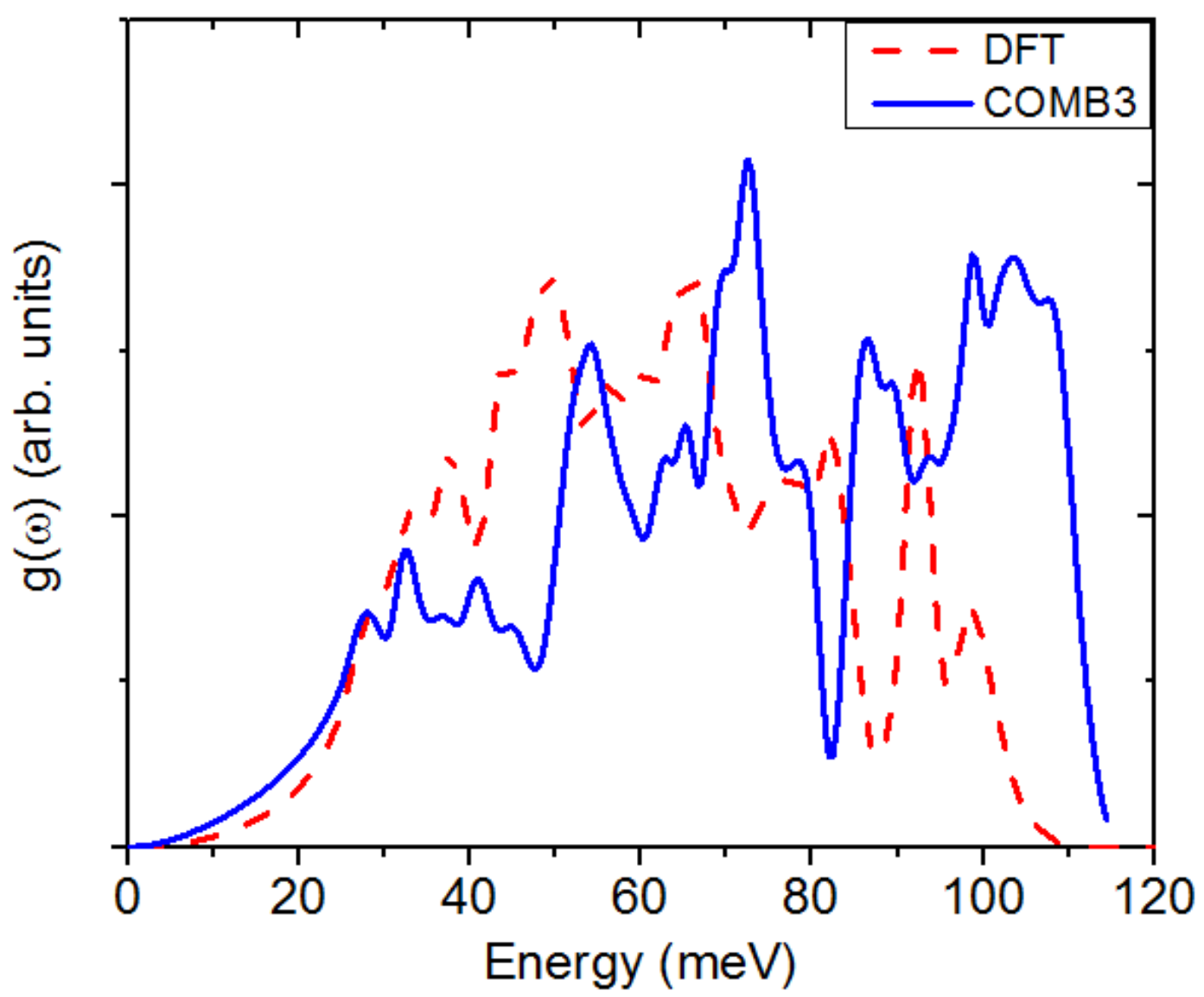

Figure 4-7. Phonon density of states for *α*-alumina using COMB3 and DFT.

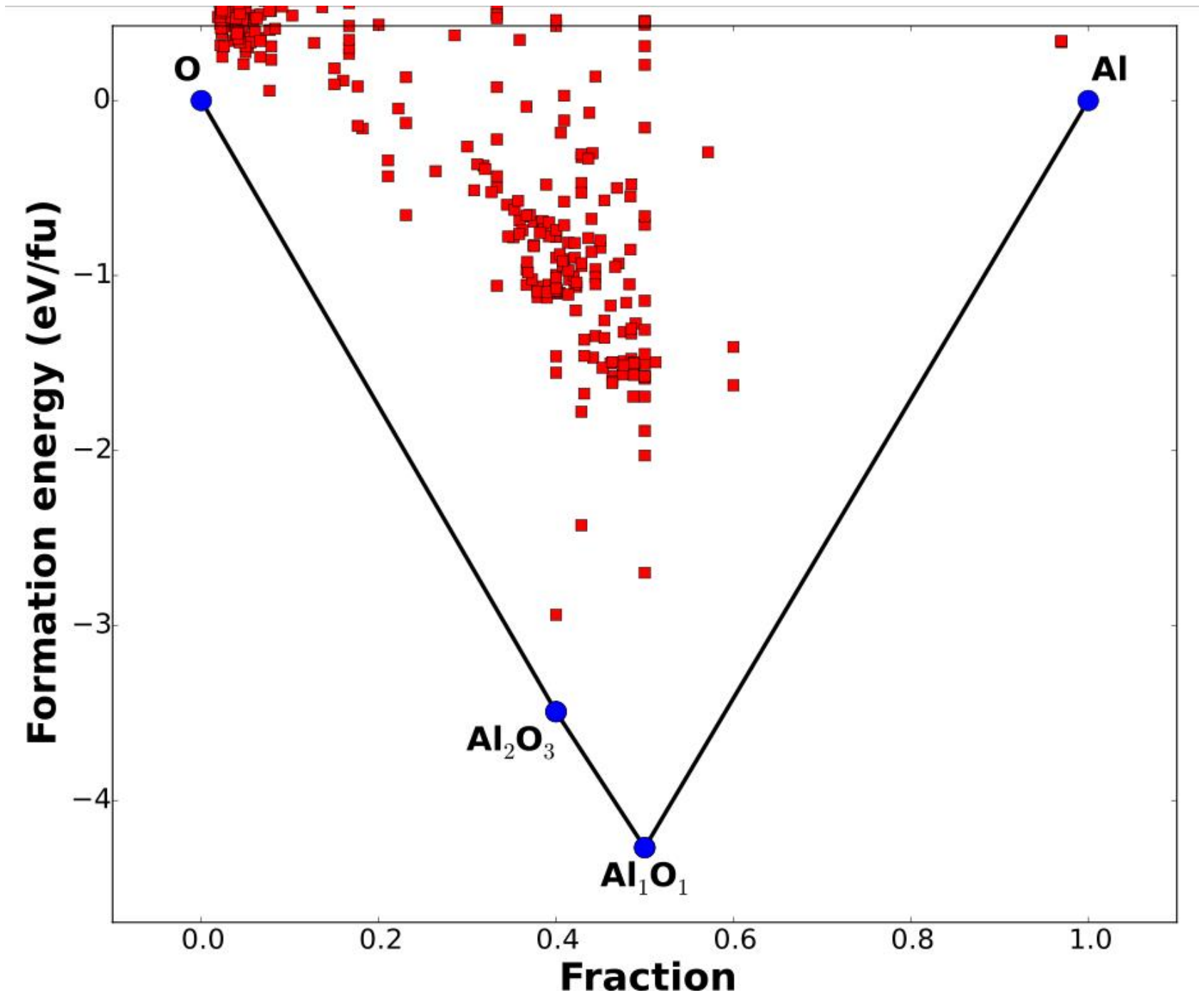

A

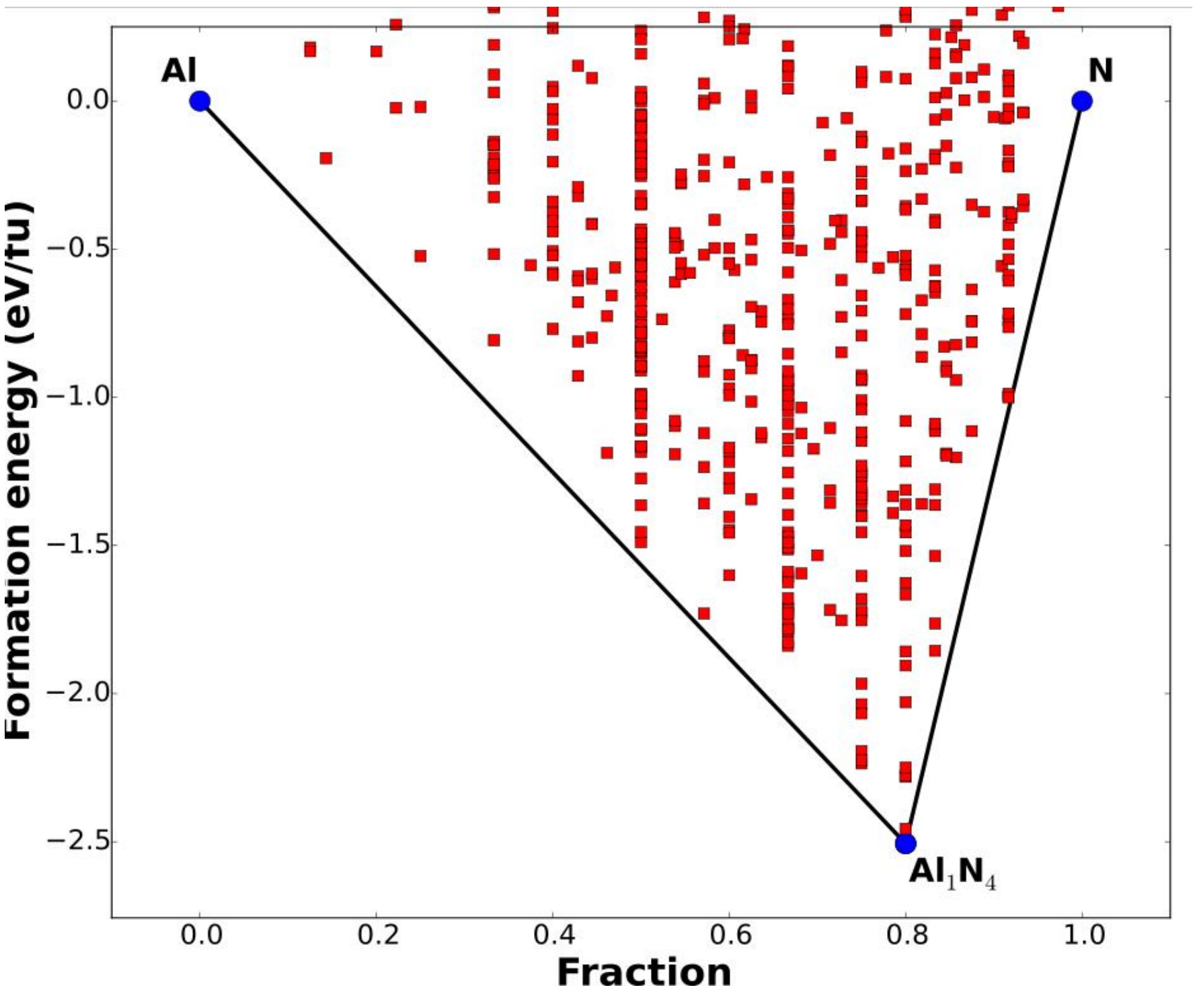

B

Figure 4-8. Genetic algorithm search results for COMB3. A) Al-O, B) Al-N system.

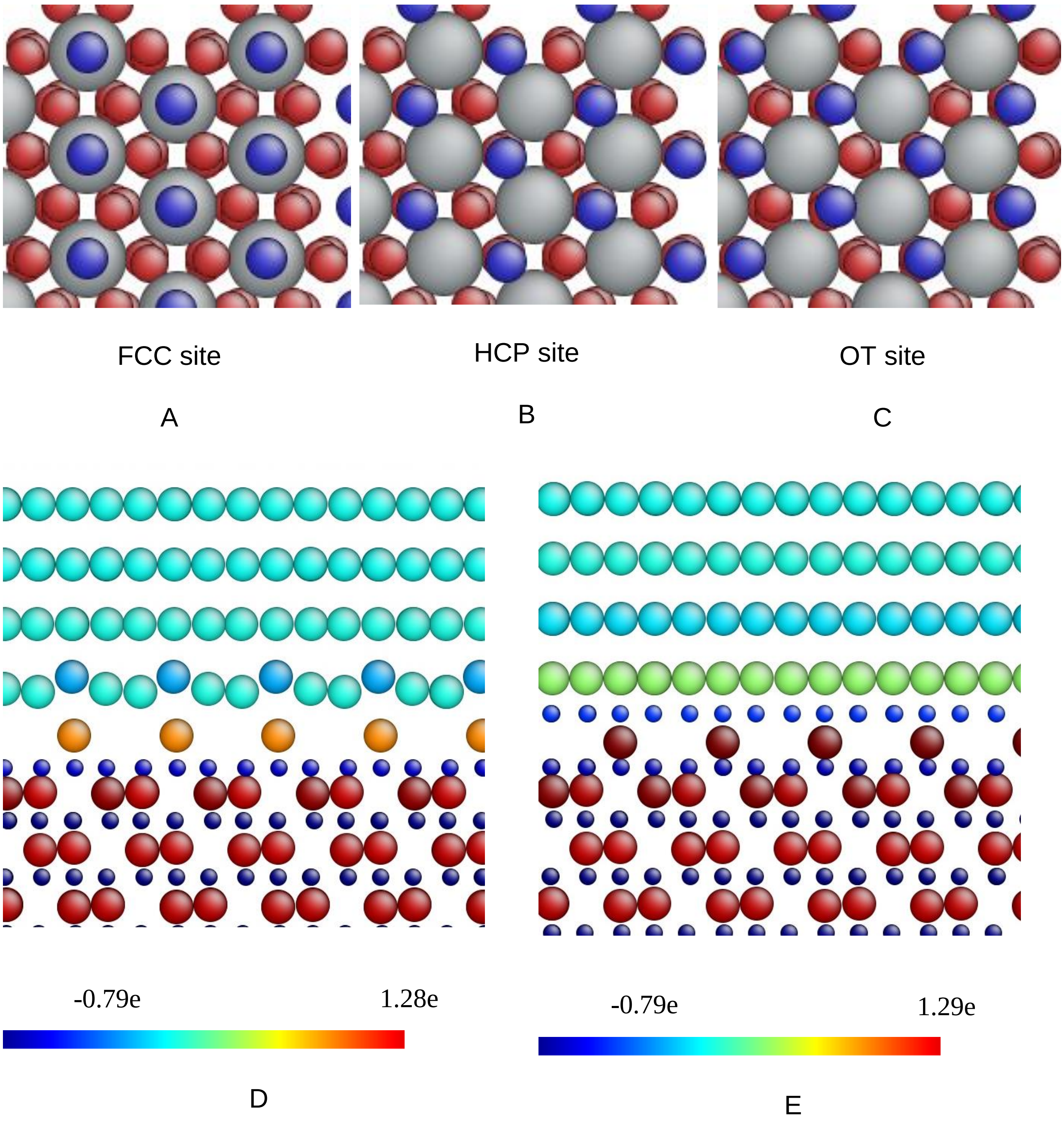

Figure 4-9. Snapshots of Al (111)-$Al_2O_3$ (0001) interface. A-C) Various sites for Al (111) surface atoms on $Al_2O_3$ (0001) surface is shown. The grey and red spheres are aluminum and oxygen from $Al_2O_3$ and the blue spheres are Al for Al (111) surface. D) Corresponds to case A in Table 1 – E) Corresponds to case D in Table 4-5, illustrate snapshots from energy minimized structures using the COMB3 potential. The atoms are color coded by atomic charge as indicated by the color bar.

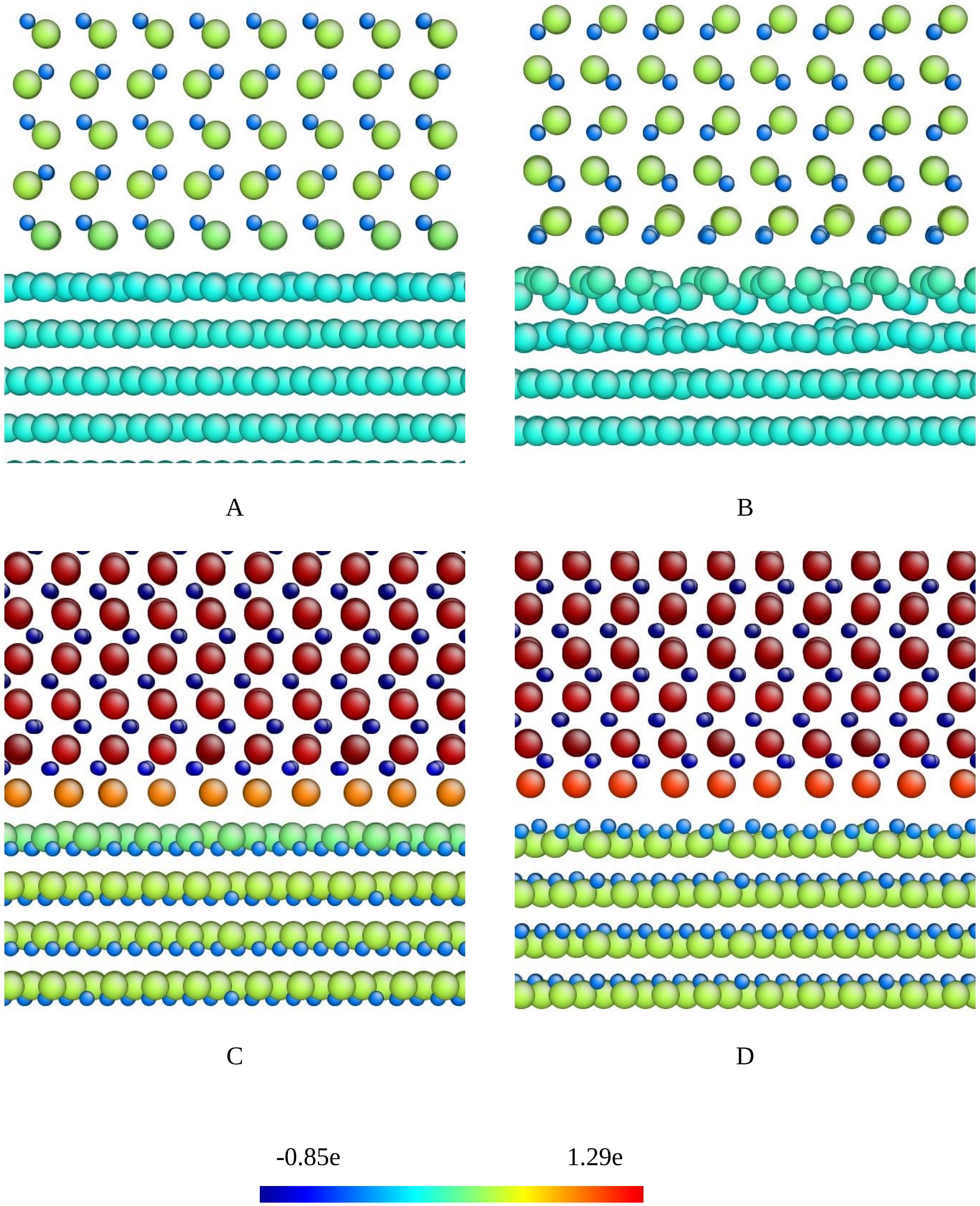

Figure 4-10. Snapshots of Al(111)-$Al_2O_3$(0001)/AlN(0001) interfaces: A) Al(111)-Al-terminated AlN(0001), B) Al(111)-N-terminated AlN(0001), C) Al-terminated-AlN(0001)-Al-terminated $Al_2O_3$(0001), D) N-terminated-AlN(0001)-Al-terminated $Al_2O_3$(0001). The atoms are color-coded according to charge using scale shown. Much rearrangement is observed for case B.

# CHAPTER 5
# MECHANICS OF NANOSTRUCTURES

## 5.1 Deformation of Metal and Ceramics: Al, $Al_2O_3$ Nanowires

Both aluminum [232] and alumina nanowires [154,226] are of great technological importance, such as for ultraviolet to infrared polarizers [233] and for high temperature nano-composites [234]. In the present work, nanowire structures were constructed by carving cylinders with diameters of 4 nm and lengths of 10.5 nm from bulk structures; these dimensions are comparable to the dimensions of samples prepared experimentally by electro-deposition [235]. Periodic boundary conditions were applied only in *z*-direction, in which stress was applied. The metallic and ceramic nanowires consisted of 7950 atoms and 15080 atoms respectively. The nanowires were then subjected to a tensile test at a strain rate of $1 \times 10^8$ $s^{-1}$ ; such a high rate is typical for MD simulations [236,237]. The system was subjected to NPT ensemble with the Nose-Hoover thermostat and a 0.1 fs time-step.

The results of the simulations are illustrated in Figure 5-1. As expected, the Al nanowire displayed the low strength, high ductility behavior [238] observed in previous simulations of FCC nanowire deformation [237], while the $Al_2O_3$ nanowire exhibited high strength and low ductility [177]. Specifically, the yield point for the Al nanowire was 0.05, at which it started to deform plastically. Snapshots for some of the points on the stress-strain plot are given in Figure 5-1A. Necking was observed after point B in the plot (strain of 0.05). A linear fit to the stress-strain curve for the Al-nanowire revealed a predicted Young's modulus of 72 GPa, which is comparable to the Young's modulus of bulk aluminum (69 GPa) [106] and to previous experimental and molecular dynamics results of Sen et al. [177]. Sudden drops of the total stress associated with major atomic rearrangements were observed after yield point of Al-nanowire. Such behavior is also observed in previous MD studies [239-241]. The Young's modulus of $Al_2O_3$ nanowire obtained from the linear fit to the stress-strain curve had value of 414 GPa, which was close to Young's modulus of bulk *α*-alumina. The actual plot deviated from the fit at a strain of 0.15, which could be regarded as the elastic limit. Snapshots of the deformed structure revealed brittle behavior of alumina as it went through abrupt rupture at point C (strain corresponding to 0.21) in Figure 5-1B. Slight buckling of the nanowire was also observed at point B, which was close to elastic limit of the nanowire. Interestingly, we observed high stress concentration at atoms (as shown in Figure 5-1C) on the basal plane (0001) before notch formation (in snapshot B and C, respectively). These behaviors were consistent with mechanism of brittle fracture in materials.

Next, we compared the maximum stress $\sigma_{th}$ and associated strain $\varepsilon_{th}$ to conventional theory [242] in which these quantities are expressed in terms of elastic constant $E$, surface energy $\gamma$ and interlayer distance d:

$$\sigma_{th} = \sqrt{\frac{E\gamma}{d}} \tag{5-1}$$

$$\varepsilon_{th} = \frac{\pi\sigma_{th}}{2E} \tag{5-2}$$

Based on our empirical potential fitting database and using eq. 5-1 and 5-2, the calculated $\sigma_{th}$ and $\varepsilon_{th}$ for Al and $Al_2O_3$ were 22.5 GPA, 0.31 and 106 GPa, 0.32 respectively while we observed these values as 4.1 GPa, 0.3 and 72 GPa, 0.21 for Al and $Al_2O_3$ respectively from our nanowire simulations. These differences could be explained based on the fact that stress-strain behavior was calculated form MD simulations rather than theoretical sine curves.

In addition to investigating perfect alumina nanowires, we also investigated the effect of vacancies on the mechanical responses of nanowires of diameter 2 nm. In experiments, the defects can be created by radiation and/or temperature [243]. Considering the mechanical response of defective nanowires, it was predicted that the ultimate strength of the alumina nanowire was more sensitive to oxygen defects compared to aluminum defects (as shown in Figure 5-2). The slopes of the curves, and hence the Young's moduli, of the defective nanowires remained almost unchanged because of the presence of the vacancies. The maximum change in the nanowire strength was predicted from the simulations to be between the 2% O and 7% O cases, which again suggested that oxygen defects play a more important role in influencing the mechanical responses of the nanostructures than do aluminum point defects. These results were consistent with results obtained by Kulkova et al. [244] for role of oxygen defects in metal-alumina interfaces. It is to be noted that the breaking point or ultimate tensile strength could be statistical in nature [237]. Hence, we carried out a separate set of simulations for the 7% Al case and found that there was not a significant difference in the results.

In a separate set of simulations, we examined the mechanical response of an Al-$Al_2O_3$ core shell model nanowire by cladding the Al-nanowire with alumina. This was done by keeping the net diameter of the nanowire of 4 nm fixed as in the previous numerical experiments. Aluminum cores of 2 and 3 nm diameter with rest being the alumina shell structure were considered. The mechanical response of these nanowire systems were dominated by the properties of the $Al_2O_3$ shell, as illustrated in Figure 5-3. This is similar to the behavior predicted in MD simulations of Si-$SiO_2$ systems [245,246] where the mechanical properties of the core-shell system was governed by mainly $SiO_2$. The slope of the stress-strain curve, and hence the Young's modulus, was predicted to increase as the percentage alumina content increased as shown in Figure 5-3A. Representative snapshots for Al with 3 nm diameter system are presented in Figure 5-3B to illustrate the brittle and metallic deformation of alumina and aluminum respectively. The ultimate tensile strains for the core-shell model nanowires were higher than in the case of the pure alumina nanowires due to Al-core. Hence, in both the cases brittle deformation for alumina-shell and metallic deformation behavior for Al-core (similar to that shown in Figure 5-3A) were predicted.

## 5.2 AlN Nanowires and Nanotubes

Next, we study the tensile properties of AlN nanowires and nanotubes using COMB AlN potential. Nanowire structures were constructed by carving cylinders with diameters of 4 nm and lengths of 10.5 nm from bulk structures; these dimensions are comparable to the dimensions of samples prepared experimentally by electro-deposition. Periodic boundary conditions were applied only in z-direction, in which stress was applied. The nanowire and nanotube consisted of 4352 atoms and 5900 atoms respectively. The nanostructures were then subjected to a tensile test at a strain rate of 1x108 $s^{-1}$ ; such a high rate is typical for MD simulations [236,237]. The system was subjected to NPT ensemble with the Nose-Hoover thermostat and a 0.1 fs time-step. The results of the simulations are illustrated in Figure 5-4. The nanowire is found to elastically deform upto 0.05 strain and then atomic rearrangement takes place upto point C. We also observe formation of chain like molecule formation at point D. Such behavior is also observed for $TiO_2$ nanowires using Morse empirical potential [247] deformation process.

Next, we deform the AlN nanotube with similar MD settings for nanowire discussed above. The Young's modulus found was comparable to that of experiments. Brittle fracture behavior was noticed for the nanotube deformation with notch formation (as shown in Figure 5-5 snapshot C). Next, we compared the maximum stress $\sigma_{th}$ and associated strain $\varepsilon_{th}$ to conventional theory [242] in which these quantities are expressed in terms of elastic constant $E$, surface energy $\gamma$ and interlayer distance *d*:

Based on our empirical potential fitting database and using eq. 5-1 and 5-2, the calculated $\sigma_{th}$ and $\varepsilon_{th}$ for AlN was 113 GPa, 0.42 respectively while we observed these values as 25 GPa, 70 GPa and 0.15 and 0.17 for nanowire and nanotube respectively. These differences could be explained based on the fact that stress-strain behavior as calculated form MD simulations rather than theoretical sine curves. In a nutshell, the results suggest that the COMB3 potential can be used to elucidate the mechanical responses for AlN nanostructure.

### 5.3 Summary

In this section it was shown that COMB potential can be used to reproduce experimental data for mechanical properties for nanostructures of metal and ceramics in one formalism. Oxygen defects were found to play more important role in $Al_2O_3$ nanowires than Al defects. The deformation behavior of core-shell model was found to be consistent with the metallic and brittle nature of fracture in the system. The elastic constants for AlN nanowire and nanotubes were also found to agree well with experiments. The snapshots provided the atomistic phenomenon that occurred during tensile loading. Results from the mechanical deformation data were reasonably comparable with the conventional Griffith’s theory data.

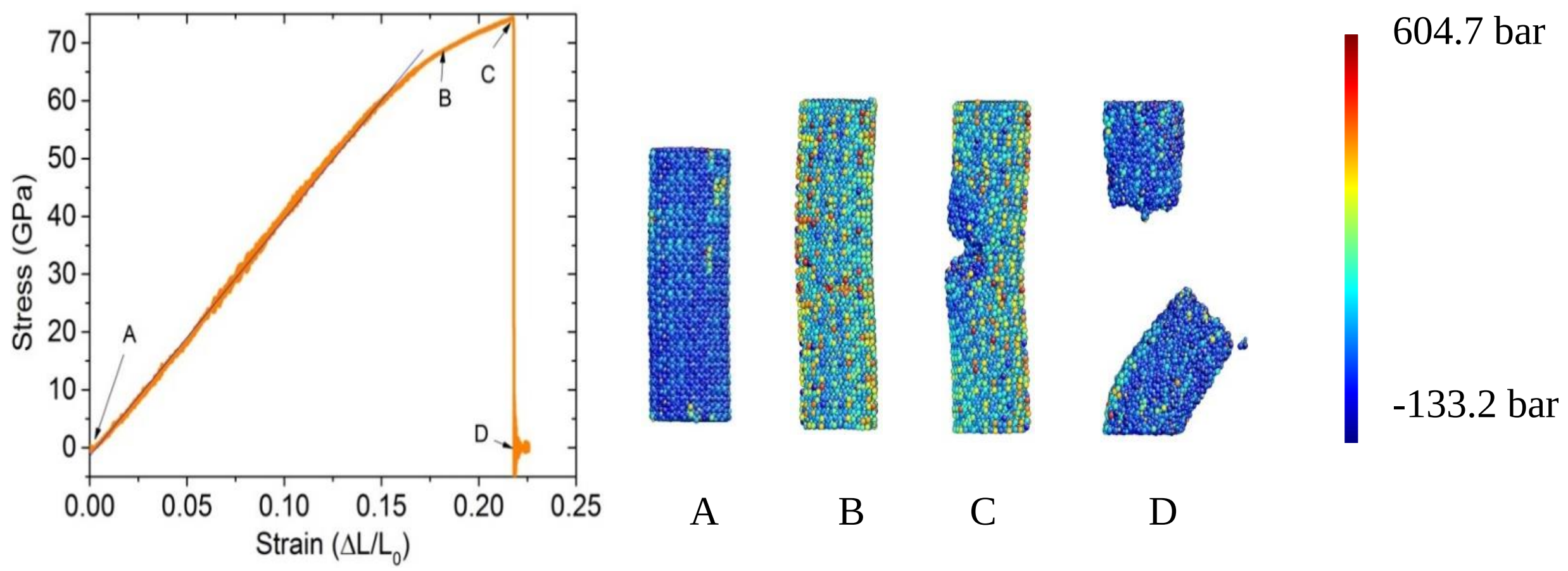

B

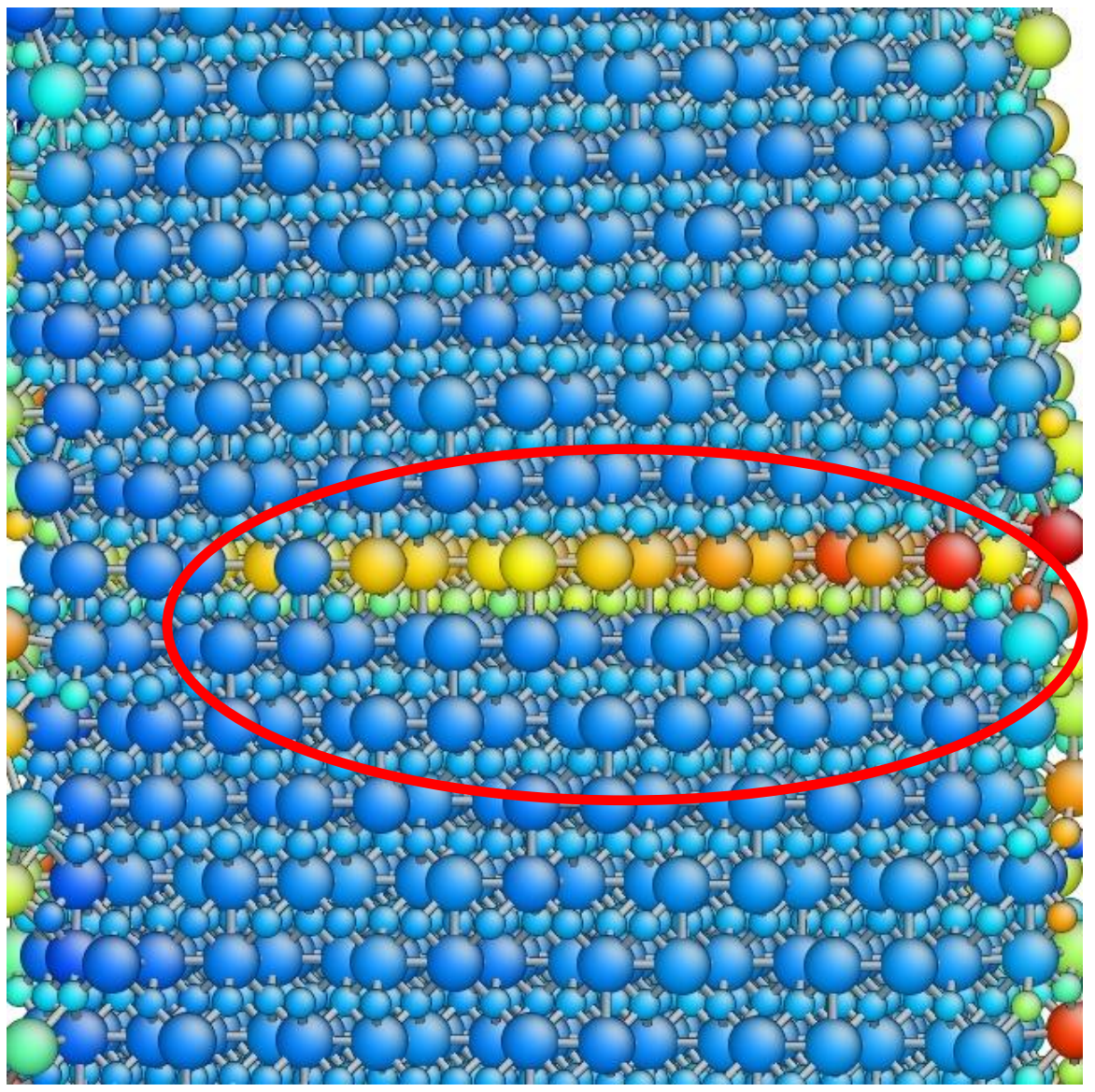

C

Figure 5-1. Stress-strain plots for nanowires. A) Al and B) $Al_2O_3$ nanowires of 10.5 nm length and 2 nm in diameter that were predicted to exhibit ductile and brittle deformation mechanisms, respectively. Stress is applied in the direction parallel to the long axis of the nanowires and is color-coded in the snapshots. C) Zoomed-in, cross-sectional view of snapshot B, where localized stress concentration is predicted to occur.

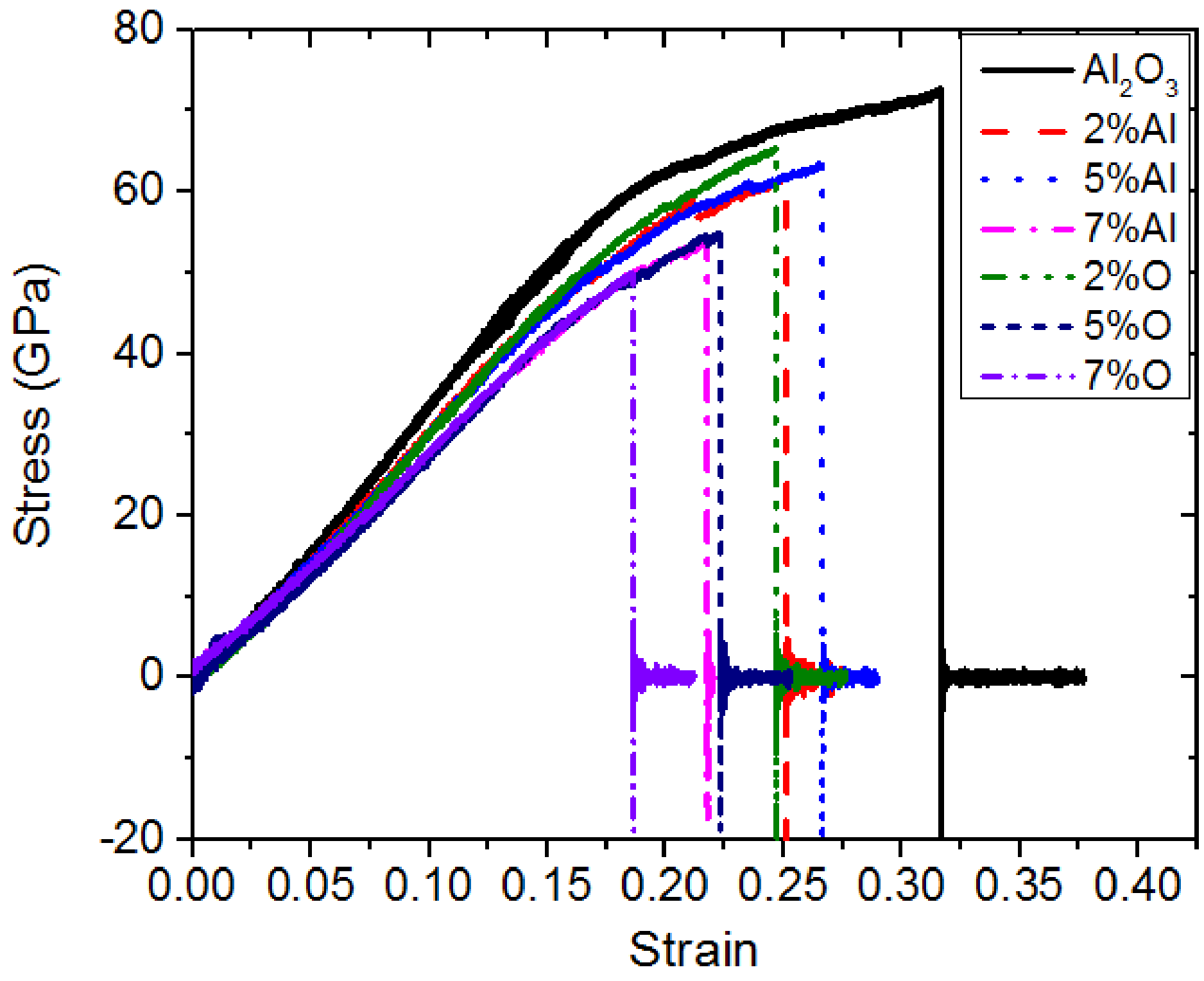

Figure 5-2. Effect of point defects on the mechanical properties of alumina nanowires. The tensile strength and Young's modulus decreased more due to oxygen vacancies than aluminum vacancies.

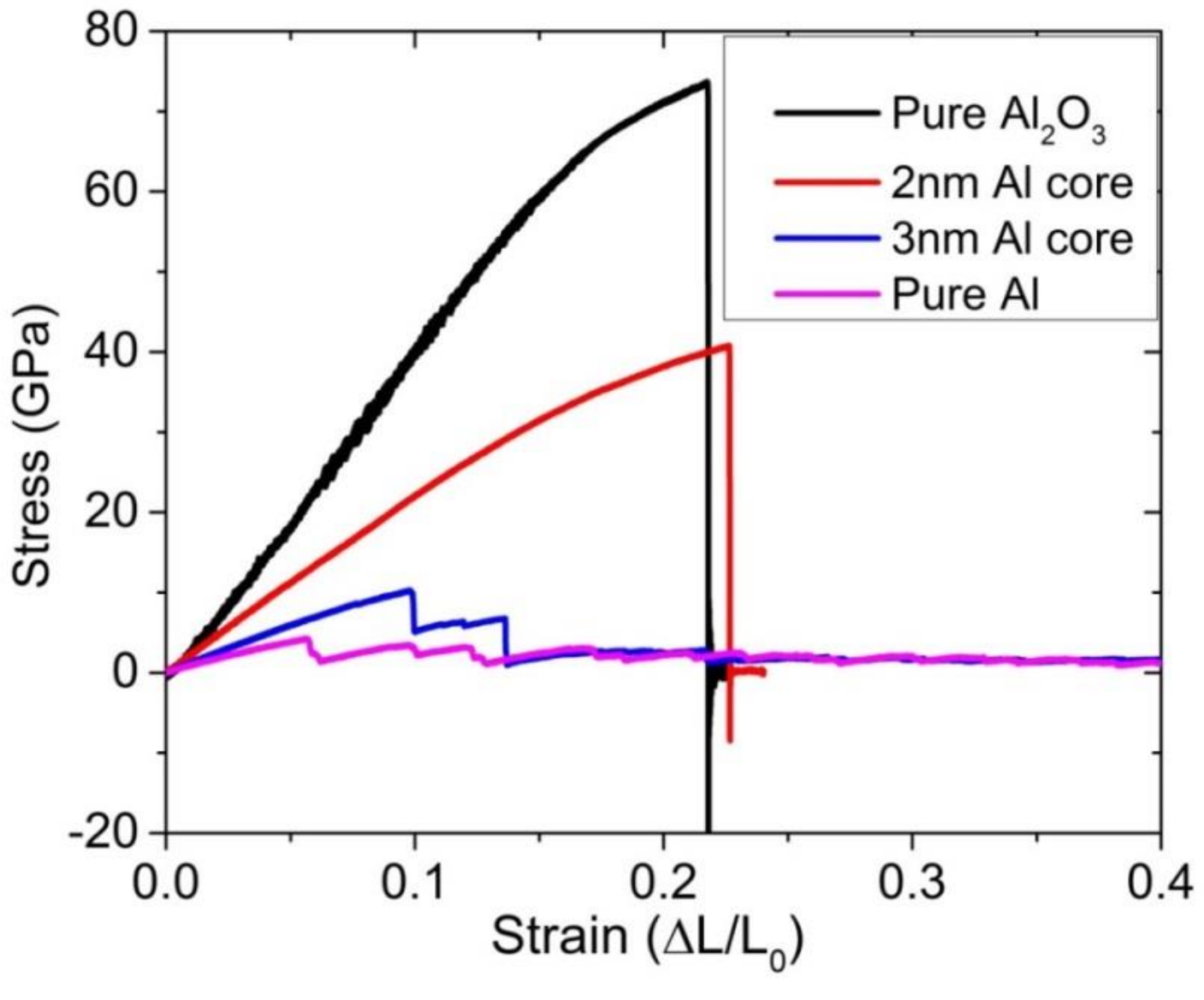

A

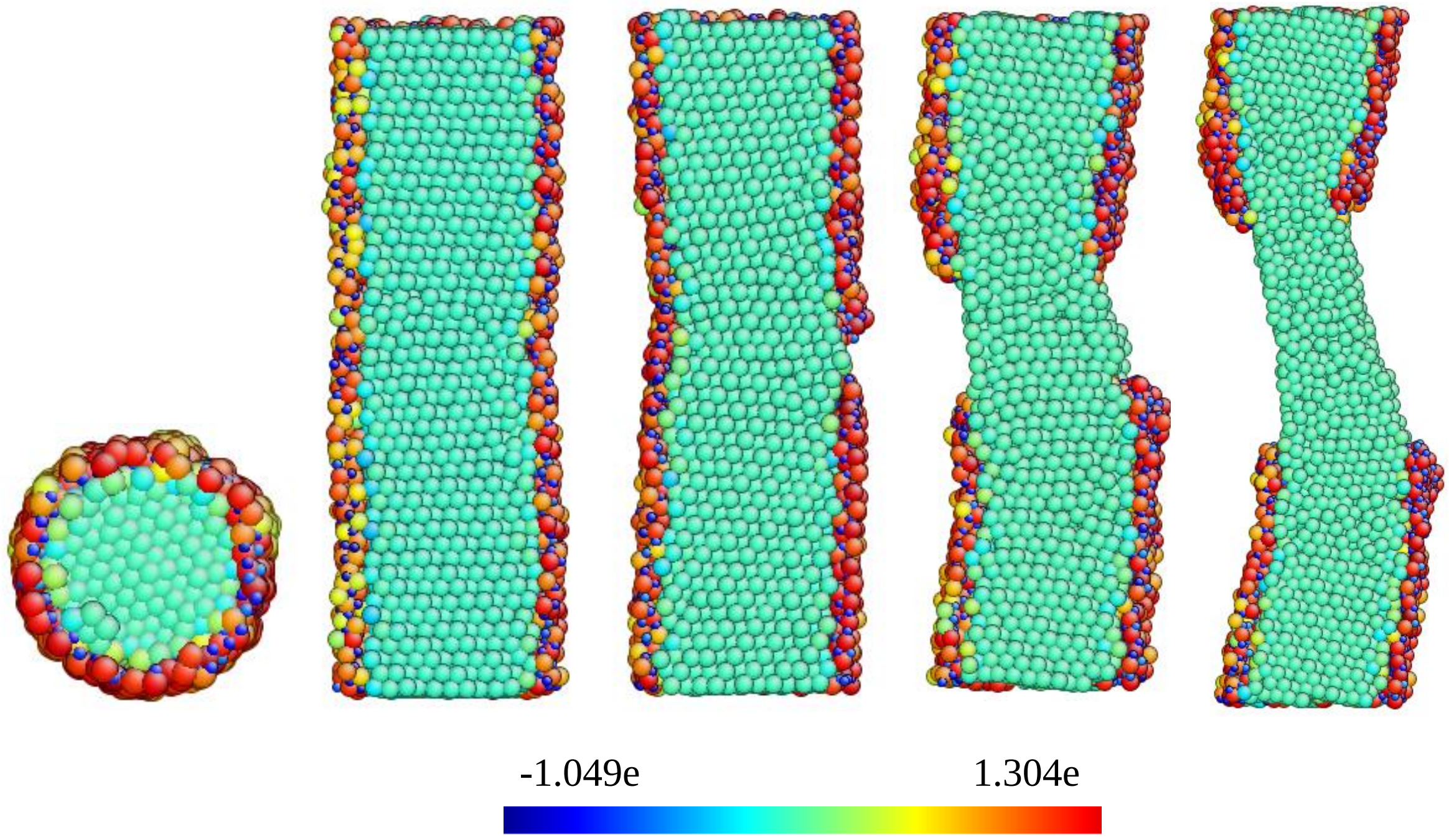

B

Figure 5-3. Tensile test of Al-$Al_2O_3$ core-shell model. A) Stress-strain curve for Al-$Al_2O_3$ core-shell model along with pure Al and $Al_2O_3$ nanowires. B) Snapshots of the 3 nm Al core model where the atoms are color-coded by charge.

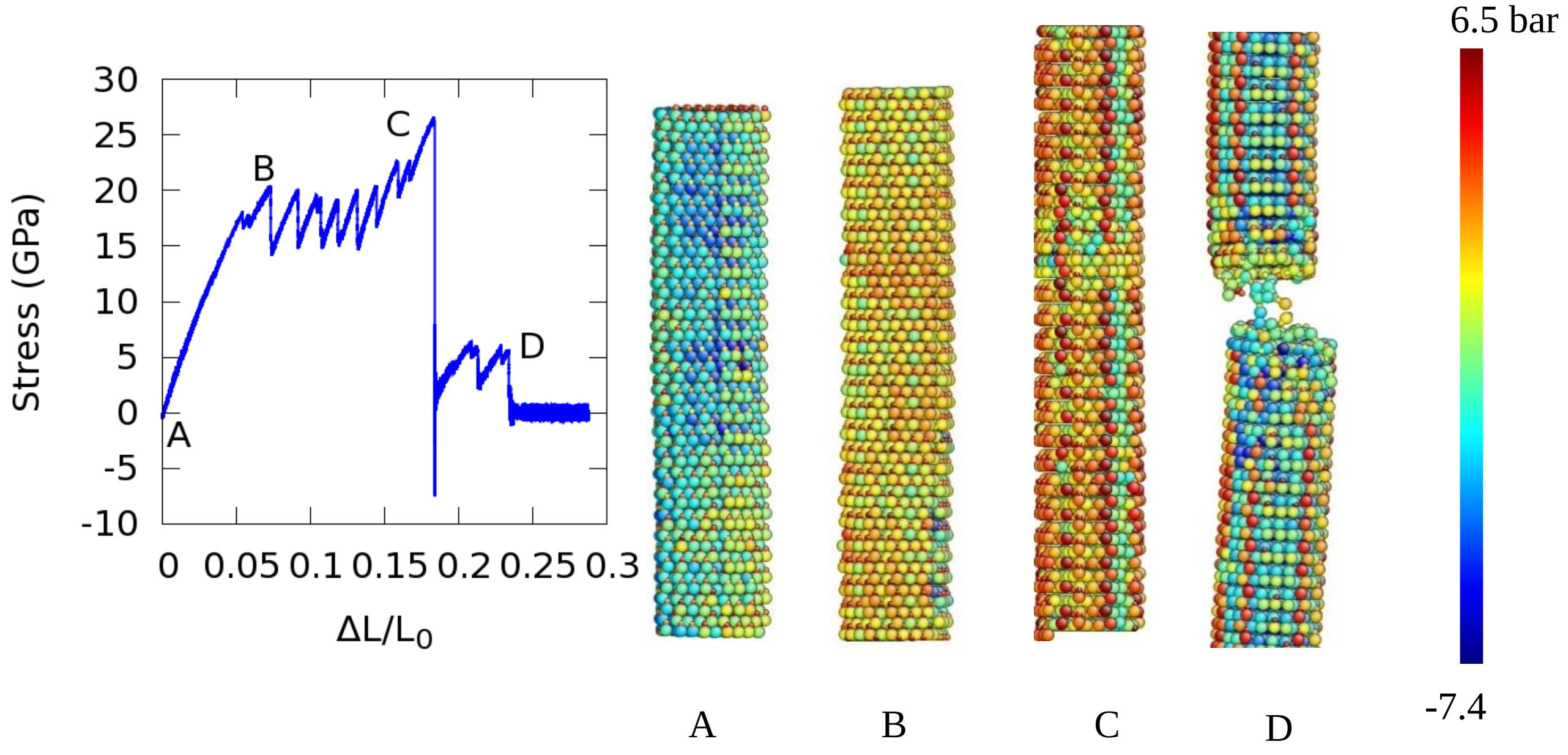

Figure 5-4. Deformation of AlN nanowire. Molecular chain are observed during deformation process

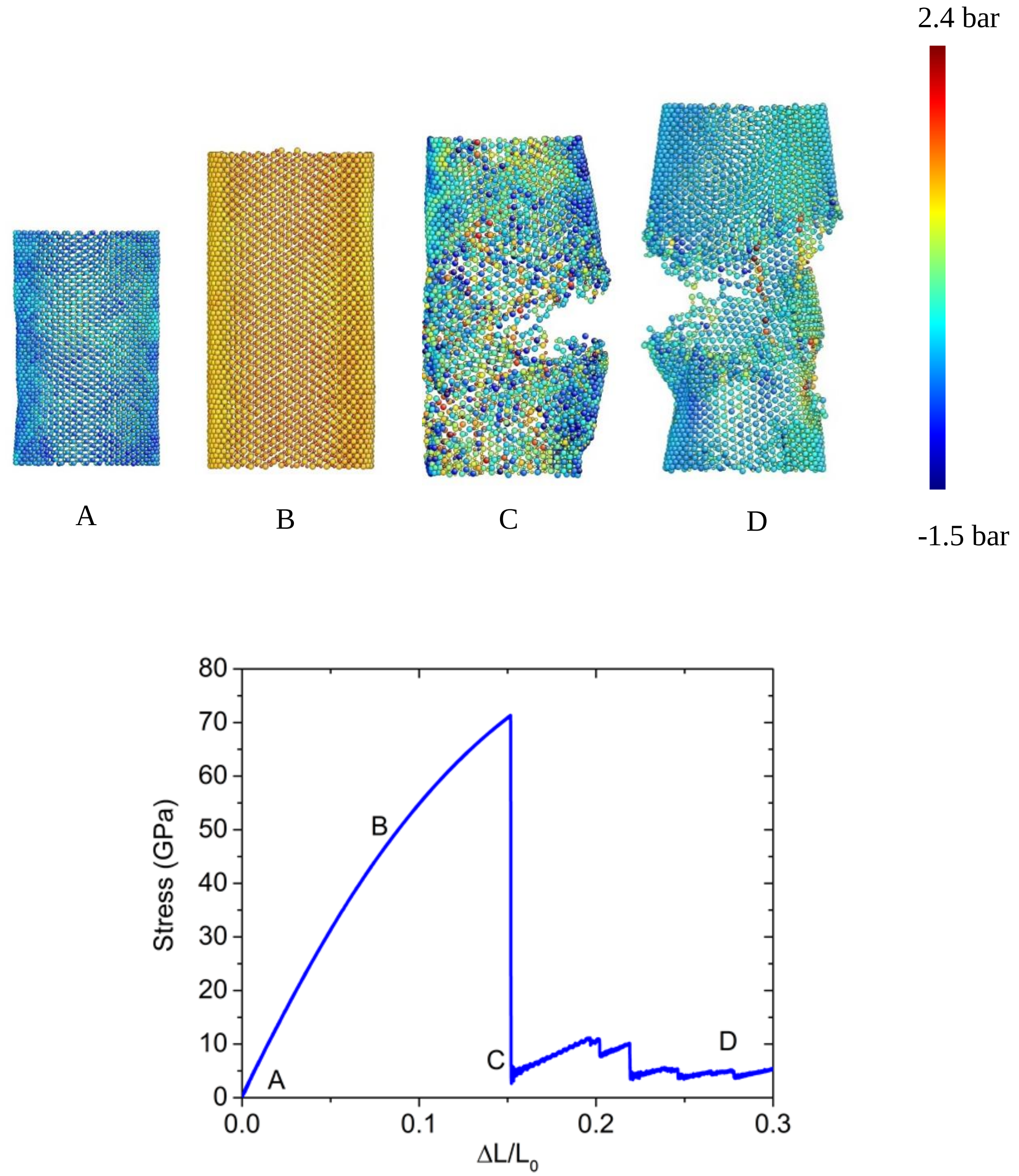

Figure 5-5. Deformation behavior of AlN nanotubes.

# CHAPTER 6
# SCREENING OF OPTOELECTRONIC MATERIALS

## 6.1 Lanthanide Doped YAG

### 6.1.1 Introduction

Lanthanide (Ln) doped and co-doped yttrium aluminum garnet (YAG) has a wide range of applications including light emitting diodes (LED) [248], lasers [249], phosphors [250], infrared pressure sensors [251] and spectral convertors for improving solar cell performance [252,253]. Among these, spectral conversion has recently received increased attention [252,254], with YAG-based materials such as Ce:YAG and Er:YAG [255] having high quantum efficiency. Interestingly, experimental efforts to co-dope Er:YAG with Ce have resulted in an increase in the quantum efficiency via enhanced infrared optical absorption. However, there is still a strong need for even higher quantum efficiencies materials, [252,254] for which a better understanding of the governing optoelectronic phenomenon is required. Recent, density-functional theory (DFT) calculations with semi-local exchange-correlation functionals revealed the atomic and electronic structure as well as optical transition characteristics for pure YAG [256], Ce-La:YAG [257] and Cr:YAG [258] . However, the semi-local exchange-correlation functionals used are problematic because they underestimate the band gap and predict an incorrect position of the *f*-states of the lanthanides [256,257,259-265]. Improved approximations to DFT, such as nonlocal hybrid functionals [266,267] and many-body approaches, such as the GW approximation, [268] are generally more accurate for electronic structure than semi-local exchange-correlation functionals. Thus, there is a need to apply these methods to address the drawbacks of conventional DFT and to develop the systematics in optoelectronic properties trends that can be used to optimize performance of materials.

In this work, we determine the optoelectronic properties of lanthanide doped and co-doped Ln:YAG using the Heyd-Scuseria-Ernzerhof hybrid functional (HSE06) [269,270], which establishes an accurate description of the electronic properties of materials [271,272] at a reasonable computational cost. We find through comparison with experimental values that the band-gap and the frequency dependent dielectric function of the undoped YAG host material as well as the defect levels and optical transitions of the Ln-doped YAG are well described by the HSE06 functional. We also find that the co-doping of Ce-doped YAG with any lanthanide except Eu and Lu lowers the energy of the optical transitions [273]. We compare materials based on presence of low energy infrared peaks, which are again related to the quantum efficiency of devices. To understand the origin of the change in optical transition energy in Ce co-doped system compared to Ce:YAG, we characterize the charge and bonding environment of the dopant atoms. We attribute the spectral shift in the co-doped materials to the volume change induced by the Ln atoms. We predict that that co-doping with Tb and Tm results in infrared optical transitions, making Tm,Ce:YAG and Tb,Ce:YAG candidate functional materials for efficient spectral up-conversion devices.

### 6.1.2 Computational Details

YAG has a body centered cubic garnet structure (space group *Ia3d* or $O_h^{10}$) that is illustrated in Figure 6-1. The structure consists of octahedrons, tetrahedrons, dodecahedrons and oxygen shared at corners and may be considered as $Y_3Al_2Al_3O_{12}$ i.e. $A_3B_2'B_3''O_{12}$ structure, in which the A ions are dodecahedral coordinated, B' is octahedral coordinated, B'' is tetrahedral coordinated, and oxygen are at corners. One cubic primitive cell of YAG consists of 160 atoms. The Y atoms occupy 24 dodecahedral sites, O atoms occupy 96 sites, aluminum is found at two different sites- octahedral (16) and tetrahedral (24). The DFT calculations are performed using the Vienna Ab-initio Simulation Package (VASP) [122,123] and the projector-augmented wave (PAW) method [274]. The doping is modeled by substituting a lanthanide for an yttrium atom in a 160-atom cubic simulation cell of YAG with a chemical formula $Y_{(3-x)}Ln_xAl_5O_{12}$ ($x$ = 0.125).

This concentration lies approximately in the middle of the range of experimental concentrations: $x$ = 0.03 to 0.3 [275,276]. The Ce co-doped system is considered with an effective composition of $Y_{2.75}Ce_{0.125}Ln_{0.125}Al_5O_{12}$. The calculations are done in two steps: first, the atomic structure is optimized using the semi-local exchange-correlation functional by Perdew, Burke and Ernzerhof (PBE) [277]. Second, the electronic structure is calculated using the HSE06 exchange-correlation functional [269,270]. The inclusion of 25% of exact exchange for short distances in the HSE06 functional improves the band gap by recovering the derivative discontinuity of the Kohn-Sham potential for integer electron numbers [278,279]. The orbitals are expanded in a plane-wave basis with an energy cutoff of 400 eV. A 4×4×4 k-point mesh is used for the Brillouin zone integration for the PBE functional and the $\Gamma$ point for the computationally more demanding HSE06 functional.

To obtain the optical properties of the doped materials, we calculate the imaginary part of the dielectric function from the Bloch wavefunctions and eigenvalues [26] :

$$\varepsilon_{\alpha\beta}(\omega)=\frac{4\pi^2e^2}{\Omega}\lim_{q\to 0}\frac{1}{q^2}\sum_{c,v,\vec{k}}2w_{\vec{k}}\delta(\zeta_{ck}-\zeta_{vk}-\omega)\left\langle\psi_{c\vec{k}+e_\alpha q}\middle|\psi_{v\vec{k}}\right\rangle\left\langle\psi_{c\vec{k}+e_\beta q}\middle|\psi_{v\vec{k}}\right\rangle^*, \quad (6\text{-}1)$$

where $e$ is electron charge, $\Omega$ is the cell volume, $w_{\vec{k}}$ is the Fermi-weight of each $k$-point, $e_\alpha$ are unit vectors along the three Cartesian directions, $|\psi_{n\vec{k}}\rangle$ is the cell-periodic part of the pseudopotential wavefunction for band $n$ and $k$-point $k$, $q$ stands for the Bloch vector of an incident wave, $c$ and $v$ stand for conduction and valence bands, $\zeta$ stands for eigenvalues of the corresponding bands respectively. The matrix elements on the right side of Eq. (1) capture the transitions allowed by symmetry and selection rules [280]. Furthermore, the energy conservation described by the $\delta$-function is usually approximated by a Gaussian-type smearing function for numerical reasons. However, it is difficult to resolve the specific transitions with such smearing due to the lanthanide dopants in the dielectric function. We therefore calculate directly the transition strength [258] for the Bloch wavefunctions and plot it against $(\zeta_{ck}-\zeta_{vk})$ at the $\Gamma$ point:

$$\eta_{\alpha\beta,cv}=\frac{8\pi^2e^2}{\Omega}\lim_{q\to 0}\frac{1}{q^2}\left\langle\psi_{c+e_\alpha q}\middle|\psi_v\right\rangle\left\langle\psi_{c+e_\beta q}\middle|\psi_v\right\rangle^* \quad (6\text{-}2)$$

### 6.1.3 Results and Discussions

Based on the transition strength spectrum, we identify the various transitions that are induced by the presence of the dopants in the material.

For pure YAG, the lattice constant obtained is 11.993 Å, within 1% of the experimental value of 12.02 Å [261,262] using PBE. The volumes of the doped Ln:YAG systems decrease along the lanthanide row ( Figure 6-2), following the well-known 'lanthanide contraction' of decreasing ionic radii along the row [281].

Furthermore, the elastic constants obtained using density functional perturbation theory for YAG are $C_{11}$=329 GPa, $C_{12}$=113 GPa, and $C_{44}$=109 GPa, which agree closely with the experimental values of 339, 114, and 116 GPa respectively [282]. This confirms that the PBE functional is well suited for the structural and mechanical properties of YAG.

Next, we assess the electronic and optical properties of pure YAG obtained from the PBE and HSE06 functional to establish the accuracy of our approach. Figure 6-3A shows that the PBE functional predicts a band-gap for pure YAG of 4.6 eV underestimating the experimental value of 6.4 eV. The HSE06 functional on the other hands predicts a value of 6.2 eV, in good agreement with experiment. The projected density of states (PDOS) shows that the conduction band is dominated by the yttrium *d*-states, while the valence band is dominated by the oxygen 2*p* states, in agreement with electron-loss near-edge spectroscopy (see Figure 5 of Ref. [283]). Figure 6-3B shows the imaginary part of the dielectric function for pure YAG, which agrees well with the measurements by Tomiki *et al*. [262,284]. The overall shape of the dielectric function for the PBE and HSE06 functional are very similar. However, the peak positions are shifted by the underestimated band-gap energy in PBE. We also verify the electronic and optical properties of the Ce doped YAG. Figure 6-3C shows that for Ce:YAG, the *f*-states of Ce are outside the band-gap region with PBE, which is inconsistent with experiments [285]. On the other hand, with the HSE06 functional the Ce *f*-states are located in the gap region of YAG. The transitions strengths, $\eta$ , for Ce:YAG using PBE and HSE06 are presented in Figure 6-3D. The PBE functional displays various unphysical transitions at low energies, while the HSE06 functional shows a first peak at 2.66 eV in close agreement with the experimental value of 2.7 eV [286]. We further characterize the electronic states responsible for the transition using the angular (*l*) and magnetic (*m*) quantum number-projected density of states. We find that the states involved in this transition are dominated by occupied mixture of *f*-orbitals (*m* = -1, 1, 2, 3) and unoccupied $d_{yz}$ (*m* = 1) orbitals-delocalized across Ce and Y ions. The *f-d* is an allowed transition due to selection rule $\Delta l = \pm 1$ (*l* = 2 for *d* and *l* = 3 for *f*-orbitals). In addition, $\Delta m = 0$ has to be satisfied for an allowed transition between *f-d* states.

We now compare the electronic and optical properties of other lanthanide-doped YAG materials. Figure 6-4 compares the transition strength for the lanthanide-doped YAG with that for pure YAG (see Figure 6-5 in the supplemental material for the density of states). Interestingly, we observe the strongest transitions from occupied *f* to unoccupied *d* states for both the Ce and Pr dopants, which is consistent with experimental observations of luminescence in these materials [250]. Similar to Ce:YAG, the predicted transition energy of 4.13 eV for Pr:YAG agrees well with the experimental value of 3.9 eV [287]. Furthermore, we establish the effects of co-doping with other lanthanides on the electronic and optical properties of LnCe:YAG. Figure 6-6 shows the transitions strength spectra for the co-doped LnCe:YAG and Table 6-1 provides the energies and associated states for the most important optical transitions. We find the co-doping of Ce:YAG with La, Pr, Nd, Pm, and Gd red-shifts, while Lu blue-shifts the *d* to *f* transitions of 2.66 eV. The predicted shifts are consistent with the experimentally observed red shift for La and Gd and blue shift for Lu co-doping of Zhang et al. [273]. For this series of lanthanide dopants (La, Pr, Nd, Pm, and Gd), the states dominating the optical transition are cerium *d* and *f* states. For the other co-dopants, the states dominating the optical transition appear to have strong contributions from the *f* and *d* orbitals of the lanthanide co-dopants.

Peaks at 2.6 eV of Ce and 4.13 eV of Pr are comparable to experiments.

To identify the cause of the shift of the optical transition with co-doping across the lanthanide series, we calculate the charge of the ions using the Bader method [288] and determine the change in volume and in Ce-O bond length for the different co-dopants. We observe that the optical transition energy is correlated with the Ce-O bond length, which changes with the lanthanide co-dopant.

To further elucidate this phenomenon in Figure 6-9 we show how the volumetric strain and the resulting change in Ce-O bond length affects the transition energies for the Ce:YAG system, for a range of volumes comparable to that of the co-doped systems. The small compressive strains induced by the La, Pr and Lu dopants result in a similar shift to that predicted for the volumetric strain. For the smaller Nd, Pm, and Gd dopants, we observe that the increased compressive strains enhances the red shift compared to the volume strain. We speculate it could be attributed to the anisotropy of the dopant-induced strain.

Interestingly, we observe infrared transitions for the cases of doping with TbCe, ErCe and TmCe. These dopant combinations result in available occupied and unoccupied *f* orbitals within the infrared range (see density of states shown in supplemental material). All other lanthanide co-doped systems only exhibit allowed *f-d* transitions. In the case of Er,Ce:YAG, experiments have already shown this material to be a high quantum efficiency up-conversion material [254]. Hence, due to the presence of these infrared transitions, which are related to quantum efficiency[280], we suggest Tb,Ce:YAG and Tm,Ce:YAG may also be useful up-conversion materials.

## 6.2 Pb Replacement in $CH_3NH_3PbI_3$

### 6.2.1 Introduction

Methylamine lead iodide ($CH_3NH_3PbI_3$ or $MAPbI_3$) has revolutionized the area of photovoltaic applications due to its high efficiency and low cost compared to conventional inorganic material based technology [289]. However, the toxicity of lead hinders the commercial utilization of the material at large scale. Lead poisoning can cause a variety of diseases to various organs and tissues, including the heart, bones, intestines, kidneys, and reproductive and nervous systems [290]. Recently, tin (Sn) was proposed as the non-toxic replacement of Pb, but it suffers from the instability issue due to its existence in multiple oxidation state; +2 and +4 [291,292]. Hence, a high throughput screening of material based on a reliable and computationally efficient method such as density functional theory (DFT) [20] is necessary to guide the experimental exploration for solving the problem. As mentioned above, it is necessary for the replacement material to remain stable in a +2 oxidation state in its halide form as the hybrid material is generally prepared via solution route using the halide of the material and methylamine [293,294]. Hence, lists of elements from the periodic table with +2 oxidation state were chosen for DFT calculations. Important criteria used during screening of the material was the band-gap ($E_g$) of the material satisfying Shockley and Queisser (SQ) [295] ($E_g$ ~ 1.7 eV to 3.3 eV), Goldschmidt's tolerance [296] factor ( $t = \frac{r_{MA} + r_I}{\sqrt{2}(r_X + r_I)}$, for $MAXI_3$, r represents ionic radii), absorption coefficient (used in calculation for maximum solar efficiency [297]), and carrier effective mass [298] (for electron, $m_{e;}$ for hole, $m_h$).

Although originally introduced by Weber [299], Kojima et al. [300] introduced $MAPbI_3$ for solar cell applications in 2009, there is still much to be done in terms of DFT study for the hybrid material such as $MAPbI_3$. To list a few of them, Mosconi e al.[301] studied the structural and optical properties of $CH_3NH_3PbX_3$ and $CH_3NH_3PbI_2X$ perovskites (X = Cl, Br, I). They showed DFT can successfully be used to calculate both lattice constant and band-gap of the material, comparable to experiments. Jishi et al. [302] compared the structural and electronic properties of $CH_3NH_3PbI_3$, $CH_3NH_3PbBr_3$, $CsPbX_3$ (X=Cl, Br, I), and rubidium lead iodide ($RbPbI_3$) using TB-mBJ exchange-potential. Feng et al. [303] studied the tetragonal and orthorhombic structures of the material and showed the effective masses were anisotropic in nature and also calculated the theoretical absorption spectra of the material. Furthermore, they showed that van der Waals interactions were important for obtaining the accurate equilibrium cell volumes from DFT. In a later work, Feng et al. [304] showed that using Sn as substituent of Pb, absorption efficiency could be increased due to reduction of bandgap of the material and they confirmed that the orthorhombic representation [305] for the hybrid material is justifiable over tetragonal and cubic structures. Amat et al. [296] showed the importance of Goldshmidt factor for the hybrid material and showed its importance in determining structural stability using first-principle calculations. Umari et al. [306] studied the effect of various levels of DFT (spin-orbit coupling, GW-methods) on the electronic properties such as band-gap and effective mass for $CH_3NH_3$ (Sn, Pb) $I_3$ to confirm the necessity of these high-level calculations. Haruyama et al. [307] showed the effect of surface terminations on stability and electronic properties of $MAPbI_3$. They showed flat terminations under the $PbI_2$ rich condition were advantageous in regards to solar cell performance. While there are so many ongoing works for optimizing the stability and efficiency of the hybrid pervoskite materials, not much has been done for lead replacement elements options.

It is to be noted that the lead iodide crystallizes into rhombohedral, P-3m1 space group. Hence, iodides with same space group with band-gap in the visible region, appropriate Goldschmidt's factor, absorption coefficient comparable to $MAPbI_3$ within visible range and low effective mass of carriers should be preferred over others as potential replacement candidates. Goldschmidt factor is an important factor in studying the stability of pervoskite materials of type $ABX_3$. For a stable pervoskite the factor is closer to 1.0. However, for the present case of $MAXI_3$ (X = replacing element of Pb), MA is a molecular specie instead of a monoatomic ion. Hence, the ionic radius of MA is controversial. Amat et al. [296] calculated the effective ionic radius of the molecule based on DFT to be 2.7 Å for $MAPbI_3$, while Yin et al. calculated it to be 2.37 Å [297]. Here, we take the data by Amat et al. [296] for the calculation of Goldschmidt's factor. Based on this data, and the ionic radii of the divalent substituent ion, the factor was calculated and shown in Table. 6-2.

### 6.2.2 Computational Details

All calculations were carried out using Vienna Ab initio Simulation Package (VASP) package [122,308]. An structure consisting of 48 atoms was taken and its complete structural optimization was performed with a 4x4x4 k-mesh using Perdew, Burke and Ernzerhof (PBE) functional [277] and with van der Waal's interaction (PBE+D2) [309] separately. A 2x1x2 supercell for the structure is shown in Figure 6-10. The projector-augmented wave (PAW) approach, developed by Bloch in VASP was used for the description of the electronic wavefunctions. Plane waves have been included up to an energy cut-off 600 eV. The energy criterion convergence was $10^{-6}$ eV. Grimme's D2 parameters [309] were taken for including van der Waal's interactions. Scalar relativistic effects were taken into account in the pseudopotential. After the geometric optimization, hybrid functional to Heyd-Scuseria-Ernzerhof (HSE06) based electronic optimization was carried out on the geometrically optimized structure from D2 correction at gamma point only. Hybrid functional is employed because it is proven to give better understanding of the electronic properties of material [271,272] at much reasonable computational cost. Hybrid functional mix 25 % of exact nonlocal exchange of Hartree-Fock theory [310] with the density functional exchange. Subsequently, effect of spin-orbit coupling was also monitored along with hybrid functional. The effect of spin-orbit coupling [311] becomes necessary especially for the heavy atoms such as transition elements and group-IV elements. Absorption coefficient was obtained from calculating dielectric function for materials, which is obtained from eq. (6-1):

Here the indices *c* and *v* refer to conduction and valence band states respectively, $u_{ck}$ and $w_k$ are the cell periodic part of the orbitals and weight of k-point respectively. Ω is cell volume and *e* is electronic charge. The effective mass of the hole and electron are obtained by fitting parabolic expression eq. 6-3 around gamma point in Brillouin zone, diagonalizing the matrix *A* then taking inverse of the matrix as in eq. 6-3 at the band extremes.

$$E(\vec{k}) = E(\vec{k}_\Gamma) \pm \frac{\hbar^2}{2}\left[\vec{k}_i - \vec{k}_\Gamma\right]\left[A_{ij}\right]\left[\vec{k}_i - \vec{k}_\Gamma\right]^T \qquad (6\text{-}3)$$

where $E(\vec{k}_\Gamma)$ represents the eigenenergies at the band extremes, that is, the values at the gamma point, $\vec{k}_\Gamma$ in the Brillouin zone. The tensor $A_{ij}$ is related to the effective mass tensor $m_{ij}^*$ by

$$[A_{ij}] = \sum_{i=1}^{3}\sum_{j=1}^{3}\frac{1}{m_{ij}^*} \tag{6-4}$$

### 6.2.3 Results and Discussions

Contrary to the expectations that PBE would underestimate the bandgap [257,259], the bandgap value obtained is closer to experimental value of 1.6 eV. In addition, the reduced mass obtained from the PBE calculation was $0.18*m_e$, which is in fair agreement with experimental data of $0.15*m_e$. It indicates DFT can describe well the electronic properties of $MAPbI_3$. On the other hand, the bandgap obtained for Sn-based pervoksite is 0.80 eV with PBE while the experimental value is 1.2 eV, which is close to our HSE calculation of 1.15 eV. Due to such unbalanced description, the band-gap values are presented for all the proposed materials with various level of DFT theory as shown in Table. 6-2 and 6-3. It shows the bandgap values of obtained from various theories vary in a reasonable range. However, HSE+SO should be considered more accurate calculations compared to others. The lattice constants are shown with D2 correction. The lattice constants decrease due to inclusion of D2 corrections, which is in agreement with work by Feng et al. [303], discussed above. The effective mass values in x [100], y [010] and z [001] directions were calculated using PBE only as it is computationally expensive to calculate dense k-mesh energies using hybrid functionals. Some of the materials have very high effective masses compared to others, which is undesirable. Anisotropic effective masses were discovered in these materials, as discussed above, but anisotropic materials are also not suited for photovoltaics during uniform wafer fabrication.

Alkali metals iodide are generally in +1 oxidation state, hence they were excluded in the present search. Alkaline earth materials were selected as they have iodides in +2 state. Among them, magnesium shows promising feature of band-gap within the visible region, low effective mass values and reasonable absorption coefficient (Figure 6-11A) compared to $MAPbI_3$ and $MASnI_3$. Different effective masses in different directions indicate the material could have anisotropic conductivity in different directions. In addition $MgI_2$ has crystallizes into P3m1 group which is similar to $PbI_2$, hence Mg is predicted as a potential replacement for Pb. Based on Goldscmidt's factor, the Ca based material should be more stable than the Mg, however Ca doesn't have band gap in visible region as shown in Table. 6-2. While alkaline earth metals are generally in +2 oxidation state, the transition elements can exist in multiple oxidation states, which is undesirable due to instability issue. Iodides of Cobalt (Co), Palladium (Pd), Zinc (Zn), Cdmium (Cd) are generally in +2 state, while Ti and V is prone to go to +4 state. However, $TiI_2$, $VI_2$, $CoI_2$ and $PdI_2$ have same space group as $PbI_2$. All of the transition metal organo halides studied here have the bandgap within visible region, however, based on the effective mass calculation Pd and Zn have lower effective masses. In addition, the Goldschmidt's factor for these materials is comparatively higher than the alkaline earth materials signifying their lesser stability. Among the transition metal series, Ti, V and Cd has relatively less Goldschmidt factor, but Ti has very high carrier effective mass. Based on the absorption coefficient (Figure 6-11) Zn, Cd and Co have relatively higher absorption compared to other transition metals discussed here.

Furthermore, although the iodides of Ge and Si, which are in same periodic table group as Pb, have their iodides in +4 state, it was academically interesting to study their behavior using DFT. Both of the materials have bandgap in the visible region and they have reasonable better effective masses compared to other materials. Based on the absorption coefficient plot (Figure 6-11C), these materials are predicted to have higher absorption compared to even Pb based compounds. For Si, there are no divalent iodides, however, for Ge the Goldschnmidt factor is not very high and based on its absorption behavior it is compelling to form this compound. Here, although the material is studied in orthorhombic group, it may lead to other phases such as tetragonal and cubic phase, but previously it has been that the band-gap and other relevant properties are not much affected after the phase change.

Further to the above-mentioned information, the absorption coefficient, the relative band alignment of the materials is also an important issue. Hence, the conduction band minima (CBM) and valence band maxima (VBM) data are compared for the materials as shown in Figure 6-12. Among alkaline materials, Mg has closest resemblance to the VBM and CBM of $MAPbI_3$. Ca, Sr has very large differences of the CBM and VBM, which is also consistent with their bandgap. Hence they might not be suitable at all for solar cell applications. Among transition elements Zn, Cd and Co are better candidates for applications. Among group IVA, Ge and Si are possible replacements for Sn. However, as discussed earlier both of these materials are unstable due to their existence in +4 oxidation state.

### 6.3 Summary

In conclusion, we have demonstrated that semi-local exchange-correlation functional PBE accurately reproduces the structural and elastic properties of the YAG, while the hybrid exchange-correlation functional HSE06 can, in addition, accurately describe the experimental band-gap and *f*-band positions. We predict that co-doping Ce:YAG with Tm or Tb results in optical transitions very close or even lower in energy than ErCe:YAG; we thus anticipate these materials will have quantum efficiencies comparable or higher than ErCe:YAG. We find that the lanthanide co-doping of Ce:YAG with La, Pr, Nd, Pm, Gd, and Lu results in a redshift of the optical transition frequency that is mainly attributed to strains of the bonding environment around the Ce-dopant. The transition strength methodology used here is directly applicable to more general problems involving for instance, semiconductors, perhaps leading to better understanding of underlying optoelectronic phenomena. We believe the findings of the work can be utilized in building systematics and making experimental samples with required material properties.

Next, major problems in substituting Pb from $CH_3NH_3PbI_3$ are discussed and then elements from periodic table are identified that might be suitable for photovoltaic applications. From alkaline earth Mg, from transition elements Zn, Cd, Co and from group IVA Ge is predicted to be a replacement material apart from Sn, which has already been studied. These results can guide experiments and save time during their search for lead-replacement material.

Table 6-1. Predicted major electronic transitions below 3 eV.

| *Name* | *f-d transition (eV)* | *f-f transition (eV)* |
|---|---|---|
| LaCe | 2.61,2.92 | |
| PrCe | 2.59,2,91 | |
| NdCe | 2.52,2.83 | |
| PmCe | 2.44,2.75 | |
| EuCe | 2.77,2.98 | |
| GdCe | 2.2,2.54,2.92 | |
| TbCe | 2.2,2.5,2.96 | 1.24,1.38,1.55 |
| DyCe | 2.96 | 2.16,2.2,2.46,2.83 |
| HoCe | 2.46,2.95 | 2.13,2.17,2.21 |
| ErCe | 1.67,2.19,2.96 | 1.29,1.37 |
| TmCe | 2.15,2.44,2.93,2.98 | 0.73,1.02 |
| LuCe | 2.68,2.97 | |

Table 6-2. Band-gap, Goldschmidt's factor, lattice constant for various $MAXI_3$. PBE represents Perdew, Burke and Ernzerhof (PBE) functional, PBE+D2 is PBE with Grimme's D2 correction term. HSE stands for Heyd-Scuseria-Ernzerhof functional. +SO represents HSE along with spin orbit coupling. PBE and PBE+D2 are used for separate geometric optimization. However, for electronic calculations structures from PBE+D2 are used.

| X | Gap | | | | Gold | a | b | c |
|---|---|---|---|---|---|---|---|---|
| | PBE | +D2 | HSE | +SO | | | | |
| Pb | 1.73 | 1.67 | 2.30 | 1.38 | 0.99 | 8.74 | 12.53 | 8.46 |
| Sn | 0.80 | 0.60 | 1.15 | 0.87 | - | 8.74 | 12.29 | 8.40 |
| Be | 2.70 | 2.53 | 3.54 | 3.39 | 1.28 | 8.49 | 12.39 | 7.40 |
| Mg | 1.47 | 1.76 | 2.68 | 2.49 | 1.16 | 8.39 | 11.78 | 7.93 |
| Ca | 3.77 | 3.82 | 4.99 | 4.73 | 1.05 | 8.67 | 12.33 | 8.32 |
| Sr | 3.84 | 3.94 | 5.02 | 4.72 | 1.04 | 8.82 | 12.74 | 8.75 |
| Ti | 1.52 | 0.54 | 1.91 | 2.26 | 1.10 | 8.27 | 11.36 | 7.84 |
| V | 0.10 | 0.09 | 0.64 | 2.72 | 1.13 | 8.15 | 11.28 | 7.71 |
| Co | 0.13 | 0.26 | 1.43 | 2.28 | 1.25 | 8.07 | 11.21 | 7.51 |
| Pd | 1.65 | 1.55 | 0.80 | 0.70 | 1.16 | 8.25 | 11.32 | 7.70 |
| Zn | 1.36 | 1.12 | 2.07 | 1.88 | 1.21 | 8.58 | 11.90 | 7.68 |
| Cd | 0.96 | 0.93 | 1.81 | 1.53 | 1.12 | 8.54 | 12.07 | 7.94 |
| Si | 0.08 | 0.04 | 0.22 | 0.56 | - | 8.43 | 11.55 | 7.98 |
| Ge | 0.84 | 0.72 | 1.25 | 1.09 | 1.15 | 8.48 | 11.84 | 8.06 |

Table 6-3. Electron and hole effective mass in different directions for various $MAXI_3$ with PBE+D2

| *X* | | $m_h$ | | $m_e$ | | |
|---|---|---|---|---|---|---|
| | [010] | [001] | [100] | [010] | [001] | [100] |
| Pb | 0.23 | 0.25 | 0.26 | 0.84 | 0.67 | 0.10 |
| Sn | 0.04 | 0.07 | 0.08 | 0.75 | 0.46 | 0.03 |
| Be | 0.56 | 1.62 | .6.79 | 0.74 | 0.47 | 0.47 |
| Mg | 0.56 | 1.24 | 1.83 | 0.36 | 0.28 | 0.28 |
| Ca | 0.47 | 2.12 | 5.61 | 0.53 | 1.30 | 0.94 |
| Sr | 0.69 | 2.37 | 11.96 | 0.39 | 0.38 | 0.31 |
| Ti | 137.2 | 0.83 | 0.68 | 31.2 | 0.55 | 0.51 |
| V | 48.7 | 0.9 | 0.8 | 4.12 | 3.97 | 0.31 |
| Co | 2.24 | 6.5 | 61.5 | 6.6 | 13.4 | 13.9 |
| Pd | 0.590 | 1.080 | 1.552 | 2.03 | 0.62 | 0.57 |
| Zn | 0.598 | 1.794 | 2.634 | 0.51 | 0.47 | 0.35 |
| Cd | 0.635 | 2.71 | 6.92 | 0.3 | 0.3 | 0.3 |
| Si | 0.029 | 0.049 | 0.1 | 0.03 | 0.81 | 0.52 |
| Ge | 0.08 | 0.11 | 0.13 | 0.78 | 0.61 | 0.05 |

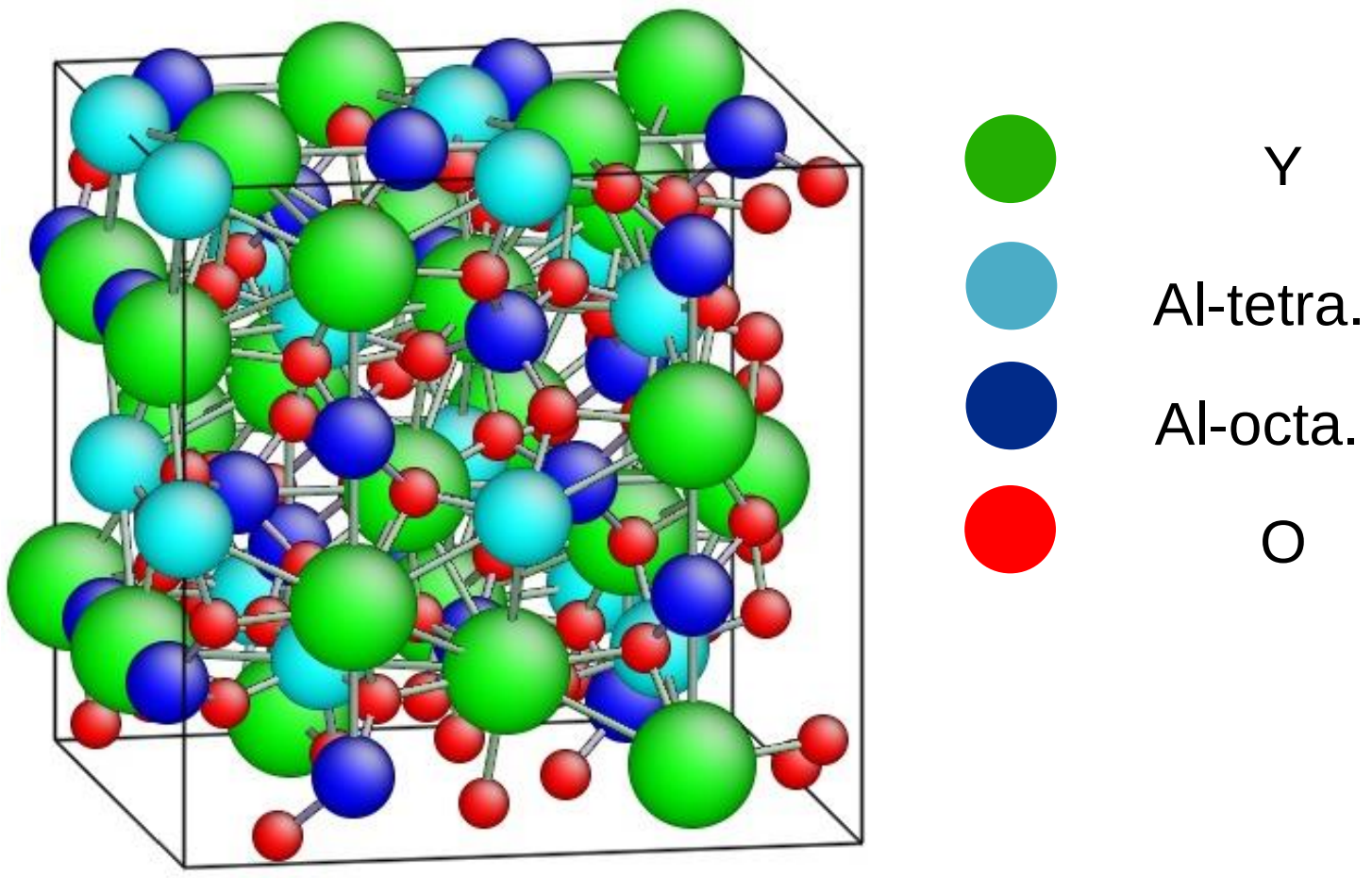

Figure 6-1. Unit cell of YAG with different sites are shown.

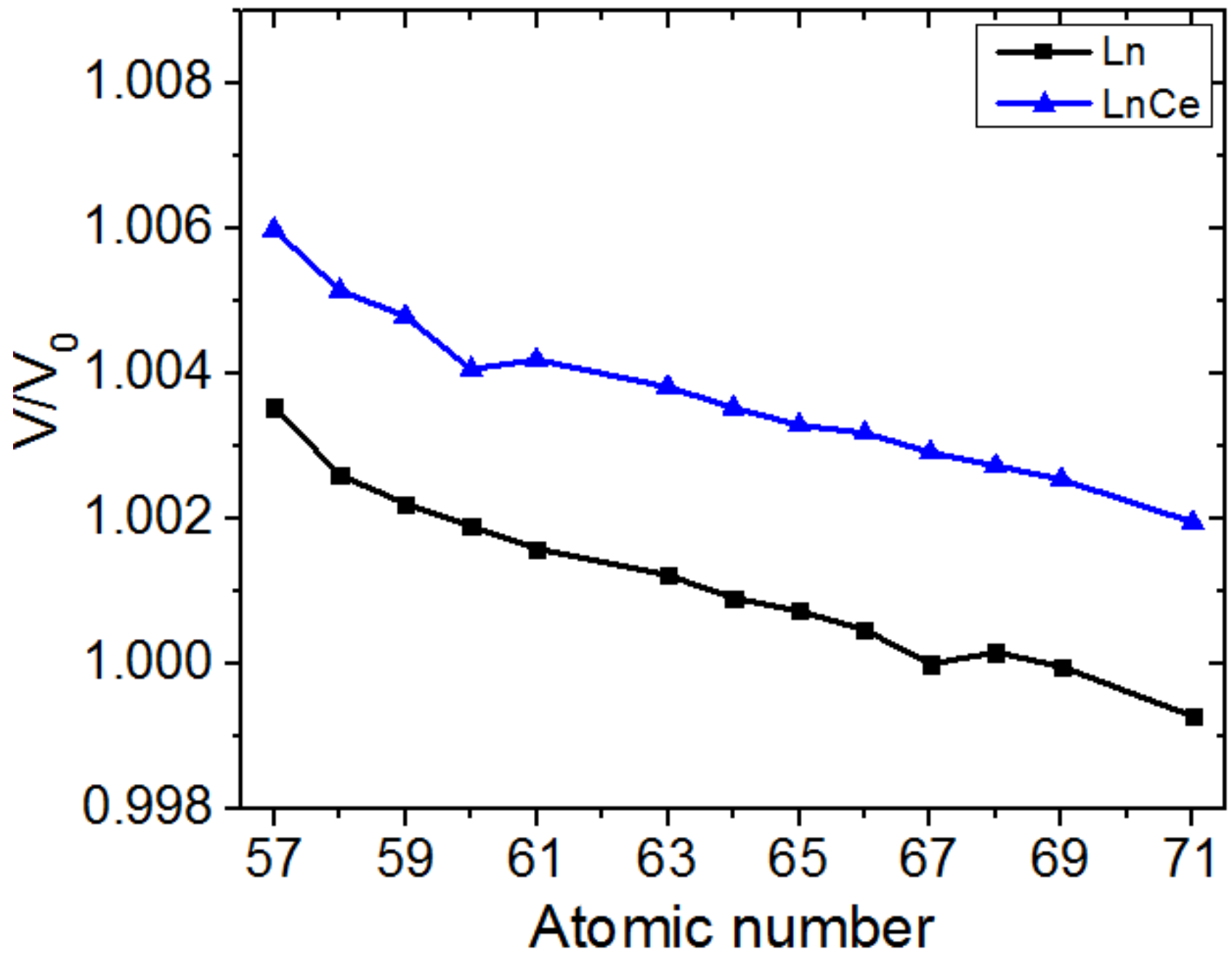

Figure 6-2. Normalized volume vs. atomic number plot for Ln:YAG and LnCe:YAG. The volumes of the system decrease along the lanthanide row, which can be explained by lanthanide contraction. $V_0$ is the volume of pure YAG.

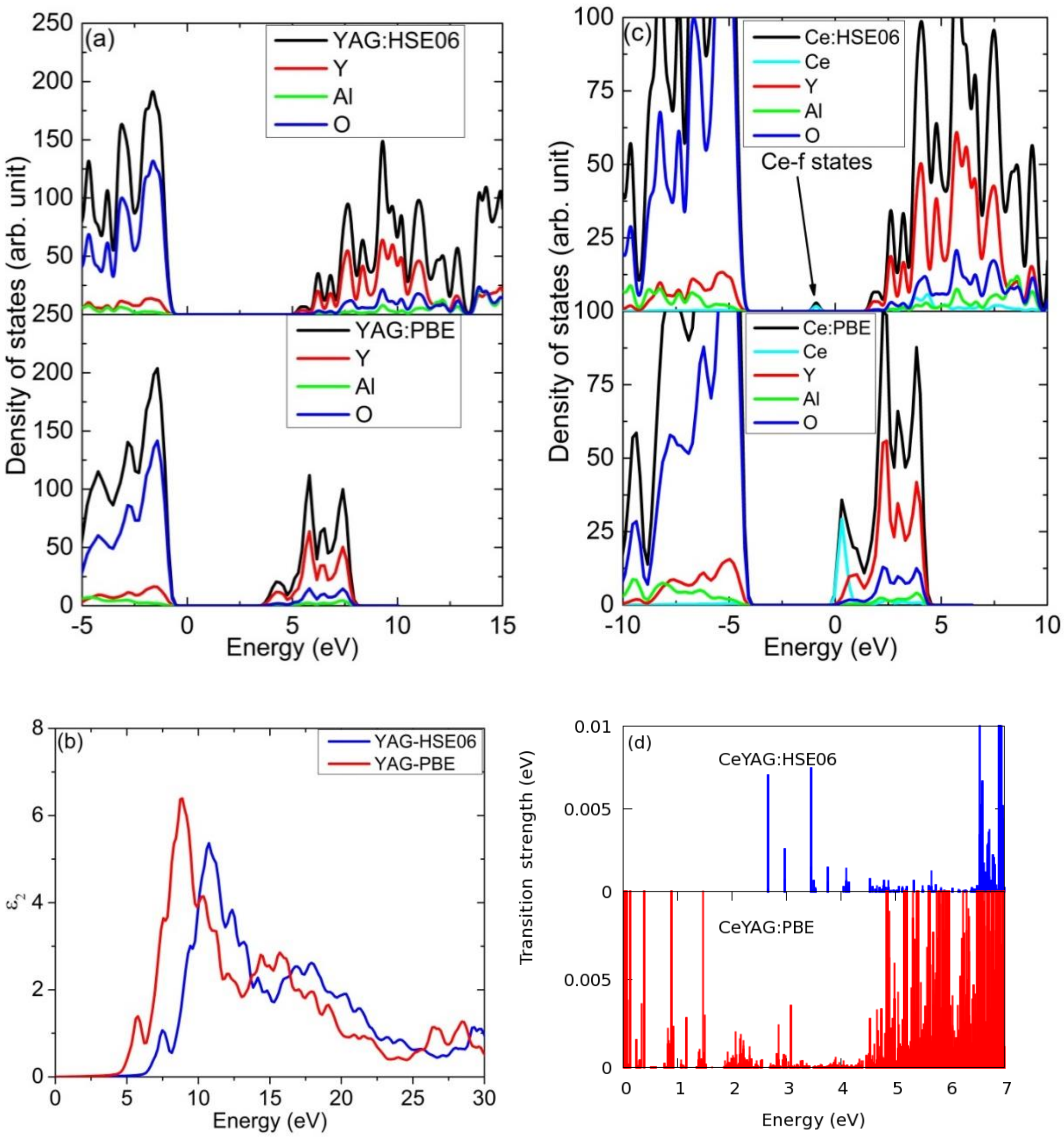

Figure 6-3. Optoelectronic properties of pure YAG. A) The projected density of states for pure YAG using the PBE and HSE06 functional show that HSE06 predicts a band gap in agreement with experiments, B) The dielectric function for pure YAG using PBE and HSE06, C) The projected density of states for Ce-doped YAG shows that the Ce-*d* and *f*-states are located inside the gap-region of YAG using HSE06, D) All possible transitions near the band-gap show that HSE06 reproduces the experimental emission wavelength for Ce:YAG. The Fermi energy is set to zero for density of states plot.

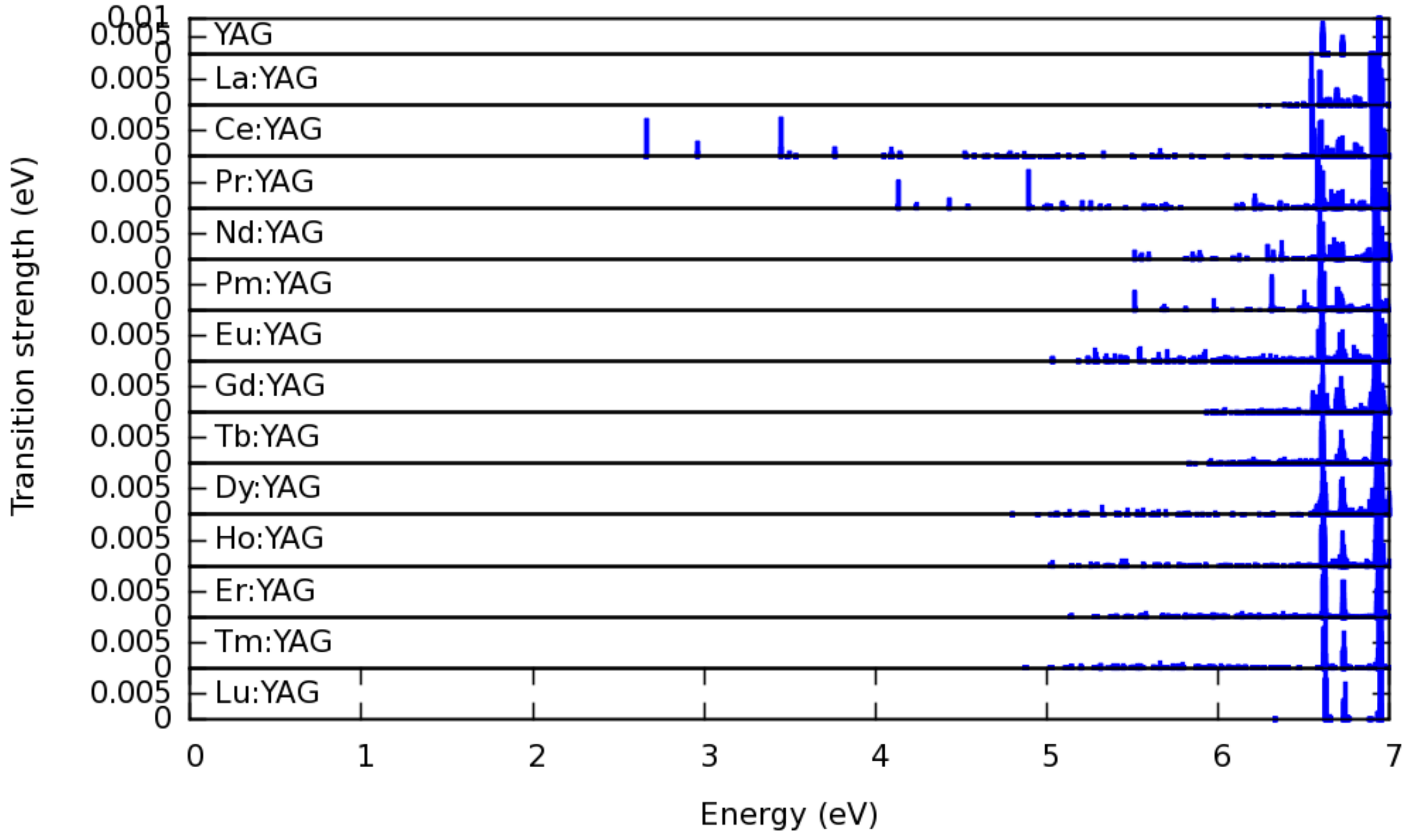

Figure 6-4. Transition strengths calculated using Eq.6-2 are shown for all lanthanides. Peaks at 2.6 eV of Ce and 4.13 eV of Pr are comparable to experiments.

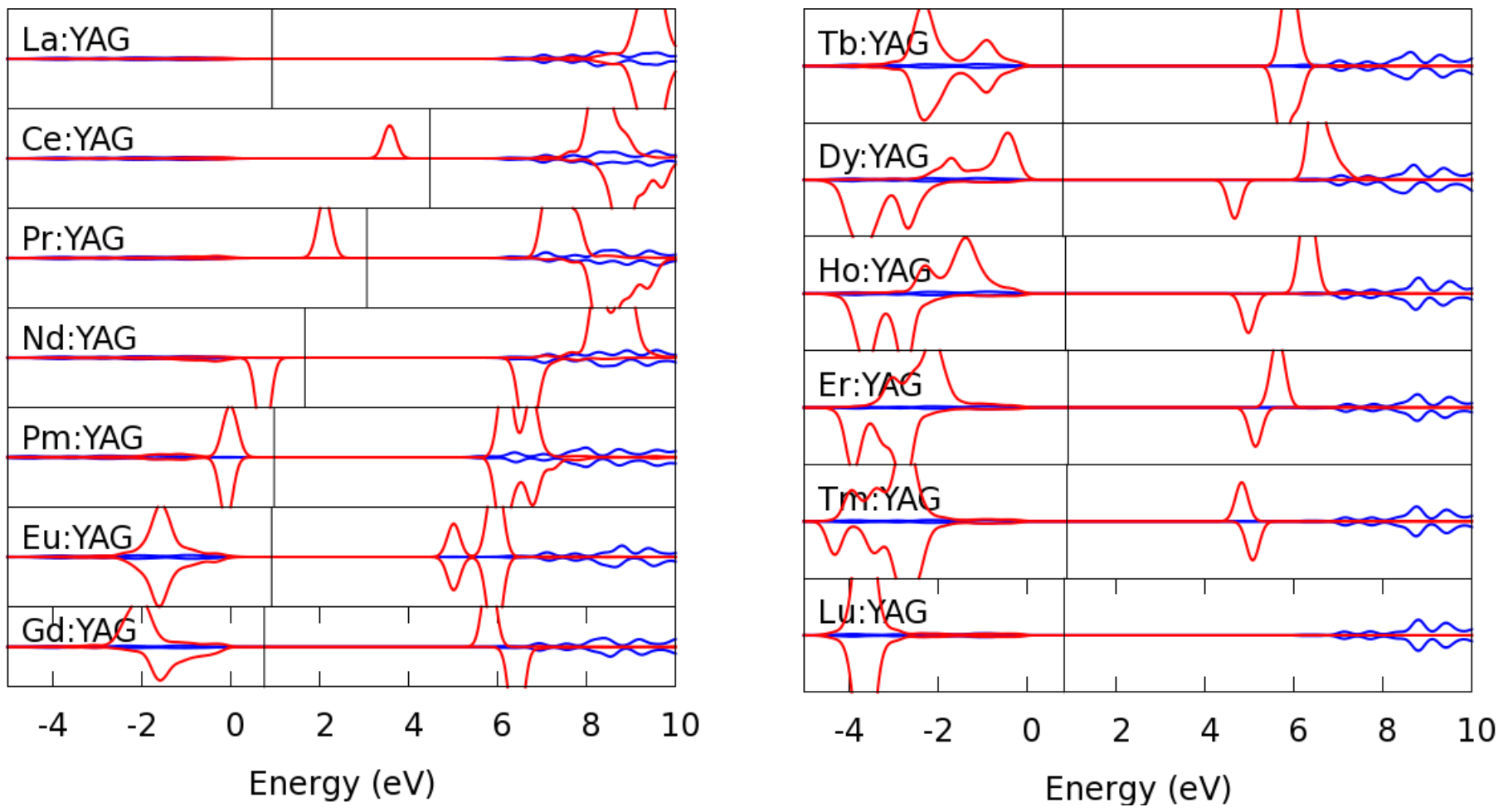

Figure 6-5. Spin- up-down, *lm*-projected density of states showing *d* (blue) and *f* (red) states for Ln in Ln:YAG. The energy axis is shifted based on the maximum occupied *d* bands of the Ln. Corresponding Fermi-energies are shown by black vertical lines. Occupied *f* and unoccupied *d* states are near the visible region only for Ce, Pr dopants.

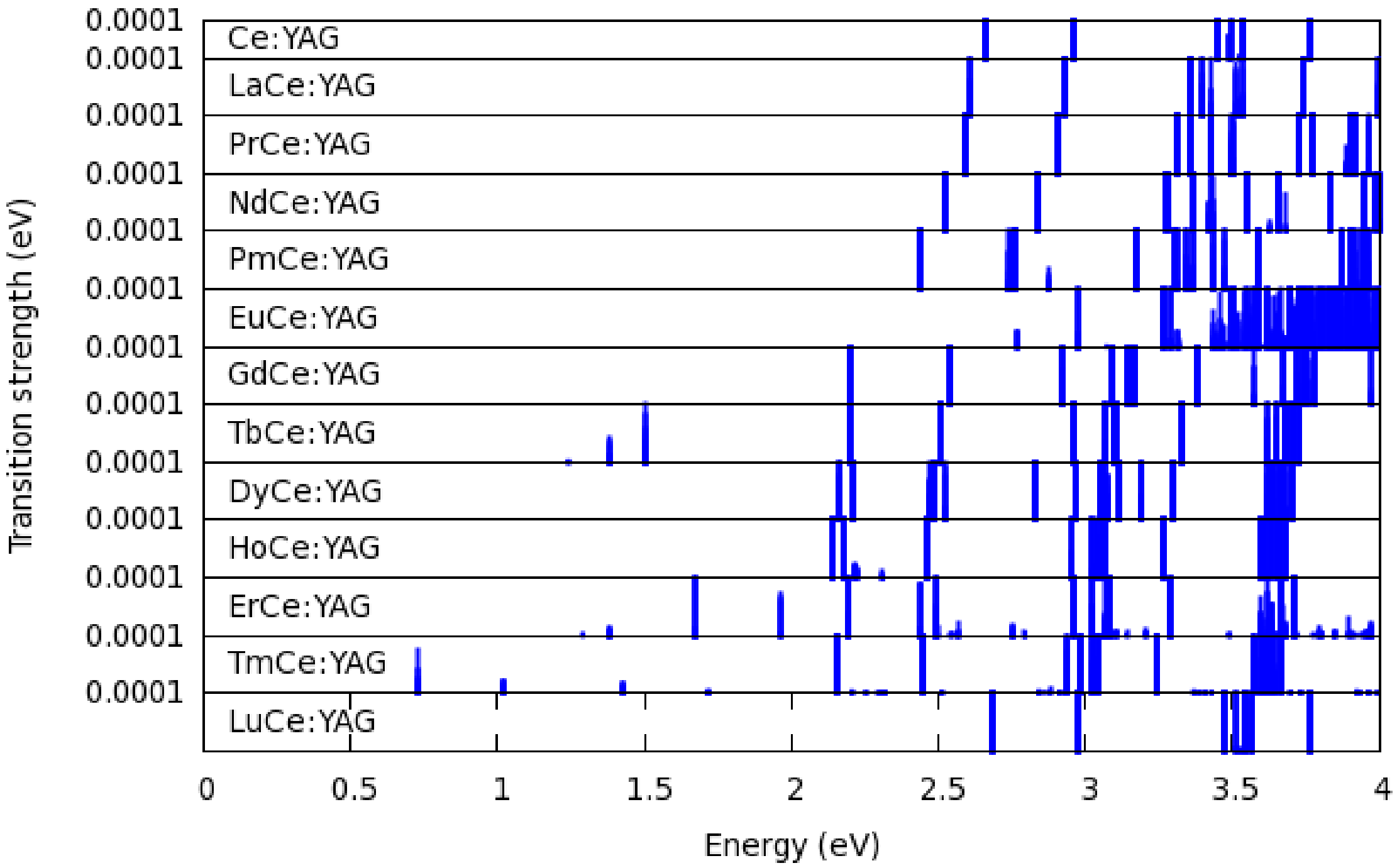

Figure 6-6. Transitions near visible spectrum for co-doped systems. For co-doping of Ce with other lanthanides, a blue shift is observed for Eu and Lu, while a red shift is observed for other lanthanides.

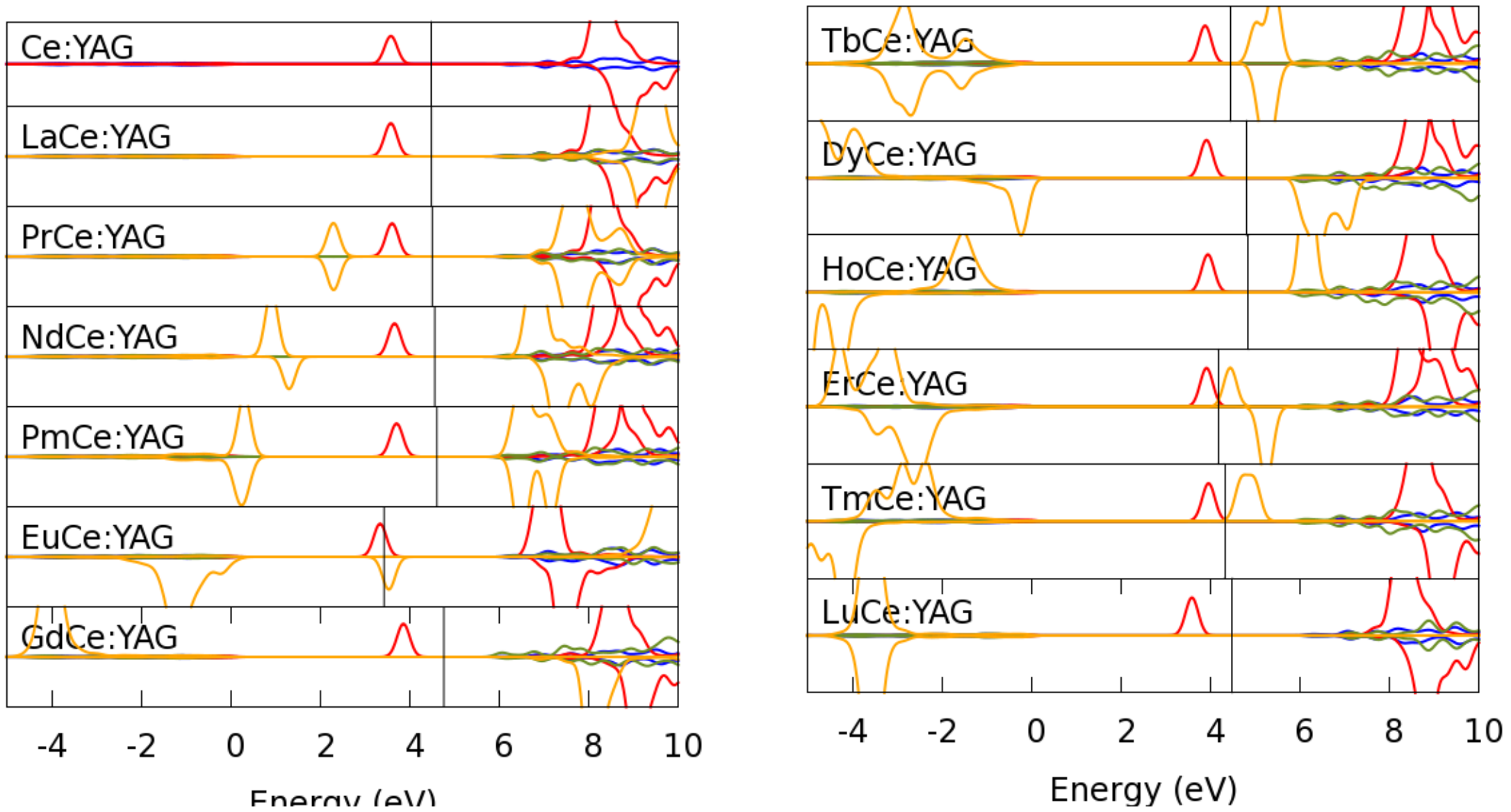

Figure 6-7. Density of states for Ce *d* (blue) and *f* (red) states, lanthanide's *d* (green), *f* (orange) in co-doped systems. Shift in *f*-states are observed for Eu and Lu case. Occupied Ce-*f* states and unoccupied lanthanide *f* states are available for TbCe, ErCe and TmCe only.

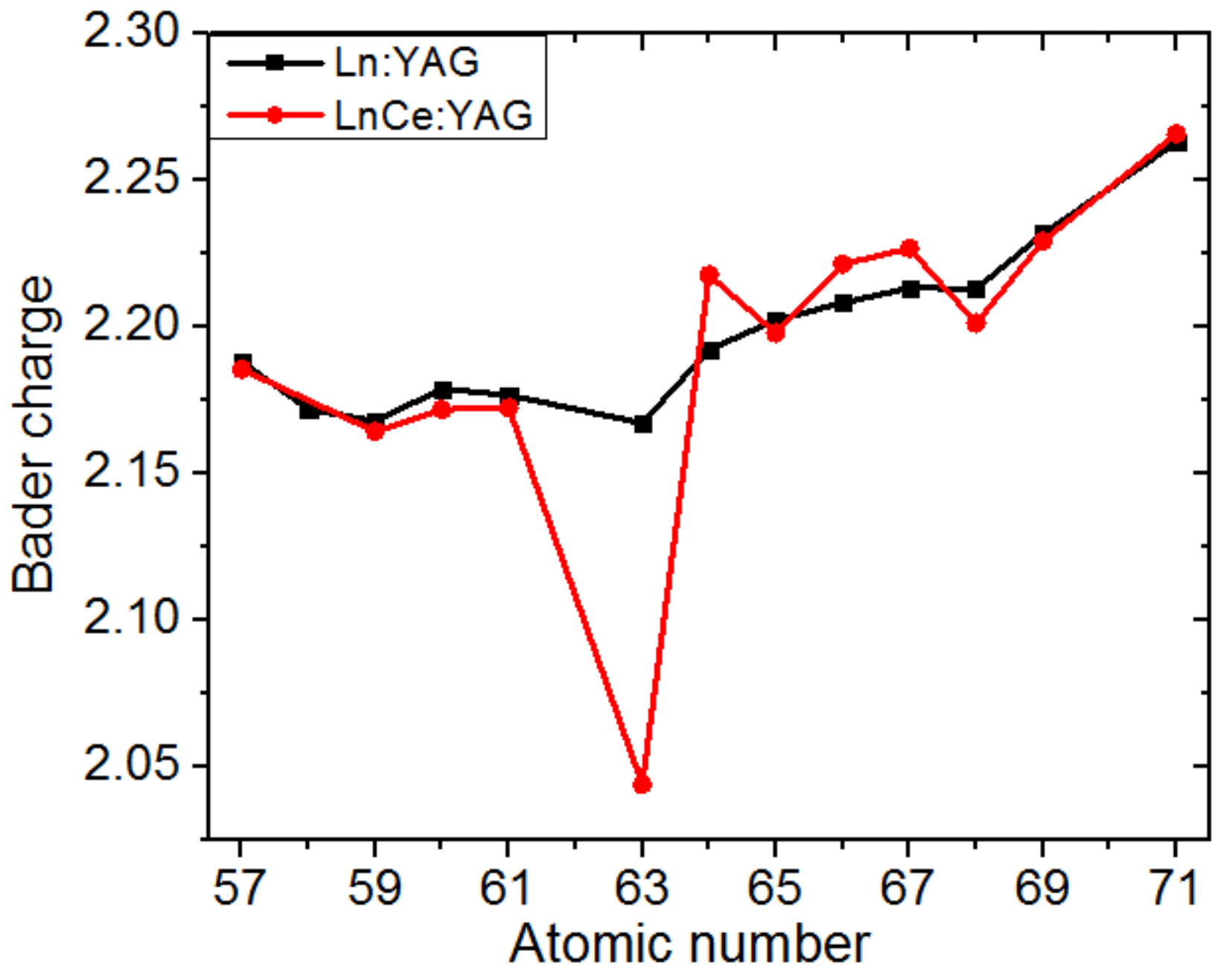

Figure 6-8. Bader charge analysis for Ln:YAG and LnCe:YAG showing the Bader charge increases along the lanthanide row, with a dip that suggests that Eu prefers the 2+ charge state.

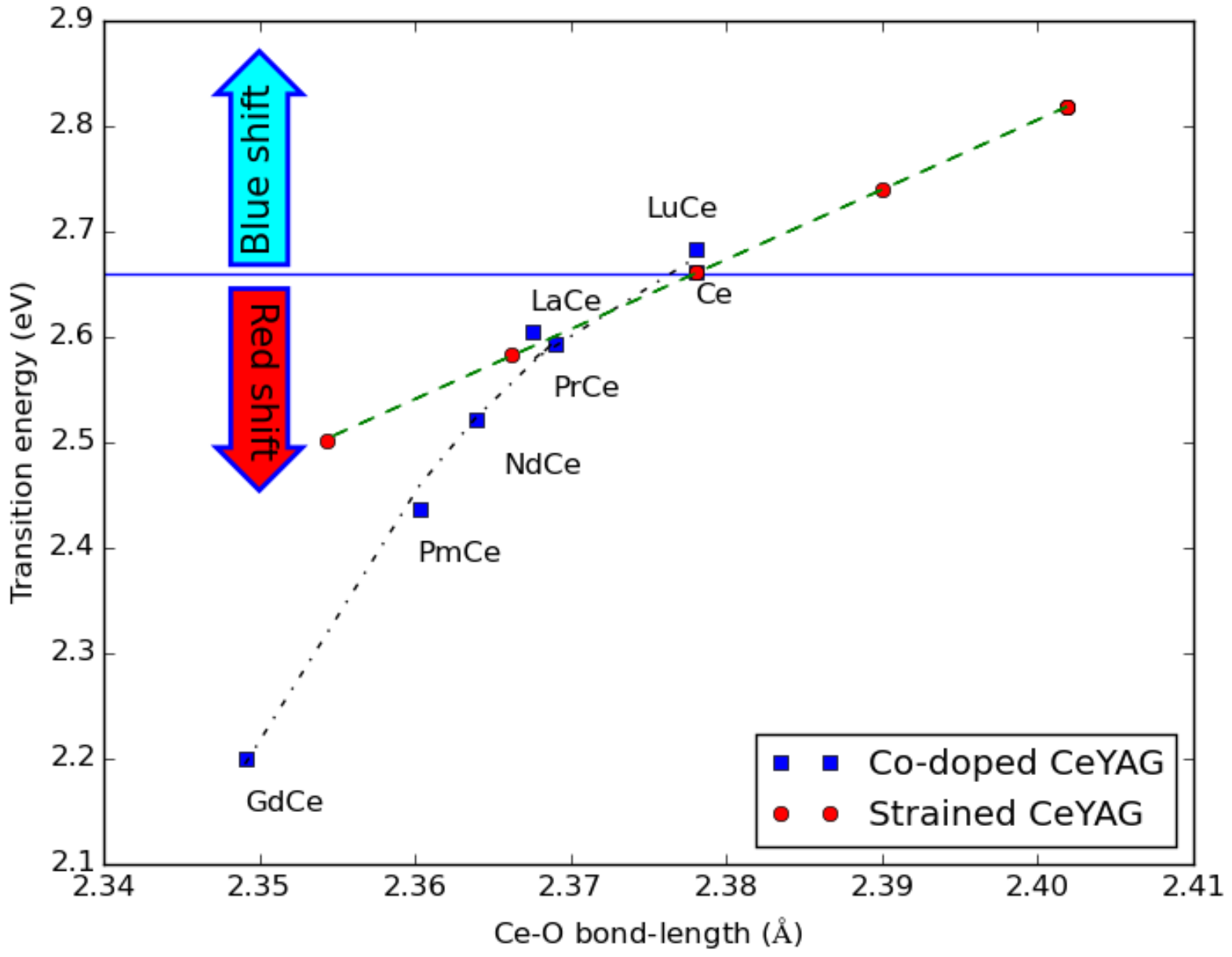

Figure 6-9. Transition energies versus minimum in distribution of Ce-O bond lengths for volume strained and co-doped CeYAG. Both of the curves follow similar trends.

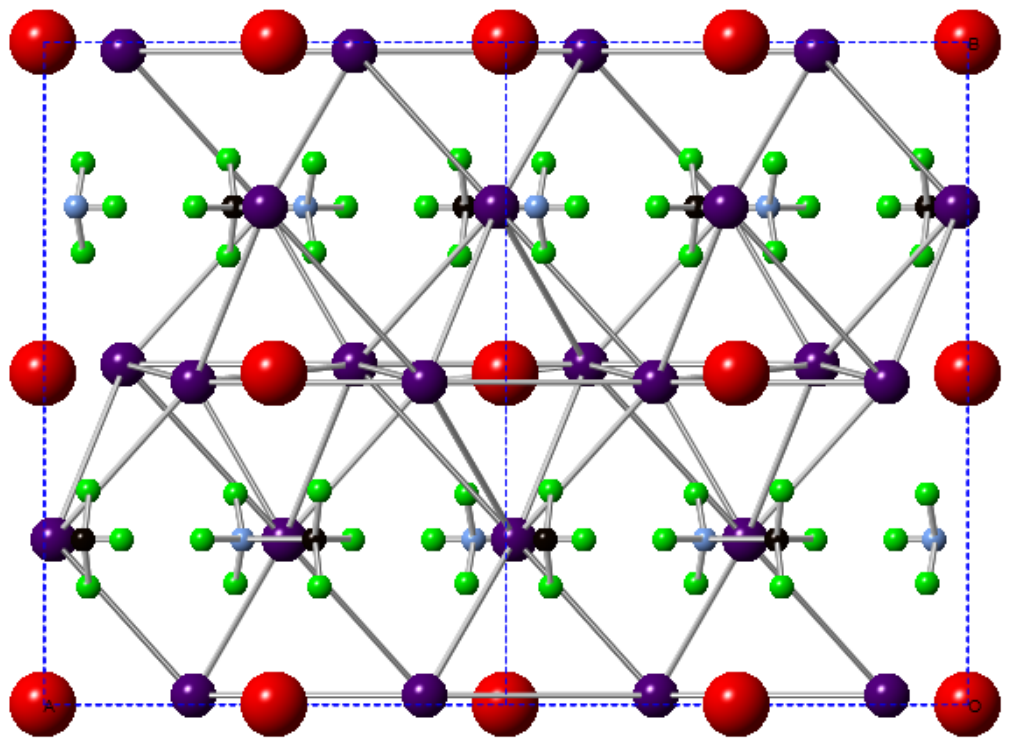

Figure 6-10. showing 2x1x2 supercell for orthorhombic $CH_3NH_3PbI_3$ . Red, purple, green, black, blue balls in the figure represent lead, iodine, hydrogen, carbon and nitrogen atoms respectively.

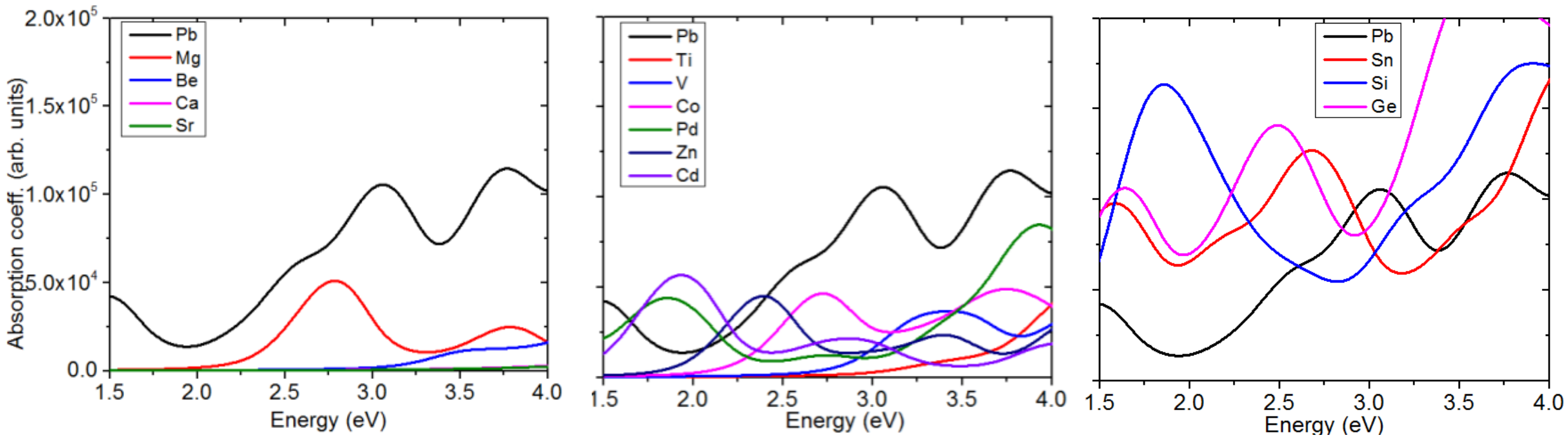

A B C

Figure 6-11. Absorption coefficient obtained from HSE+SO calculation for the various materials discussed above. Among alkaline earth elements Mg, among transition elements Zn, Cd, Co and among group IVA Ge have comparatively high absorption among respective groups. Absorption of $MAPbI_3$ is also given for reference.

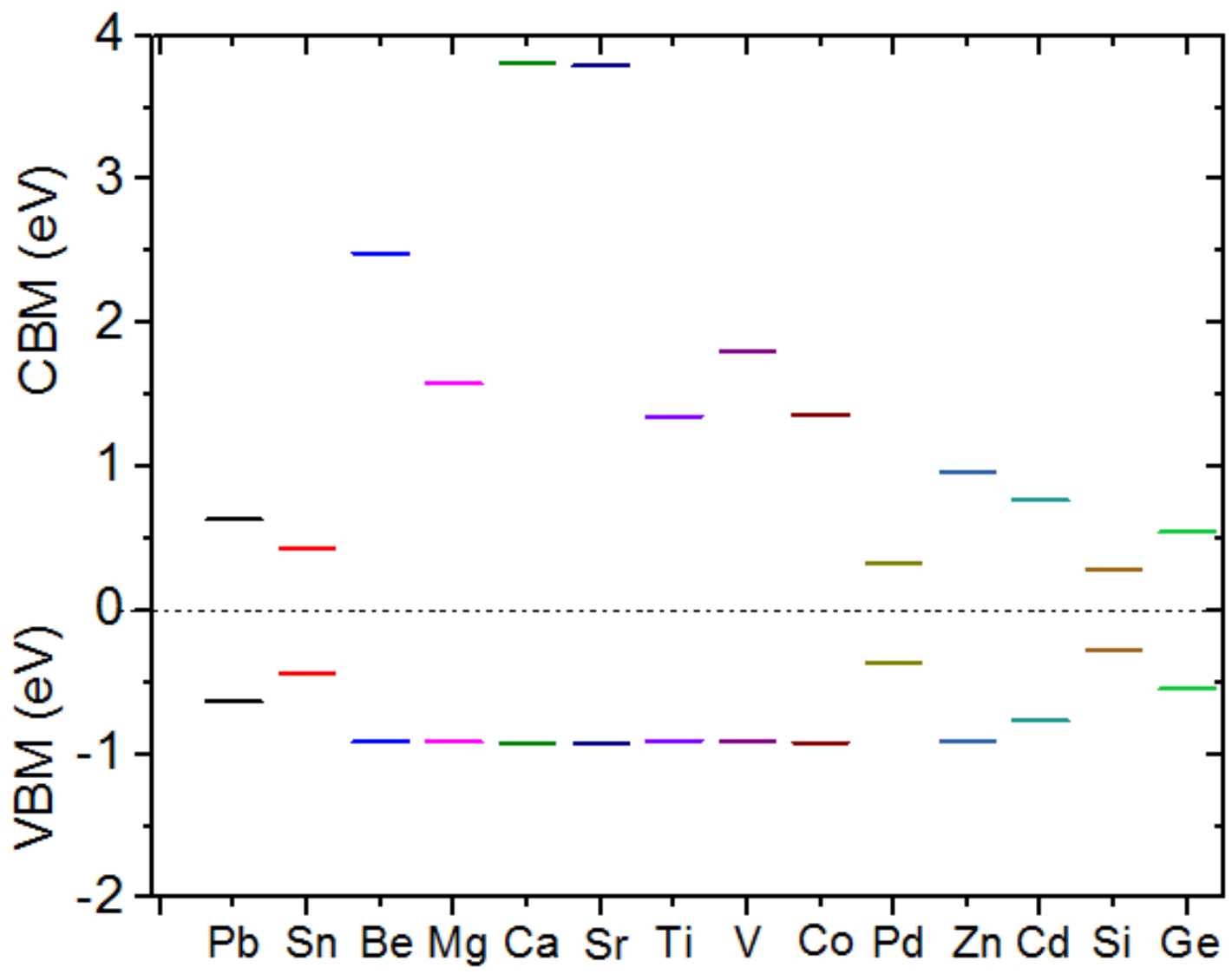

Figure 6-12. Valence band maxima (VBM) and conduction band minima (CBM) positions for the proposed materials compared to $MAPbI_3$.Sn, Ti, Co, Zn, Cd, Ge are comparatively closer to $MAPbI_3$.

# CHAPTER 7
# CONCLUSIONS

In this thesis, computational materials modeling methods were used to investigate multiple problems related to defects of various types associated with materials of different types. For instance, empirical potential based molecular dynamics simulations were used examine the surface modification of polymers due to energetic polyatomic ion deposition, as well as the thermodynamics and mechanics of metal-ceramic interfaces and nanostructures. In addition, density functional theory was used to screen substituents in optoelectronic materials.

First, surface modification of amorphous polymers PS and PMMA with polyatomic ion beams was carried out with REBO potential. The primary goal of the work was to quantify the atomistic phenomenon based on the density profile analysis, mass-spec analysis of products formed and sputtered, surface roughness and specification of preferred sites on monomers. The results clearly indicated the preferred site on monomer changes with incident beam energy and number of saturated bonds in the ion beam. It also quantified the upper limits for constituents atoms in the computational mass-spec analysis of products formed and sputtered after deposition events. The depth profile showed that for a given energy and chemistry of the ion beam, the maximum depth that the ion beam could be modify, and the nature of the surface modification, could be predicted computationally.

Second, a third-generation COMB potential was developed for the $Al/Al_2O_3/AlN$ system to provide understanding of the atomistic phenomena that take place at the interface and at nanostructures. This COMB3 potential could not only reproduce the experimental and ab-inito data to which it was fit, but also predicted the work of adhesion of the interface structures and the elastic modulus of the nanostructures comparable to the experiments. The snapshots for interfaces and nanostructures during calculations and simulations demonstrated the atomistic events that occur during the processes. The potential has the appeal to extend it across other elements in the periodic table for fulfilling the long cherished dream of a charge based reactive unified force-field for materials for computational material science investigations. It is possible to extend the Al-O-N potentials for Ni, Pt, Zn and Ti elements, work that is being undertaken by others.

Third, the effect of defects on the electronic and optical properties of materials were carried out for $Y_3Al_5O_{12}$ and $CH_3NH_3PbI_3$. The results clearly indicated use of hybrid functional as a suitable candidate for optoelectronic studies because of its excellent agreement with experimental data. The red-shift and blue-shift of absorption energy peak due to co-doping in Ce:YAG was correlated with the change in bonding environment around Ce-atom. The results also demonstrated possibility of new highly efficient materials using co-doped YAG. Computational prediction for substitution of toxic Pb with other non-toxic elements in $CH_3NH_3PbI_3$ were predicted. These predictions are guiding on-going experimental work..

Summing up, this thesis established the importance of defects in materials and elucidated how computational materials methods can be used to tailor the properties of defect-mediated materials.

# APPENDIX
# POTENTIAL PARAMETERS FOR COMB POTENTIALS DEVELOPED IN THIS WORK

The atomic and electrostatic parameters of COMB potentials for pure elements (Al, O, and N) are given in Table A-1, while the bond-typed parameters for pure elements and binary systems ($Al_2O_3$, and AlN) are given in Tables A-2 and A-3.

Table A-1. Atomic and electrostatic parameters of pure elements (Al, O, N) for COMB3 potentials.

| Parameters | Al | O | N |
|---|---|---|---|
| X ($eVe^{-1}$) | 3.10166 | 6.59963 | 6.209731 |
| J $(eVe^{-1})^2$ | 3.53243 | 3.10166 | 9.29225 |
| K $(eVe^{-1})^3$ | 0.48166 | 0.76043 | -1.25446 |
| L $(eVe^{-1})^4$ | 0.0 | 0.0 | 0.286350 |
| ξ (Ặ$^{-1}$) | 2.1433 | 2.14331 | 1.438711 |
| Z (e) | 0.052709 | -1.53917 | -0.552136 |
| $P^{\chi}$($eV.e^{-3}.r^{-3}$) | 1.5063 | 3.25885 | 3.7385490 |
| $P^{J}$($eV.e^{-3}.r^{-3}$) | 0.2 | 0.30569 | 0.5167863 |
| $D_U$ (Ặ) | -0.03512360 | -1.21395 | 0.5167863 |
| $D_L$ (Ặ) | 0.0105133 | 0.0076644 | 0.1969380 |
| $Q_U$ (e) | 3.00000000 | 6.0000000 | 5.000000 |
| $Q_L$ (e) | -5.000 | -2.00000000 | -3.000000 |

Table A-2. COMB3 pairwise and bond-order parameters for $Al_2O_3$.

| Parameters | Al-Al | Al-O | O-Al | O-O |
|---|---|---|---|---|
| $A_{ij}(eV)$ | 473.3379 | 1818.689 | 1818.689 | 4956.339 |
| $B_{ij}^{1}(eV)$ | 242.8881 | 22.189 | 22.189 | 688.163500 |
| $B_{ij}^{2}(eV)$ | 0.0000 | 0.0 | 0.0 | 0.0 |
| $B_{ij}^{3}(eV)$ | 0.0000 | 0.0 | 0.0 | 0.0 |
| $\lambda$ (Å$^{-1}$) | 1.8617 | 4.177 | 4.177 | 5.295 |
| $\alpha_{ij}^{1}$ (Å$^{-1}$) | 1.4993 | 1.0115 | 1.0115 | 3.259 |
| $\alpha_{ij}^{2}$ (Å$^{-1}$) | 1.1 | 0.0 | 0.0 | 0.0 |
| $\alpha_{ij}^{3}$ (Å$^{-1}$) | 0.6858 | 0.0 | 0.0 | 0.0 |
| $\beta$ (Å$^{-1}$) | 1.4993 | 0.0 | 4.31 | 3.259 |
| $b_6$ | 0.0000 | 0.0 | 0.0 | -1.0257 |
| $b_5$ | 0.0000 | 0.0 | 0.0 | -0.1700 |
| $b_4$ | 0.0000 | 0.0 | 0.0 | 2.0153 |
| $b_3$ | 0.0000 | 0.0 | 0.0 | -2.2384 |
| $b_2$ | 0.0000 | 0.0 | 0.0 | 1.1852 |
| $b_1$ | 0.0000 | 0.0 | 0.0 | 2.7092 |
| $b_0$ | 0.0327 | 0.0005 | 0.3225 | 2.18446021 |
| R (Å) | 3.4 | 2.4 | 2.4 | 1.7 |
| S (Å) | 3.7 | 2.7 | 2.7 | 2.1 |
| $LP_0$ | 0.0 | -0.03 | 0.0 | 0.0 |
| $LP_1$ | -0.00687 | 1.0 | 0.0 | 0.0 |
| $LP_2$ | -0.00446 | 1.0 | 0.0 | 0.0 |
| $LP_3$ | -0.04001 | 1.6 | 0.0 | 0.0 |
| $LP_4$ | -0.01405 | 0.0 | 0.0 | 0.0 |
| $LP_5$ | 0.003549 | 0.0 | 0.0 | 0.0 |

Table A-3. COMB3 pairwise and bond-order parameters for AlN.

| Parameters | Al-Al | Al-N | N-Al | N-N |
|---|---|---|---|---|
| $A_{ij}(eV)$ | 473.3379 | 1677.4790 | 1677.4790 | 7654.97300 |
| $B_{ij}^{1}(eV)$ | 242.8881 | 241.88890 | 241.88890 | 2102.2960 |
| $B_{ij}^{2}(eV)$ | 0.0000 | 0.0 | 0.0 | 0.0 |
| $B_{ij}^{3}(eV)$ | 0.0000 | 0.0 | 0.0 | 0.0 |
| $\lambda$ (Å$^{-1}$) | 1.8617 | 3.56577 | 3.56577 | 5.21803 |
| $\alpha_{ij}^{1}$ (Å$^{-1}$) | 1.4993 | 1.81353 | 1.81353 | 3.73854 |
| $\alpha_{ij}^{2}$ (Å$^{-1}$) | 1.1 | 0.0 | 0.0 | 0.0 |
| $\alpha_{ij}^{3}$ (Å$^{-1}$) | 0.6858 | 0.0 | 0.0 | 0.0 |
| $\beta$ (Å$^{-1}$) | 1.4993 | 1.16517 | 0.0 | 3.738549 |
| $b_6$ | 0.0000 | 0.0 | 0.0 | -1.0257 |
| $b_5$ | 0.0000 | 0.0 | 0.0 | -0.1700 |
| $b_4$ | 0.0000 | 0.0 | 0.0 | 2.0153 |
| $b_3$ | 0.0000 | 0.0 | 0.0 | -2.2384 |
| $b_2$ | 0.0000 | 0.0 | 0.0 | 1.1852 |
| $b_1$ | 0.0000 | 0.0 | 0.0 | 2.7092 |
| $b_0$ | 0.0327 | 0.074639 | 1.5931860 | 1.6913250 |
| R (Å) | 3.4 | 2.56 | 2.56 | 2.0 |
| S (Å) | 3.7 | 2.96 | 2.96 | 2.3 |
| $LP_0$ | 0.0 | -0.03 | 0.0 | 0.0 |
| $LP_1$ | -0.00687 | 1.1 | 0.0 | 0.0 |
| $LP_2$ | -0.00446 | 1.0 | 0.0 | 0.0 |
| $LP_3$ | -0.04001 | 0.0 | 0.0 | 0.0 |
| $LP_4$ | -0.01405 | 0.16 | 0.2 | 0.0 |
| $LP_5$ | 0.003549 | 0.0 | 1.1 | 0.0 |

LIST OF REFERENCES

[1] W. D. Callister and D. G. Rethwisch, Materials science and engineering: an introduction (Wiley New York, 2007), Vol. 7.

[2] M. Finnis, Journal of Physics: Condensed Matter **8**, 5811 (1996).

[3] M. Segall, P. J. Lindan, M. a. Probert, C. Pickard, P. Hasnip, S. Clark, and M. Payne, Journal of Physics: Condensed Matter **14**, 2717 (2002).

[4] J. Hafner, Journal of computational chemistry **29**, 2044 (2008).

[5] W. K. Liu, E. G. Karpov, and H. S. Park, Nano mechanics and materials: theory, multiscale methods and applications (John Wiley & Sons, 2006).

[6] A. Dowling et al., London: The Royal Society & The Royal Academy of Engineering Report, 61 (2004).

[7] J. Thijssen, Computational physics (Cambridge University Press, 2007).

[8] R. P. Feynman, Engineering and science **23**, 22 (1960).

[9] S. O. Kasap, Principles of electronic materials and devices (McGraw-Hill New York, NY, 2006), Vol. 81.

[10] N. W. Ashcroft and N. D. Mermin, There is no corresponding record for this reference (2005).

[11] D. Tildesley and M. Allen, Clarendon, Oxford (1987).

[12] F. Jensen, Introduction to computational chemistry (John Wiley & Sons, 2013).

[13] C. R. Brundle, C. A. Evans, and S. Wilson, Encyclopedia of materials characterization: surfaces, interfaces, thin films (Gulf Professional Publishing, 1992).

[14] F. H. Allen, Acta Crystallographica Section B: Structural Science **58**, 380 (2002).

[15] R. T. Downs and M. Hall-Wallace, American Mineralogist **88**, 247 (2003).

[16] S. P. Ong, S. Cholia, A. Jain, M. Brafman, D. Gunter, G. Ceder, and K. A. Persson, Computational Materials Science **97**, 209 (2015).

[17] S. Gražulis et al., Journal of Applied Crystallography **42**, 726 (2009).

[18] Y. Xu, M. Yamazaki, and P. Villars, Japanese Journal of Applied Physics **50**, 11RH02 (2011).

[19] S. Plimpton, Journal of Computational Physics **117**, 1 (1995).

[20] D. S. a. J. A. Steckel, Density Functional Theory: A Practical Introduction ( Wiley-Interscience, 2009).

[21] M. Marder, Nature **386**, 219 (1997).

[22] D. W. Brenner, O. A. Shenderova, J. A. Harrison, S. J. Stuart, B. Ni, and S. B. Sinnott, Journal of Physics: Condensed Matter **14**, 783 (2002).

[23] T. W. Kemper and S. B. Sinnott, The Journal of Physical Chemistry C **115**, 23936 (2011).

[24] R. Sanderson, Science **114**, 670 (1951).

[25] A. Zur and T. McGill, Journal of applied physics **55**, 378 (1984).

[26] M. Gajdoš, K. Hummer, G. Kresse, J. Furthmüller, and F. Bechstedt, Physical Review B **73**, 045112 (2006).

[27] J. Pan, C. Leygraf, D. Thierry, and A. M. Ektessabi, Journal of Biomedical Materials Research **35**, 309 (1997).

[28] B. Adhikari, D. De, and S. Maiti, Progress in Polymer Science **25**, 909 (2000).

[29] A. M. N. Ch S. S. R. Kumar, Resonance **2(4)**, 55 (1997).

[30] I. Utke, P. Hoffmann, and J. Melngailis, Journal of Vacuum Science & Technology B **26**, 1197 (2008).

[31] B. Adhikari and S. Majumdar, Progress in Polymer Science **29**, 699 (2004).

[32] U. Lange, N. V. Roznyatouskaya, and V. M. Mirsky, Analytica Chimica Acta **614**, 1 (2008).

[33] S. F. Y. P. Wu, Y. Jang, R. Holze, Journal of Power Sources **108**, 245 (2002).

[34] X. Ji, K. T. Lee, and L. F. Nazar, Nature Materials **8**, 500 (2009).

[35] J. Q. Sun, I. Bello, S. Bederka, W. M. Lau, and Z. D. Lin, J. Vac. Sci. Technol. A-Vac. Surf. Films **14**, 1382 (1996).

[36] J. Roncali, Chemical Reviews **92**, 711 (1992).

[37] R. D. McCullough, Advanced Materials **10**, 93 (1998).

[38] I. Jang, B. Ni, and S. B. Sinnott, J. Vac. Sci. Technol. A-Vac. Surf. Films **20**, 564 (2002).

[39] I. Jang, R. Phillips, and S. B. Sinnott, Journal of Applied Physics **92**, 3363 (2002).

[40] I. Jang and S. B. Sinnott, Applied Physics Letters **84**, 5118 (2004).

[41] F. A. Akin et al., Journal of Physical Chemistry B **108**, 9656 (2004).

[42] C. Y. Huang, H. M. Ku, and S. Chao, Applied Optics **48**, 69 (2009).

[43] W. D. Hsu, I. Jang, and S. B. Sinnott, Chemistry of Materials **18**, 914 (2006).

[44] C. F. Wen-Dung Hsu, Sharon Pregler , Susan B Sinnott, Journal of Physical Chemistry. C **113**, 17860 (2009).

[45] D. J. T. M. P. Allen (1986).

[46] I. K. Jang and S. B. Sinnott, Journal of Physical Chemistry B **108**, 18993 (2004).

[47] E. T. Ada, O. Kornienko, and L. Hanley, Journal of Physical Chemistry B **102**, 3959 (1998).

[48] L. Hanley, Y. S. Choi, E. R. Fuoco, F. A. Akin, M. B. J. Wijesundara, M. Li, T. B. Aleksey, and M. Schlossman, Nuclear Instruments & Methods in Physics Research Section B-Beam Interactions with Materials and Atoms **203**, 116 (2003).

[49] Y. Karade, S. A. Pihan, W. H. Bruenger, A. Dietzel, R. Berger, and K. Graf, Langmuir **25**, 3108 (2009).

[50] D. W. Brenner, O. A. Shenderova, J. A. Harrison, S. J. Stuart, B. Ni, and S. B. Sinnott, Journal of Physics-Condensed Matter **14**, 783, Pii s0953-8984(02)31186-x (2002).

[51] T. W. Kemper and S. B. Sinnott, Journal of Physical Chemistry C **115**, 23936 (2011).

[52] S. J. V. Frankland and D. W. Brenner, Chemical Physics Letters **334**, 18 (2001).

[53] H. Paul, Nuclear Instruments & Methods in Physics Research Section B-Beam Interactions with Materials and Atoms **247**, 166 (2006).

[54] J. T. Padding and W. J. Briels, Journal of Chemical Physics **117**, 925 (2002).

[55] C. Pastorino, T. Kreer, M. Mueller, and K. Binder, Physical Review E **76**, 026706 (2007).

[56] D. J. Tobias, G. J. Martyna, and M. L. Klein, Journal of Physical Chemistry **97**, 12959 (1993).

[57] Z.-H. Hong, S.-F. Hwang, and T.-H. Fang, Computational Materials Science **48**, 520 (2010).

[58] G. F. Meyers, B. M. DeKoven, and J. T. Seitz, Langmuir **8**, 2330 (1992).

[59] J. J. Vegh et al., Journal of Applied Physics **104** (2008).

[60] S. Tepavcevic, Y. Choi, and L. Hanley, Journal of the American Chemical Society **125**, 2396 (2003).

[61] Y. Ikada, Biomaterials **15**, 725 (1994).

[62] C. Oehr, Nuclear Instruments and Methods in Physics Research Section B: Beam Interactions with Materials and Atoms **208**, 40 (2003).

[63] P. Favia and R. d'Agostino, Surface & Coatings Technology **98**, 1102 (1998).

[64] W. Jacob, Thin Solid Films **326**, 1 (1998).

[65] S. Muhl and J. M. Mendez, Diam. Relat. Mat. **8**, 1809 (1999).

[66] A. Anders, Surface & Coatings Technology **93**, 158 (1997).

[67] T. Schwarz-Selinger, A. von Keudell, and W. Jacob, J. Appl. Phys. **86**, 3988 (1999).

[68] K. Ichiki, S. Ninomiya, Y. Nakata, H. Yamada, T. Seki, T. Aoki, and J. Matsuo, Surface and Interface Analysis **43**, 120 (2011).

[69] L. Hanley, E. Fuoco, M. B. J. Wijesundara, A. J. Beck, P. N. Brookes, and R. D. Short, Journal of Vacuum Science & Technology a-Vacuum Surfaces and Films **19**, 1531 (2001).

[70] A. von Keudell, Thin Solid Films **402**, 1 (2002).

[71] A. Anderson and W. R. Ashurst, Langmuir **25**, 11541 (2009).

[72] E. M. Liston, L. Martinu, and M. R. Wertheimer, J. Adhes. Sci. Technol. **7**, 1091 (1993).

[73] K. L. Choy, Prog. Mater. Sci. **48**, 57 (2003).

[74] K. Upadhya and T. C. Tiearney, Jom-Journal of the Minerals Metals & Materials Society **41**, 6 (1989).

[75] J. L. V. Maurice H. Francombe Academic Press (1994).

[76] E. M. Bringa and R. E. Johnson, Nucl. Instrum. Methods Phys. Res. Sect. B-Beam Interact. Mater. Atoms **180**, 99 (2001).

[77] E. Salonen, K. Nordlund, J. Tarus, T. Ahlgren, J. Keinonen, and C. H. Wu, Physical Review B **60**, 14005 (1999).

[78] C. O. Reinhold, P. S. Krstic, and S. J. Stuart, Nucl. Instrum. Methods Phys. Res. Sect. B-Beam Interact. Mater. Atoms **267**, 691 (2009).

[79] Y. Lemeur, F. Meyer, C. Pellet, C. Schwebel, P. Moller, A. Buxbaum, A. Raizman, and M. Eizenberg, Thin Solid Films **222**, 180 (1992).

[80] D. J. Li, F. Z. Cui, H. Q. Gu, and W. Z. Li, Vacuum **56**, 205 (2000).

[81] M. B. J. Wijesundara, E. Fuoco, and L. Hanley, Langmuir **17**, 5721 (2001).

[82] E. R. Fuoco, G. Gillen, M. B. J. Wijesundara, W. E. Wallace, and L. Hanley, Journal of Physical Chemistry B **105**, 3950 (2001).

[83] Y. T. Su, T. R. Shan, and S. B. Sinnott, Nucl. Instrum. Methods Phys. Res. Sect. B-Beam Interact. Mater. Atoms **267**, 2525 (2009).

[84] M. Prasad, P. F. Conforti, B. J. Garrison, and Y. G. Yingling, Applied Surface Science **253**, 6382 (2007).

[85] P. F. Conforti, M. Prasad, and B. J. Garrison, Applied Surface Science **253**, 6386 (2007).

[86] P. F. Conforti, M. Prasad, and B. J. Garrison, Journal of Physical Chemistry C **111**, 12024 (2007).

[87] A. Licciardello, M. E. Fragala, G. Foti, G. Compagnini, and O. Puglisi, Nucl. Instrum. Methods Phys. Res. Sect. B-Beam Interact. Mater. Atoms **116**, 168 (1996).

[88] H. Ryssel, K. Haberger, and H. Kranz, Journal of Vacuum Science & Technology **19**, 1358 (1981).

[89] P. F. Conforti, M. Prasad, and B. J. Garrison, Accounts of Chemical Research **41**, 915 (2008).

[90] C. C. Tian and C. R. Vidal, Journal of Physics B-Atomic Molecular and Optical Physics **31**, 895 (1998).

[91] J. Benedikt, Journal of Physics D-Applied Physics **43** (2010).

[92] J. Polvi, P. Luukkonen, K. Nordlund, T. T. Jarvi, T. W. Kemper, and S. B. Sinnott, Journal of Physical Chemistry B **116**, 13932 (2012).

[93] E. M. Bringa, M. Jakas, and R. E. Johnson, Nucl. Instrum. Methods Phys. Res. Sect. B-Beam Interact. Mater. Atoms **164**, 762 (2000).

[94] J. J. Vegh and D. B. Graves, Plasma Processes and Polymers **6**, 320 (2009).

[95] M. P. Allen Computer Simulation of Liquids (1986).

[96] M. J. Donachie, ASM International; 2 edition (August 1, 2002).

[97] M. S. Daw and M. I. Baskes, Physical Review B **29**, 6443 (1984).

[98] H. W. Sheng, M. J. Kramer, A. Cadien, T. Fujita, and M. W. Chen, Physical Review B **83**, 134118 (2011).

[99] S. B. Sinnott and D. W. Brenner, MRS Bulletin **37**, 469 (2012).

[100] F. Ercolessi and J. B. Adams, Europhysics Letters **26**, 583 (1994).

[101] A. F. V. a. S. P. Chen, MRS Symposia Proceedings No. 82 (Materials Research Society, Pittsburgh (1987).

[102] J. M. Winey, K. Alison, and Y. M. Gupta, Modelling and Simulation in Materials Science and Engineering **17**, 055004 (2009).

[103] M. I. Mendelev, M. J. Kramer, C. A. Becker, and M. Asta, Philosophical Magazine **88**, 1723 (2008).

[104] D. J. Oh and R. A. Johnson, Journal of Materials Research **3**, 471 (1988).

[105] C. L. Rohrer, Modelling and Simulation in Materials Science and Engineering **2**, 119 (1994).

[106] Y. Mishin, D. Farkas, M. J. Mehl, and D. A. Papaconstantopoulos, Physical Review B **59**, 3393 (1999).

[107] L. Xiang-Yang, E. Furio, and B. A. James, Modelling and Simulation in Materials Science and Engineering **12**, 665 (2004).

[108] B. Jelinek, S. Groh, M. F. Horstemeyer, J. Houze, S. G. Kim, G. J. Wagner, A. Moitra, and M. I. Baskes, Physical Review B **85**, 245102 (2012).

[109] F. Apostol and Y. Mishin, Physical Review B **83**, 054116 (2011).

[110] X. Y. Liu, C. L. Liu, and L. J. Borucki, Acta Materialia **47**, 3227 (1999).

[111] M. I. Mendelev, D. J. Srolovitz, G. Ackland, and S. Han, Journal of Biomedical Materials Research Part A **20**, 208 (2005).

[112] M. I. Mendelev, M. Asta, M. J. Rahman, and J. J. Hoyt, Philosophical Magazine **89**, 3269 (2009).

[113] D. Schopf, P. Brommer, B. Frigan, and H.-R. Trebin, Physical Review B **85**, 054201 (2012).

[114] R. R. Zope and Y. Mishin, Physical Review B **68**, 024102 (2003).

[115] J. E. Angelo, N. R. Moody, and M. I. Baskes, Modelling and Simulation in Materials Science and Engineering **3**, 289 (1995).

[116] B.-J. Lee, W.-S. Ko, H.-K. Kim, and E.-H. Kim, Calphad **34**, 510 (2010).

[117] W. Sekkal and A. Zaoui, Physica B: Condensed Matter **404**, 335 (2009).

[118] F. H. Streitz and J. W. Mintmire, Physical Review B **50**, 11996 (1994).

[119] Q. Zhang, T. Çağın, A. van Duin, W. A. Goddard, Y. Qi, and L. G. Hector, Physical Review B **69**, 045423 (2004).

[120] G. Kresse and J. Hafner, Physical Review B **47**, 558 (1993).

[121] G. Kresse and J. Hafner, Physical Review B **49**, 14251 (1994).

[122] G. Kresse and J. Furthmüller, Computational Materials Science **6**, 15 (1996).

[123] G. Kresse and J. Furthmüller, Physical Review B **54**, 11169 (1996).

[124] J. A. Martinez, D. E. Yilmaz, T. Liang, S. B. Sinnott, and S. R. Phillpot, Current Opinion in Solid State and Materials Science **17**, 263 (2013).

[125] H. T. C.R. Rao, A. Fieger, C. Heumann, T. Nittner and S. Scheid Springer Series in Statistics (1999).

[126] C. Kittel, Wiley-Interscience,New York (1986).

[127] R. C. Weast, CRC, Boca Raton, FL (1984).

[128] G. S. a. H. Wang, MIT Press, Cambridge, MA (1977).

[129] Y. Mishin, D. Farkas, M. J. Mehl, and D. A. Papaconstantopoulos, Physical Review B **59**, 3393 (1999).

[130] W. R. Tyson and W. A. Miller, Surface Science **62**, 267 (1977).

[131] A. M. Rodríguez, G. Bozzolo, and J. Ferrante, Surface Science **289**, 100 (1993).

[132] F. L. Tang, X. G. Cheng, W. J. Lu, and W. Y. Yu, Physica B-Condensed Matter **405**, 1248 (2010).

[133] L. E. Murr, (Addison-Wesley, Reading, MA, 1975).

[134] R. H. Rautioaho, Phys. Status Solidi B **112**, 83 (1982).

[135] K. H. W. a. R. L. Peck, Philos. Mag. **23**, 611 (1971).

[136] G. Lu, N. Kioussis, V. V. Bulatov, and E. Kaxiras, Physical Review B **62**, 3099 (2000).

[137] N. M. Rosengaard and H. L. Skriver, Physical Review B **47**, 12865 (1993).

[138] S. Kibey, J. B. Liu, D. D. Johnson, and H. Sehitoglu, Acta Materialia **55**, 6843 (2007).

[139] H. E. Schaefer, physica status solidi (a) **102**, 47 (1987).

[140] A. Devita and M. J. Gillan, Journal of Physics-Condensed Matter **3**, 6225 (1991).

[141] J. Yu, S. B. Sinnott, and S. R. Phillpot, Philosophical Magazine Letters **89**, 136 (2009).

[142] M. Jahnátek, J. Hafner, and M. Krajčí, Physical Review B **79**, 224103 (2009).

[143] V. Yamakov, D. Wolf, S. R. Phillpot, and H. Gleiter, Acta Materialia **51**, 4135 (2003).

[144] J. Schiotz, F. D. Di Tolla, and K. W. Jacobsen, Nature **391**, 561 (1998).

[145] H. Van Swygenhoven, P. M. Derlet, and A. G. Froseth, Nature Materials **3**, 399 (2004).

[146] X. Z. Liao, F. Zhou, E. J. Lavernia, S. G. Srinivasan, M. I. Baskes, D. W. He, and Y. T. Zhu, Applied Physics Letters **83**, 632 (2003).

[147] K. M. H. a. J. L. Olson, Springer, Berlin **13** (1981).

[148] S. J. Plimpton and A. P. Thompson, MRS Bulletin **37**, 513 (2012).

[149] E. D. a. H. Hubner, Alumina (Springer-Verlag, 1984).

[150] A. Cox et al., The Surface Science of Metal Oxides (Cambridge University Press, New York,, 1994).

[151] M. Ksiazek, N. Sobczak, B. Mikulowski, W. Radziwill, and I. Surowiak, Materials Science and Engineering: A **324**, 162 (2002).

[152] Y. Yoshino and T. Shibata, Journal of the American Ceramic Society **75**, 2756 (1992).

[153] W. Zhang, J. Smith, X.-G. Wang, and A. Evans, Physical Review B **67**, 245414 (2003).

[154] C. N. R. Rao, F. L. Deepak, G. Gundiah, and A. Govindaraj, Progress in Solid State Chemistry **31**, 5 (2003).

[155] G. Pilania, B. J. Thijsse, R. G. Hoagland, I. Lazic, S. M. Valone, and X.-Y. Liu, Scientific Reports **4**, 4485 (2014).

[156] A. T. Alpas, J. D. Embury, D. A. Hardwick, and R. W. Springer, Journal of Materials Science **25**, 1603 (1990).

[157] A. H. Graham, C. R. Bowen, J. Robbins, and J. Taylor, Sensors and Actuators B: Chemical **138**, 296 (2009).

[158] M.-G. Ma, Y.-J. Zhu, and Z.-L. Xu, Materials Letters **61**, 1812 (2007).

[159] Y. Wang, A. Santos, G. Kaur, A. Evdokiou, and D. Losic, Biomaterials **35**, 5517 (2014).

[160] D. Medlin, K. McCarty, R. Hwang, S. Guthrie, and M. Baskes, Thin Solid Films **299**, 110 (1997).

[161] P. Vashishta, R. K. Kalia, A. Nakano, and J. P. Rino, Journal of Applied Physics **103**, 083504 (2008).

[162] M. W. Finnis, Journal of Physics-Condensed Matter **8**, 5811 (1996).

[163] H. Momida, T. Hamada, Y. Takagi, T. Yamamoto, T. Uda, and T. Ohno, Physical Review B **73**, 054108 (2006).

[164] Y. H. Li et al., Chemical Physics Letters **350**, 412 (2001).

[165] D. J. Siegel, L. G. Hector, and J. B. Adams, MRS Online Proceedings Library **654**, AA4.2.1 (2000).

[166] D. J. Siegel, L. G. Hector Jr, and J. B. Adams, Physical Review B **65**, 085415 (2002).

[167] W. Zhang and J. R. Smith, Physical Review Letters **85**, 3225 (2000).

[168] J. Kang, J. Zhu, C. Curtis, D. Blake, G. Glatzmaier, Y.-H. Kim, and S.-H. Wei, Physical review letters **108**, 226105 (2012).

[169] Y. K. Kaoru Ohno et al., Computational Materials Science: From Ab Initio to Monte Carlo Methods (Springer, Berlin, 1999), Springer Series in Solid-State Sciences

[170] J. Gale, C. Catlow, and W. Mackrodt, Modelling and simulation in materials science and engineering **1**, 73 (1992).

[171] M. W. Finnis and J. E. Sinclair, Philosophical Magazine a-Physics of Condensed Matter Structure Defects and Mechanical Properties **50**, 45 (1984).

[172] S. Alavi, J. W. Mintmire, and D. L. Thompson, The Journal of Physical Chemistry B **109**, 209 (2005).

[173] T. Campbell, R. K. Kalia, A. Nakano, P. Vashishta, S. Ogata, and S. Rodgers, Physical Review Letters **82**, 4866 (1999).

[174] X. Zhou, H. Wadley, J.-S. Filhol, and M. Neurock, Physical Review B **69**, 035402 (2004).

[175] I. Lazić and B. J. Thijsse, Computational Materials Science **53**, 483 (2012).

[176] G. Pilania, B. J. Thijsse, R. G. Hoagland, I. Lazic, S. M. Valone, and X.-Y. Liu, Sci. Rep. **4** (2014).

[177] F. G. Sen, Y. Qi, A. C. van Duin, and A. T. Alpas, Applied Physics Letters **102**, 051912 (2013).

[178] T. Liang, T.-R. Shan, Y.-T. Cheng, B. D. Devine, M. Noordhoek, Y. Li, Z. Lu, S. R. Phillpot, and S. B. Sinnott, Materials Science and Engineering: R: Reports **74**, 255 (2013).

[179] S. Plimpton, Journal of computational physics **117**, 1 (1995).

[180] Y. S. Hideo Aoki, Russell J. Hemley, Physics Meets Mineralogy: Condensed Matter Physics in the Geosciences ( Cambridge University Press, Cambridge, 2000), pp. 354.

[181] J. Sarsam, M. W. Finnis, and P. Tangney, The Journal of Chemical Physics **139** (2013).

[182] W. H. Gitzen, Alumina as a Ceramic Material (The American Ceramic Society, Columbus, OH, 1970).

[183] D. R. Lide, CRC Handbook of Chemistry and Physics (CRC Press, Boca Raton,FL, 2005).

[184] B. Holm, R. Ahuja, Y. Yourdshahyan, B. Johansson, and B. I. Lundqvist, Physical Review B **59**, 12777 (1999).

[185] N. D. M. Hine, K. Frensch, W. M. C. Foulkes, and M. W. Finnis, Physical Review B **79**, 024112 (2009).

[186] X.-G. Wang, A. Chaka, and M. Scheffler, Physical Review Letters **84**, 3650 (2000).

[187] I. G. Batirev, A. Alavi, M. W. Finnis, and T. Deutsch, Physical Review Letters **82**, 1510 (1999).

[188] J. M. McHale, A. Auroux, A. J. Perrotta, and A. Navrotsky, Science **277**, 788 (1997).

[189] M. Causà, R. Dovesi, C. Pisani, and C. Roetti, Surface Science **215**, 259 (1989).

[190] M. Baudin and K. Hermansson, Surface Science **474**, 107 (2001).

[191] J. Sun, T. Stirner, and A. Matthews, Surface & Coatings Technology **201**, 4205 (2006).

[192] C. Ruberto, Y. Yourdshahyan, and B. I. Lundqvist, Physical Review B **67**, 195412 (2003).

[193] J. Sun, T. Stirner, and A. Matthews, Surface and Coatings Technology **201**, 4205 (2006).

[194] A. Chernatynskiy and S. R. Phillpot, Physical Review B **82**, 134301 (2010).

[195] Z. Łodziana and K. Parliński, Physical Review B **67**, 174106 (2003).

[196] B. D Bloor et al., The Encyclopedia of Advanced Materials (Pergamon press, Cambridge, 1994), Vol. 1, pp. 86.

[197] G. Stan, C. Ciobanu, T. Thayer, G. Wang, J. Creighton, K. Purushotham, L. Bendersky, and R. Cook, Nanotechnology **20**, 035706 (2009).

[198] H.-C. Hsu et al., Applied Physics Letters **101**, 121902 (2012).

[199] Y.-R. Lin and S.-T. Wu, Surface science **516**, L535 (2002).

[200] C. M. Montesa, N. Shibata, T. Tohei, and Y. Ikuhara, Journal of materials science **46**, 4392 (2011).

[201] Y. Tokumoto, Y. Sato, T. Yamamoto, N. Shibata, and Y. Ikuhara, Journal of materials science **41**, 2553 (2006).

[202] S. Strite and H. Morkoç, Journal of Vacuum Science & Technology B **10**, 1237 (1992).

[203] R. Vispute, H. Wu, and J. Narayan, Applied physics letters **67**, 1549 (1995).

[204] J. Hafner, C. Wolverton, and G. Ceder, MRS bulletin **31**, 659 (2006).

[205] E. Wimmer, Journal of Computer-Aided Materials Design **1**, 215 (1994).

[206] C. J. Cramer and F. Bickelhaupt, Angewandte chemie-international edition in english **42**, 381 (2003).

[207] R. Di Felice and J. E. Northrup, Applied physics letters **73**, 936 (1998).

[208] S. Ogata and H. Kitagawa, Computational materials science **15**, 435 (1999).

[209] Y. Li, Z. Zhou, P. Shen, S. Zhang, and Z. Chen, Nanotechnology **20**, 215701 (2009).

[210] J. Chisholm, D. Lewis, and P. Bristowe, Journal of Physics: Condensed Matter **11**, L235 (1999).

[211] M. Tungare, Y. Shi, N. Tripathi, P. Suvarna, and F. S. Shahedipour-Sandvik, physica status solidi (a) **208**, 1569 (2011).

[212] F. Benkabou, H. Aourag, P. J. Becker, and M. Certier, Molecular Simulation **23**, 327 (2000).

[213] D. Powell, M. Migliorato, and A. Cullis, Physical Review B **75**, 115202 (2007).

[214] J. W. Kang and H. J. Hwang, Computational materials science **31**, 237 (2004).
[215] P. Vashishta, R. K. Kalia, A. Nakano, and J. P. Rino, Journal of Applied Physics **109**, 033514 (2011).

[216] W. W. Tipton and R. G. Hennig, Journal of Physics: Condensed Matter **25**, 495401 (2013).

[217] A. Jain et al., APL Materials **1**, 011002 (2013).

[218] M. M. Islam, A. Ostadhossein, O. Borodin, A. T. Yeates, W. W. Tipton, R. G. Hennig, N. Kumar, and A. C. van Duin, Physical Chemistry Chemical Physics **17**, 3383 (2015).

[219] A. Zoroddu, F. Bernardini, P. Ruggerone, and V. Fiorentini, Physical Review B **64**, 045208 (2001).

[220] E. Ruiz, S. Alvarez, and P. Alemany, Physical Review B **49**, 7115 (1994).

[221] A. Hung, S. P. Russo, D. G. McCulloch, and S. Prawer, The Journal of chemical physics **120**, 4890 (2004).

[222] D. Holec and P. H. Mayrhofer, Scripta materialia **67**, 760 (2012).

[223] J. Sun, T. Stirner, and A. Matthews, Surface and Coatings Technology **201**, 4205 (2006).

[224] W. Zhang and J. Smith, Physical review letters **85**, 3225 (2000).

[225] G. Pilania, B. J. Thijsse, R. G. Hoagland, I. Lazić, S. M. Valone, and X.-Y. Liu, Scientific reports **4** (2014).

[226] S. H. Oh, M. F. Chisholm, Y. Kauffmann, W. D. Kaplan, W. Luo, M. Rühle, and C. Scheu, Science **330**, 489 (2010).

[227] H. T. Li, L. F. Chen, X. Yuan, W. Q. Zhang, J. R. Smith, and A. G. Evans, Journal of the American Ceramic Society **94**, s154 (2011).

[228] J. R. Smith, T. Hong, and D. J. Srolovitz, Physical review letters **72**, 4021 (1994).

[229] R. Couturier, D. Ducret, P. Merle, J. Disson, and P. Joubert, Journal of the European Ceramic Society **17**, 1861 (1997).

[230] S. Ogata and H. Kitagawa, Japan Institute of Metals, Journal **60**, 1079 (1996).

[231] K. Choudhary, T. Liang, A. Chernatynskiy, S. R. Phillpot, and S. B. Sinnott, Journal of Physics: Condensed Matter **27**, 305004 (2015).

[232] C. Untiedt, G. Rubio, S. Vieira, and N. Agraït, Physical Review B **56**, 2154 (1997).

[233] J. J. Wang, F. Walters, X. Liu, P. Sciortino, and X. Deng, Applied physics letters **90**, 061104 (2007).

[234] Q. Zhao, X. Xu, H. Zhang, Y. Chen, J. Xu, and D. Yu, Applied Physics A **79**, 1721 (2004).

[235] Z. L. Xiao et al., Nano Letters **2**, 1293 (2002).

[236] M. R. Sørensen, M. Brandbyge, and K. W. Jacobsen, Physical Review B **57**, 3283 (1998).

[237] L. Pastor-Abia, M. Caturla, E. SanFabian, G. Chiappe, and E. Louis, Physical Review B **83**, 165441 (2011).

[238] Z.-J. Wang, Q.-J. Li, Z.-W. Shan, J. Li, J. Sun, and E. Ma, Applied Physics Letters **100**, 071906 (2012).

[239] H. Wu, European Journal of Mechanics-A/Solids **25**, 370 (2006).

[240] H. S. Park and J. A. Zimmerman, Physical Review B **72**, 054106 (2005).

[241] L. Pastor-Abia, M. Caturla, E. SanFabian, G. Chiappe, and E. Louis, physica status solidi (c) **6**, 2119 (2009).

[242] A. A. Griffith, Philosophical transactions of the royal society of london. Series A , 163 (1921).

[243] I. S. Molchan, T. V. Molchan, N. V. Gaponenko, P. Skeldon, and G. E. Thompson, Electrochemistry Communications **12**, 693 (2010).

[244] S. Kulkova, S. Eremeev, S. Hocker, and S. Schmauder, Physics of the Solid State **52**, 2589 (2010).

[245] Y. Jing and Q. Meng, Physica B: Condensed Matter **405**, 2413 (2010).

[246] J. Guénolé, J. Godet, and S. Brochard, Physical Review B **87**, 045201 (2013).

[247] L. Dai, C. Sow, C. Lim, W. Cheong, and V. Tan, Nano letters **9**, 576 (2009).

[248] N. Wei, T. Lu, F. Li, W. Zhang, B. Ma, Z. Lu, and J. Qi, Applied Physics Letters **101**, 061902 (2012).

[249] I. Shoji, S. Kurimura, Y. Sato, T. Taira, A. Ikesue, and K. Yoshida, Applied Physics Letters **77**, 939 (2000).

[250] D. Haranath, H. Chander, P. Sharma, and S. Singh, Applied physics letters **89**, 173118 (2006).

[251] S. Kobyakov, A. Kaminska, A. Suchocki, D. Galanciak, and M. Malinowski, Applied physics letters **88**, 234102 (2006).

[252] J. X. Meng, K. W. Cheah, Z. P. Shi, and J. Q. Li, Applied Physics Letters **91**, 151107 (2007).

[253] G. Zhu, X. Wang, H. Li, L. Pan, H. Sun, X. Liu, T. Lv, and Z. Sun, Chemical Communications **48**, 958 (2012).

[254] Q. Wang, L. Su, H. Li, L. Zheng, X. Xu, H. Tang, X. Guo, D. Jiang, and J. Xu, physica status solidi (a) **208**, 2839 (2011).

[255] M. Pokhrel, G. Kumar, P. Samuel, K. Ueda, T. Yanagitani, H. Yagi, and D. Sardar, Optical Materials Express **1**, 1272 (2011).

[256] A. B. Munoz-Garcia, E. Anglada, and L. Seijo, International Journal of Quantum Chemistry **109**, 1991 (2009).

[257] A. Belen Munoz-Garcia and L. Seijo, Journal of Physical Chemistry A **115**, 815 (2011).

[258] W. Ching, Y. N. Xu, and B. Brickeen, Applied physics letters **74**, 3755 (1999).

[259] A. J. Cohen, P. Mori-Sanchez, and W. Yang, Science **321**, 792 (2008).

[260] J. Graciani, A. M. Márquez, J. J. Plata, Y. Ortega, N. C. Hernández, A. Meyer, C. M. Zicovich-Wilson, and J. F. Sanz, Journal of Chemical Theory and Computation **7**, 56 (2010).

[261] F. F. Tetsuhiko Tomiki, Minoru Kaminao, Masami Fujisawa, Yoshikazu Tanahara and Tomoyoshi Futemma, J. Phys. Soc. Jpn. **58**, 1801 (1989).

[262] Y. G. Tetsuhiko Tomiki, Tohru Shikenbaru, Tomoyoshi Futemma, Masatada Yuri, Yoshihiro Aiura, Hirohito Fukutani, Hiroo Kato, Junkoh Tamashiro, Ysuneaki Miyahara and Akira Yonesu, Journal of the Physical Society of Japan **62**, 1388 (1993).

[263] H. Guo, M. Zhang, J. Han, H. Zhang, and N. Song, Physica B: Condensed Matter **407**, 2262 (2012).

[264] S. Jiang, J. Chen, Y. Long, and T. Lu, Journal of the American Ceramic Society **95**, 3894 (2012).

[265] Z. Huang, J. Feng, and W. Pan, Solid State Sciences **14**, 1327 (2012).

[266] Y. Zhang, G. Kresse, and C. Wolverton, Physical review letters **112**, 075502 (2014).

[267] J. Paier, M. Marsman, K. Hummer, G. Kresse, I. C. Gerber, and J. G. Ángyán, The Journal of chemical physics **124**, 154709 (2006).

[268] M. Shishkin, M. Marsman, and G. Kresse, Physical review letters **99**, 246403 (2007).

[269] J. Heyd, G. E. Scuseria, and M. Ernzerhof, Journal of Chemical Physics **124** (2006).

[270] J. P. Perdew, M. Ernzerhof, and K. Burke, The Journal of Chemical Physics **105**, 9982 (1996).

[271] J. Heyd and G. E. Scuseria, The Journal of Chemical Physics **121**, 1187 (2004).

[272] J. Heyd, J. E. Peralta, G. E. Scuseria, and R. L. Martin, The Journal of Chemical Physics **123** (2005).

[273] L. Zhang, H. Yang, P. D. Han, L. X. Wang, and Q. T. Zhang, in Materials Science Forum (Trans Tech Publ, 2011), pp. 129.

[274] P. E. Blöchl, Physical Review B **50**, 17953 (1994).

[275] C. Varney, D. Mackay, A. Pratt, S. Reda, and F. Selim, Journal of Applied Physics **111**, 063505 (2012).

[276] Y.-D. Huh, Y.-S. Cho, and Y. R. Do, Bull. Korean Chem. Soc **23**, 1435 (2002).

[277] J. P. Perdew, K. Burke, and M. Ernzerhof, Physical Review Letters **77**, 3865 (1996).

[278] A. D. Becke, The Journal of Chemical Physics **98**, 1372 (1993).

[279] A. J. Cohen, P. Mori-Sánchez, and W. Yang, Science **321**, 792 (2008).

[280] F. Wooten, Optical properties of solids (Academic press, 2013).

[281] U. Müller, Inorganic structural chemistry (New York, 1993).

[282] L. Mezeix and D. J. Green, International Journal of Applied Ceramic Technology **3**, 166 (2006).

[283] M. A. Gülgün, W.-Y. Ching, Y.-N. Xu, and M. Rühle, Philosophical Magazine B **79**, 921 (1999).

[284] Y. I. Tetsuhiko Tomiki, Yoshie Kadekawa, Yoshiiku Ganaha, Nagao Toyokawa, Tsutomu Miyazato, Meiko Miyazato, Tetsuya Kohatsu, Hirohumi Shimabukuro and Junkoh Tamashiro, J. Phys. Soc. Jpn. **65**, 1106 (1996).

[285] T. Tomiki et al., Journal of the Physical Society of Japan **60**, 2437 (1991).

[286] C. W. Thiel, H. Cruguel, H. Wu, Y. Sun, G. J. Lapeyre, R. L. Cone, R. W. Equall, and R. M. Macfarlane, Physical Review B **64**, 085107 (2001).

[287] P. Dorenbos, R. Visser, C. Van Eijk, N. Khaidukov, and M. Korzhik, Nuclear Science, IEEE Transactions on **40**, 388 (1993).

[288] G. Henkelman, A. Arnaldsson, and H. Jónsson, Computational Materials Science **36**, 354 (2006).

[289] M. Antonietta Loi and J. C. Hummelen, Nat Mater **12**, 1087 (2013).

[290] D. A. Gidlow, Occupational Medicine-Oxford **54**, 76 (2004).

[291] A. Muntasar, D. Le Roux, and G. Denes, Journal of Radioanalytical and Nuclear Chemistry, Articles **190**, 431 (1995).

[292] K. Yamada, K. Nakada, Y. Takeuchi, K. Nawa, and Y. Yamane, Bulletin of the Chemical Society of Japan **84**, 926 (2011).

[293] D. Bi, S.-J. Moon, L. Häggman, G. Boschloo, L. Yang, E. M. Johansson, M. K. Nazeeruddin, M. Grätzel, and A. Hagfeldt, Rsc Advances **3**, 18762 (2013).

[294] G. Xing, N. Mathews, S. Sun, S. S. Lim, Y. M. Lam, M. Grätzel, S. Mhaisalkar, and T. C. Sum, Science **342**, 344 (2013).

[295] W. Shockley and H. J. Queisser, Journal of Applied Physics **32**, 510 (1961).

[296] A. Amat, E. Mosconi, E. Ronca, C. Quarti, P. Umari, M. K. Nazeeruddin, M. Grätzel, and F. De Angelis, Nano Letters **14**, 3608 (2014).

[297] W. J. Yin, T. Shi, and Y. Yan, Advanced Materials (2014).

[298] S. Kasap, Principles of Electronic Materials and Devices (McGraw-Hill, Inc., 2006).
[299] D. Weber, Z. Naturforsch. 33b, 1443 (1978).

[300] A. Kojima, K. Teshima, Y. Shirai, and T. Miyasaka, Journal of the American Chemical Society **131**, 6050 (2009).

[301] E. Mosconi, A. Amat, M. K. Nazeeruddin, M. Grätzel, and F. De Angelis, The Journal of Physical Chemistry C **117**, 13902 (2013).

[302] R. A. Jishi, O. B. Ta, and A. A. Sharif, The Journal of Physical Chemistry C (2014).

[303] J. Feng and B. Xiao, The Journal of Physical Chemistry Letters **5**, 1278 (2014).

[304] J. Feng and B. Xiao, The Journal of Physical Chemistry C **118**, 19655 (2014).

[305] H. Mashiyama, Y. Kurihara, and T. Azetsu, Journal of the Korean Physical Society **32**, S156 (1998).

[306] P. Umari, E. Mosconi, and F. De Angelis, Scientific reports **4** (2014).

[307] J. Haruyama, K. Sodeyama, L. Han, and Y. Tateyama, The Journal of Physical Chemistry Letters **5**, 2903 (2014).

[308] G. Kresse and J. Furthmuller, Physical Review B **54**, 11169 (1996).

[309] S. Grimme, Journal of computational chemistry **27**, 1787 (2006).

[310] P. J. Stephens, F. J. Devlin, C. F. Chabalowski, and M. J. Frisch, Journal of Physical Chemistry **98**, 11623 (1994).

[311] R. Braunstein and E. Kane, Journal of Physics and Chemistry of Solids **23**, 1423 (1962).

# BIOGRAPHICAL SKETCH

Kamal Choudhary was born in West Bengal, India in 1989. He was schooled at Saraswati Sishu Vidyamandir, Champdani Arya Vidyapith and Sri Aurobindo Vidyamandir. He attended Government College of Engineering and Ceramic Technology in Kolkata from 2007 to 2011 to receive his Bachelor of Technology degree in ceramic technology. During his schooling and undergraduate studies he received guidance from Pradeep Biswas, Dr. Asis Banerjee and Dr. Sergej Flach. After this, he attended the University of Florida from 2011 to 2015 in pursuit of Master of Science and Doctor of Philosophy degrees in materials science and engineering under the supervision of Dr. Susan B. Sinnott.